\documentclass[traditabstract]{aa}  

\usepackage{graphicx}
\usepackage{rotating}
\usepackage{natbib}
\usepackage{txfonts}
\usepackage{threeparttable}
\usepackage{float}
\usepackage{xcolor}
\usepackage{pdflscape}
\usepackage{caption}

\begin{document} 

\title{The CARMENES search for exoplanets around M dwarfs}
\subtitle{Telluric absorption corrected high S/N optical and near-infrared template spectra of 382 M dwarf stars\thanks{
The template library spectra are 
accessible online (\url{http://carmenes.cab.inta-csic.es}).}}

\author{
     E. Nagel\inst{\ref{inst:hs},\ref{inst:iag},\ref{inst:tls}}
\and S.~Czesla\inst{\ref{inst:tls},\ref{inst:hs}}
\and A.~Kaminski\inst{\ref{inst:lsw}}
\and M.~Zechmeister\inst{\ref{inst:iag}}
\and L.~Tal-Or\inst{\ref{inst:ariel},\ref{inst:iag}}
\and J.\,H.\,M.\,M.~ Schmitt\inst{\ref{inst:hs}}
\and A.~Reiners\inst{\ref{inst:iag}}
\and A.~Quirrenbach\inst{\ref{inst:lsw}}
\and A.~Garc\'{i}a L\'{o}pez\inst{\ref{inst:cab},\ref{inst:isdefe}}
\and J.\,A.~Caballero\inst{\ref{inst:cab}}
\and I.~Ribas\inst{\ref{inst:ice},\ref{inst:ieec}}
\and P.\,J.\,~Amado\inst{\ref{inst:iaa}}
\and V.\,J.\,S.~B\'{e}jar\inst{\ref{inst:iac},\ref{inst:ull}}
\and M.~Cort\'{e}s-Contreras\inst{\ref{inst:ucm}}
\and S.~Dreizler\inst{\ref{inst:iag}}
\and A.\,P.~Hatzes\inst{\ref{inst:tls}}
\and Th.~Henning\inst{\ref{inst:mpia}}
\and S.\,V.~Jeffers\inst{\ref{inst:mps}}
\and M.~K\"urster\inst{\ref{inst:mpia}}
\and M.~Lafarga\inst{\ref{inst:warwick}}
\and M.~L\'{o}pez-Puertas\inst{\ref{inst:iaa}}
\and D.~ Montes\inst{\ref{inst:ucm}}
\and J.\,C.~Morales\inst{\ref{inst:ice},\ref{inst:ieec}}
\and S.~Pedraz\inst{\ref{inst:caha}}
\and A.~Schweitzer\inst{\ref{inst:hs}}
}
   
\institute{
     \label{inst:hs} Hamburger Sternwarte, Gojenbergsweg 112, 21029 Hamburg, Germany\\
\email{evangelos.nagel@hs.uni-hamburg.de}
\and \label{inst:iag} Universit\"at G\"ottingen, Institut f\"ur Astrophysik und Geophysik, Friedrich-Hund-Platz 1, 37077 G\"ottingen, Germany
\and \label{inst:tls} Th\"uringer Landessternwarte Tautenburg, Sternwarte 5, 07778 Tautenburg, Germany
\and \label{inst:lsw} Landessternwarte, Zentrum f\"ur Astronomie der Universit\"at Heidelberg, K\"onigstuhl 12, 69117 Heidelberg, Germany
\and \label{inst:ariel} Department of Physics, Ariel University, Ariel 40700, Israel
\and \label{inst:cab} Centro de Astrobiolog\'{i}a (CSIC-INTA), ESAC, Camino Bajo del Castillo s/n, 28692 Villanueva de la Ca\~{n}ada, Madrid, Spain
\and \label{inst:isdefe} ISDEFE, Beatriz de Bobadilla 3, 28040 Madrid, Spain 
\and \label{inst:ice} Institut de Ci\`encies de l'Espai (CSIC), Campus UAB, c/ de Can Magrans s/n, 08193 Bellaterra, Barcelona, Spain
\and \label{inst:ieec} Institut d'Estudis Espacials de Catalunya (IEEC), c/ Gran Capit\`{a} 2-4, 08034 Barcelona, Spain
\and \label{inst:iaa} Instituto de Astrof\'{i}sica de Andaluc\'{i}a (CSIC), Glorieta de la Astronom\'{i}a s/n, 18008 Granada, Spain
\and \label{inst:iac} Instituto de Astrof\'{i}sica de Canarias, c/ V\'{i}a L\'{a}ctea s/n, 38205 La Laguna, Tenerife, Spain
\and \label{inst:ull} Departamento de Astrof\'{i}sica, Universidad de La Laguna, Avda. Francisco S\'{a}nchez s/n, 38206 La Laguna, Tenerife, Spain
\and \label{inst:ucm} Departamento de F\'{i}sica de la Tierra y Astrof\'{i}sica and IPARCOS-UCM (Intituto de F\'{i}sica de Part\'{i}culas y del Cosmos de la UCM), Facultad de Ciencias F\'{i}sicas, Universidad Complutense de Madrid, 28040, Madrid, Spain
\and \label{inst:mpia} Max-Planck-Institut f\"ur Astronomie, K\"onigstuhl 17, 69117 Heidelberg, Germany
\and \label{inst:mps} Max-Planck-Institut f\"ur Sonnensystemforschung, Justus-von-Liebig-Weg 3, 37077 G\"ottingen, Germany
\and \label{inst:warwick} Department of Physics, University of Warwick, Gibbet Hill Road, Coventry CV4 7AL, United Kingdom
\and \label{inst:caha} Centro Astron\'omico Hispano en Andaluc\'ia (CSIC-Junta de Andaluc\'{i}a), Observatorio Astron\'omico de Calar Alto, Sierra de los Filabres, 04550 G\'{e}rgal, Almer\'{i}a, Spain
}

\date{Received 29 March 2023 / Accepted 25 September 2023}

\abstract
{
Light from celestial objects interacts with the molecules of the Earth's atmosphere, 
resulting in the production of telluric absorption lines in ground-based spectral data.
Correcting for these lines, which strongly affect red and infrared wavelengths, 
is often needed in a wide variety of scientific applications. 
Here, we present the template division telluric modeling (TDTM) technique,
a method for accurately removing telluric absorption lines in stars that 
exhibit numerous intrinsic features.
Based on the Earth's barycentric motion
throughout the year, our approach is suited for disentangling telluric and 
stellar spectral components.
By fitting a synthetic transmission model, telluric-free spectra are derived.
We demonstrate the performance of the TDTM technique in correcting 
telluric contamination
using a high-resolution optical spectral time series of the 
feature-rich M3.0 dwarf star Wolf~294 that was obtained with the CARMENES spectrograph.
We apply the TDTM approach to the CARMENES survey sample, which consists of 
382 targets encompassing 22\,357 optical and 20\,314 near-infrared spectra,
to correct for telluric absorption. 
The corrected spectra are coadded to construct 
template spectra for each of our targets. 
This library of telluric-free, high signal-to-noise ratio, high-resolution ($\mathcal{R}>80\,000$) templates comprises
the most comprehensive collection of spectral 
M-dwarf data available to date, both in terms of quantity and quality, 
and is available at the project website. 
}

\maketitle

\section{Introduction}

Cool, low-mass M dwarfs are particularly
promising targets for detecting rocky habitable-zone planets because
of their relatively short orbital periods of about $20$\,d
and the favorable planet-to-star mass ratios.
As M~dwarfs are
intrinsically faint in the visible wavelength range and
emit the bulk of their energy at $\sim 1$\,$\mu$m,
a new generation of high-resolution spectrographs, such as CARMENES \citep{Quirrenbach2020}, 
SPIRou \citep{Donati2018}, HPF \citep{Mahadevan2014}, 
IRD \citep{Kotani2018}, CRIRES$^+$ \citep{Dorn2023}, and NIRPS \citep{Wildi2017},
have been designed and built to exploit the near-infrared wavelength range for 
radial velocity (RV) planet searches.

Ground-based spectroscopic observations at optical and, in particular, at near-infrared 
wavelengths are affected by absorption and emission features produced 
by the Earth's atmosphere. 
Rotational-vibrational transitions of molecules such as water (H$_2$O),
oxygen (O$_2$), carbon dioxide (CO$_2$), and methane (CH$_4$) 
produce numerous absorption lines and broad absorption bands. 

The relative strength of individual lines exhibits a considerable range, 
extending from shallow so-called microtellurics, 
which present with flux depths of $\lesssim 1\mathrm{-}2\,\%$, 
to potent lines characterized by completely opaque line cores, 
signifying a total absence of light transmission.
These telluric lines are a common nuisance in ground-based spectroscopy,
and in particular, in precise RV measurements \citep{Cunha2014, Leet2019, Wang2022, Latouf2022}.

The atmospheric conditions at the time of observation determine the telluric contribution
to a stellar spectrum observed from the ground. 
The observed strength of the telluric lines depends on the location of the observatory, particularly its elevation, and on the airmass of the target.
Furthermore, telluric absorption lines are Doppler shifted and broadened
due to turbulent wind motions along the line of sight \citep[e.g.,][]{Caccin1985}.
The atmospheric temperature and partial-pressure structure have a direct impact
on the line profiles. The situation is further complicated by the 
temporal variability in temperature, pressure, and chemical composition
of the Earth's atmosphere on seasonal, daily, and hourly timescales, which is
particularly pronounced for the atmospheric water vapor content
\citep{Smette2015}.

To date, several elaborated approaches have been developed 
to correct for telluric lines. They can 
roughly be subdivided into empirical, data-driven, and forward-modeling approaches.
A widely used empirical technique is telluric division. Here,
telluric standard stars (TSS), which usually are rapidly rotating B- to A-type stars, 
are observed along with the science target observation, preferably 
similar to the science target in time, airmass, and direction.
However, some compromises have to be made regarding these parameters.
The drawbacks of the telluric division method 
have been extensively discussed by,
for instance, \citet{Vacca2003}, \citet{Bailey2007}, \citet{Seifahrt2010}, 
\citet{Gullikson2014}, and \citet{Smette2015}. 
Large RV surveys such as those conducted by the HARPS and CARMENES 
collaborations refrained from frequent
TSS observations because a tremendous amount of additional observing time is needed.

An empirical approach, avoiding frequent and repeated TSS observations,
was presented by \citet{Artigau2014}. These authors
created a library of TSS spectra, observed on a dense grid of airmasses and water columns.
By carrying out a principal component analysis, 
they identified independently 
varying spectral absorption patterns. Thus, telluric spectra were synthesized by using 
linear combinations of individual absorbances, which \citet{Artigau2014}
subsequently removed from HARPS measurements.

A data-driven technique involving machine-learning algorithms was
presented by \citet{Bedell2019}. Their \url{wobble} algorithm
incorporates a model for simultaneously 
deriving the stellar spectra, telluric spectra, and RVs from 
spectral time series without relying on external information on either the stellar
or the telluric transmission spectrum.

While purely data-driven techniques are highly
flexible, copious information on the atmosphere of the Earth and its spectrum is
available.
Synthetic transmission models of the spectrum of the Earth's atmosphere take advantage of this.
They can be generated by radiative transfer codes \citep[for an overview see][]{Seifahrt2010}
in combination with precise molecular line databases, and have become widely used
\citep[e.g.,][]{Bailey2007, Seifahrt2010, Lockwood2014, Husser2014, Gullikson2014, Rudolf2016, Allart2022}. 
Recently, \citet{Artigau2022} introduced a hybrid method in which they first employed
the TAPAS\footnote{\textbf{T}ransmissions of the \textbf{A}tmos\textbf{P}here for \textbf{AS}tromomical data,\\ \url{http://cds-espri.ipsl.fr/tapas/}} atmospheric model \citep{Bertaux2014} to establish a first-order correction,
and then computed an empirical correction model derived from a sample of TSSs for the residuals that remain.
This approach was implemented into the SPIRou data reduction pipeline \url{APERO} \citep{Cook2022}.

The software package 
\texttt{molecfit}\footnote{\url{http://www.eso.org/sci/software/pipelines/skytools/molecfit}},
developed by \citet{Smette2015} and \citet{Kausch2015}, implements a
synthetic telluric transmission model and allows correcting observed spectra for
telluric contamination. To this end, \url{molecfit}
incorporates the line-by-line radiative transfer code {\tt LBLRTM} \citep{Clough2005} and
the HITRAN molecular line list \citep{Rothman2009}.
To synthesize a telluric spectrum, \url{molecfit} requires a number of parameters
such as a model of the instrumental line spread function (LSF), an atmospheric profile
describing the meteorological conditions during the observation, and the column
density of the molecular species.
In turn, the observed spectrum, which contains the telluric transmission spectrum,
can be used to find best-fit values for these parameters.
\texttt{molecfit} does not consider the actual stellar spectrum, which occurs
as a contaminant of the telluric spectrum in this context. Therefore, only a suitable
subrange of the observed spectrum is commonly used in the fit, which ideally contains
moderately saturated telluric lines with a well-defined 
continuum. Based on the resulting atmospheric parameters, the telluric spectrum in
the remaining range can then be inferred. While this approach works well in many
cases, it becomes problematic in objects with ubiquitous intrinsic features
such as M~dwarfs, where the fitting becomes inaccurate because of numerous blends 
between stellar and telluric lines. 

One of the largest M-dwarf surveys has been conducted by CARMENES. 
In the course of the guaranteed time observations (GTO) program in 2016 to 2020,
more than 19\,000 high-resolution spectra were obtained in the optical 
and about the same number in the near-infrared, which is equivalent to about 5000 observing hours
\citep{Quirrenbach2020, Ribas2023}. 
Meanwhile, the survey is being continued as a legacy program with 300 additional nights until the end of 2023.
As telluric contamination is particularly severe in the wavelength range covered 
by CARMENES (5\,200--17\,100\,\r{A}), accurate telluric correction is called for.

To expand the range of applications of \texttt{molecfit} to M-dwarf spectra with strong line crowding,
we developed the template division telluric modeling (TDTM) technique.
The heart of TDTM is the disentanglement of sections of the telluric
and stellar spectra by taking advantage of the relative shift between stellar 
and telluric lines caused by the Earth's barycentric motion.
We then use \texttt{molecfit} to fit a synthetic transmission model to the
extracted telluric spectrum and apply these results
to correct for telluric absorption in the entire science spectrum.
While the TDTM technique works independently of stellar spectral type,
it is only applicable when spectroscopic time series that sample
a range of barycentric velocity shifts are available.
Time series like this have been provided by the CARMENES survey for more than 300 M-dwarf stars \citep{Reiners2018}. 
In recent years, our individual telluric-corrected spectra as well as telluric-corrected template spectra 
have been used in the context of planet detection \citep[e.g.,][]{Polanski2021, Blanco-Pozo2023, Kossakowski2023}, atmospheric characterization 
\citep[e.g.,][]{Nortmann2018, Salz2018, Alonso-Floriano2019, Palle2020, Orell-Miquel2022, Orell-Miquel2023},
magnetic field measurements \citep{Shulyak2019, Reiners2022},
photospheric parameter determinations \citep{Passegger2019, Passegger2020, Marfil2020, Marfil2021},
abundance analyses \citep{Abia2020, Shan2021}, or 
chromospheric activity studies \citep{Fuhrmeister2019, Fuhrmeister2020, Fuhrmeister2022, Fuhrmeister2023, Hintz2020, Hintz2023}.
Here, we apply TDTM, which is publicly available\footnote{\url{https://github.com/evangelosnagel/tdtm}}, to the CARMENES survey spectra
and subsequently construct telluric-free template spectra with a high signal-to-noise ratio (S/N)
for the stars in our sample. As a service to the community, we publish them along with this paper.

Our paper is structured as follows.
In Sect.~\ref{section:observations} we describe our M-dwarf sample.
The TDTM technique and the data preparation are presented in Sect.~\ref{section:method}, 
and our findings are discussed in Sect.~\ref{section:results}. Finally, we
summarize our results in Sect.~\ref{section:conclusion}.

\section{M-dwarf sample}
\label{section:observations}

The spectra used in this work were taken within the context 
of the CARMENES\footnote{\textbf{C}alar \textbf{A}lto high-\textbf{R}esolution 
search for \textbf{M} dwarfs with \textbf{E}xoearths with 
\textbf{N}ear-infrared and optical \textbf{E}chelle \textbf{S}pectrographs.} 
survey. 
Constructed by 11 German and Spanish institutions, 
CARMENES consists of a pair of cross-dispersed fiber-fed 
\'{e}chelle spectrographs, mounted on the 3.5\,m telescope 
of the Calar Alto Observatory in Spain 
\citep{Quirrenbach2020}. 

The instrument has two channels.
The visual channel (VIS) provides wavelength coverage between $5\,200\,$\r{A} and $9\,600\,$\r{A} 
with a resolution of $\mathcal{R}=94\,600$, and the near-infrared channel (NIR)
has a resolving power of $\mathcal{R}=80\,400$ and covers the spectral range 
from $9\,600\,\AA$ to $17\,100\,\AA$. 
Both channels are enclosed in vacuum vessels 
in the coud\'{e} room to ensure long-term stability.   
The data reduction was carried out using the standard CARMENES 
reduction pipeline {\tt caracal}\footnote{\textbf{CA}RMENES \textbf{R}eduction 
\textbf{A}nd \textbf{CAL}ibration} \citep{Zechmeister2014, Caballero2016b}. 

We removed 17 spectroscopic binaries and triple systems from the full CARMENES M-dwarf sample of 402 stars, as found by \citet{Baroch2018, Baroch2021}.
In addition, we excluded three targets that were observed for a period shorter than a month.
Our final sample in this work encompasses 382 stars across the full 
M-dwarf range from M0.0 to M9.0. Additionally, two K5.0 and two K7.0 dwarfs are included in the sample. 
Insights into target selection, sample characteristics, and data quality were provided by 
\citet{Reiners2018}, \citet{Quirrenbach2020}, \citet{Ribas2023},
and references therein. 

To demonstrate the use of the TDTM technique,
we selected 444 VIS and 398 NIR observations of the bright M3.0 dwarf 
Wolf~294 (HD 265866, GJ~251, J06548+332), obtained between January 2016 and May 2023.
Wolf~294 serves as a representative of an object with a feature-rich spectrum.
A typical S/N for CARMENES survey observations is 150 in the $J$ band
\citep{Reiners2018}, which translates into a median exposure time of 545\,s for Wolf~294. We refer to \citet{Stock2020} for more details on the star and its temperate super-Earth.

\section{Telluric correction}
\label{section:method}

Fitting telluric transmission models
to observed M-dwarf spectra is challenging because the stellar spectrum
consists of numerous atomic and molecular lines without a well-defined continuum.
The fundamental idea behind the TDTM approach is to construct a template of the stellar spectrum with a high S/N from a spectral time series, which can subsequently be used to eliminate the stellar
contribution and to model the residual telluric spectrum. 

\begin{figure}
\begin{center}
\includegraphics[width=0.49\textwidth]{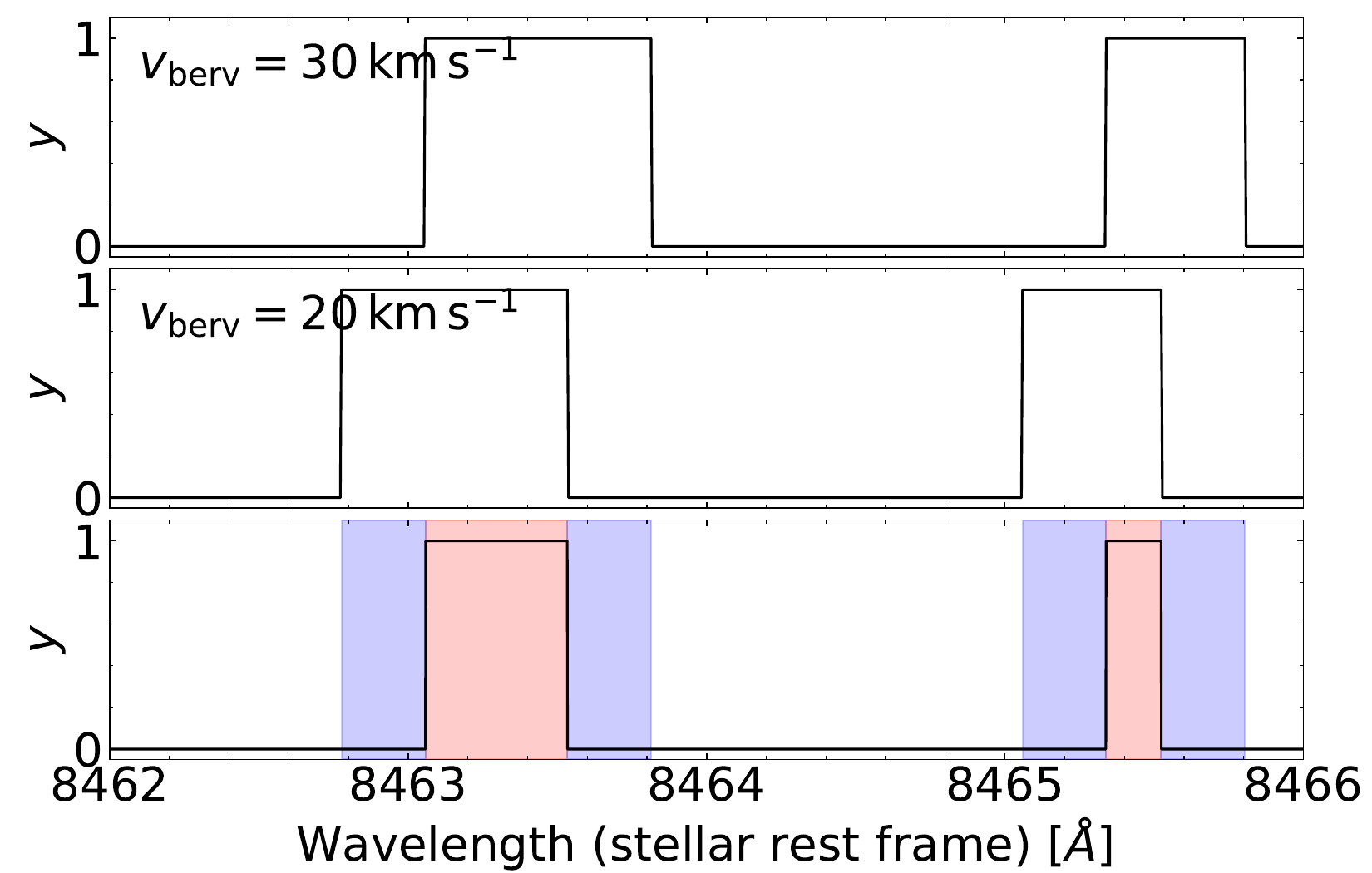}
\caption{\label{figure:mask} 
Example of template completion using a binary telluric mask. 
\textit{Upper and middle panels:} Small section of the telluric binary mask 
for two observations after the correction for Earth's barycentric motion. 
Wavelength ranges with $y=1$ are masked as tellurics.
\textit{Lower panel:} Resulting mask of the template. The red shaded wavelength ranges
correspond to template knots with $M_k = 0$ and contribute
to the total masked template fraction $\gamma$.
The blue shaded wavelength ranges are used for telluric modeling and 
contribute to $\gamma\,'$.}
\end{center}
\end{figure}

\begin{figure*}
\begin{center}
\includegraphics[width=1\textwidth]{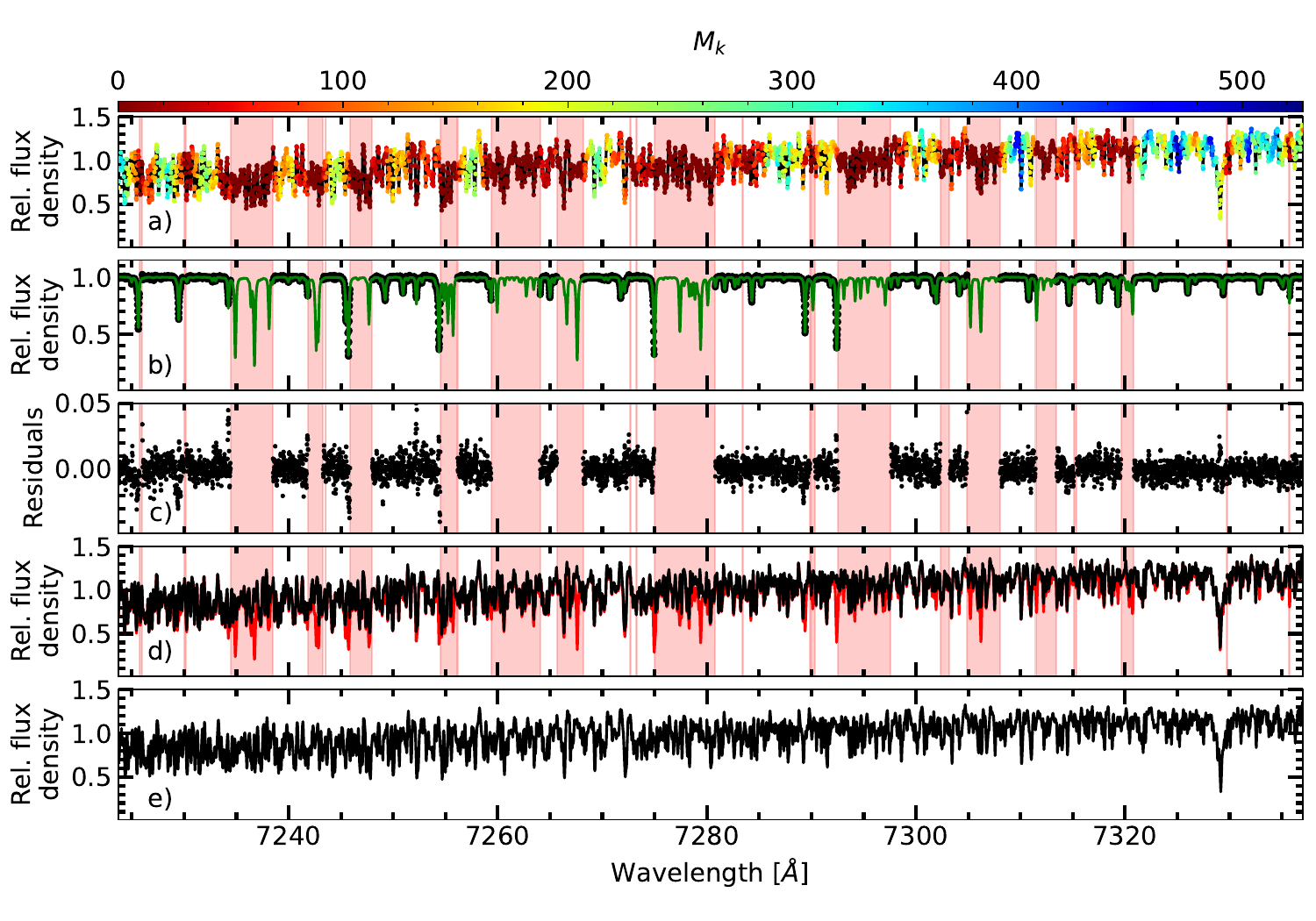}
\caption{
Illustration of the TDTM method. 
\textit{Panel a:} Segment of the VIS template spectrum of Wolf~294.
The number $M_k$ of exposure pixels that contribute to each template knot is color-coded.
The red shaded wavelength ranges mark knots with $M_k = 0$.
\textit{Panel b:}
        One residual telluric spectrum $F_{n,i} / S(\lambda_{n,i})$ (black dots) 
after the division of the science spectrum by the template, 
and the best-fit telluric model ($T(\lambda_{n,i})$, green line) derived with \texttt{molecfit}.
The red shaded wavelength ranges are excluded from the transmission model fit. 
\textit{Panel c:}
	Absolute residuals $F_{n,i}/S(\lambda_{n,i}) - T(\lambda_{n,i})$ of the fit.
\textit{Panel d:}
CARMENES spectrum before ($F_{n,i}$, red line) and after ($F_{n,i} / T(\lambda_{n,i})$) 
correction with the transmission model derived with \texttt{molecfit} (black line).
\textit{Panel e:}
	Telluric free high S/N template spectrum of Wolf~294 (black line) built using 444 telluric absorption-corrected 
CARMENES observations. The order has an S/N of 2310.
\label{figure:tdtm_vis}}
\end{center}
\end{figure*}

\subsection{Template construction}
\label{subsection:template}

We used the {\tt serval}\footnote{\textbf{S}p\textbf{E}ctrum \textbf{R}adial \textbf{V}elocity \textbf{A}na\textbf{L}yser, \\
\url{https://github.com/mzechmeister/serval}} code 
\citep{Zechmeister2018} to compute the stellar template spectrum
based on a time series of input spectra.
Following the nomenclature of \citet{Zechmeister2018}, we indexed the set of spectra by
$n=1,...,N$, and each spectrum was composed of discrete flux density measurements $f_{n,i}$ 
at pixel $i$ with uncertainties $\epsilon_{n,i}$ and 
calibrated wavelengths $\lambda_{n,i}$.
We furthermore write the continuous model for the observation in the form
\begin{equation}
f(\lambda) = [s(\lambda) \cdot t(\lambda)] \otimes L(\lambda)\, ,
\label{equation:spec}
\end{equation}
where $ s(\lambda)$ denotes the intrinsic stellar spectrum,
$ t(\lambda)$ the intrinsic telluric absorption spectrum at the time of observation,
$L(\lambda)$ the LSF, and $\otimes$ the convolution operator.

In each observation, a stellar spectrum with an a priori unknown RV shift and a telluric
spectrum with variable properties, but fixed in the rest frame of the Earth, were superimposed.
The ultimate goal of {\tt serval} is to derive precise measurements of the stellar RV.
To this end, a preferably accurate and complete template of the stellar spectrum is required.
This template was constructed by {\tt serval} from the spectral time series, 
following an iterative and sequential forward-modeling approach.

Template completeness is improved by taking advantage of the relative RV shift of the stellar and telluric
spectra induced by Earth's barycentric motion, which can uncover sections of the stellar spectrum in some observations that may be
affected by telluric lines
in others with a less favorable relative RV shift.
To identify these sections, {\tt serval} uses a binary mask $m_{n,i}$,
flagging spectral pixels that are affected by known atmospheric absorption features
that are typically deeper than 1\,\% (see Sect.~\ref{appendix:mask}).
Additionally, {\tt serval} employs
a bad-pixel map that flags saturated pixels, outliers, 
pixels providing unphysical (negative) flux densities, and pixels affected by sky emission. 
The template S/N is improved by coadding
the individual spectra after an appropriate RV shift. 
During this process, the spectra are transformed 
to the stellar rest frame, that is, they are corrected for barycentric and stellar RVs.
{\tt serval} also accounts for secular acceleration, which is
a change in the stellar radial velocity due to the high proper motion of close-by stars
\citep{Kuerster2003, Zechmeister2009}. 

As a starting point for RV and template calculation,
{\tt serval} calculates preliminary RVs using the highest S/N spectrum 
of the CARMENES spectral time series as a template. 
These RVs are subsequently used to improve the stellar template spectrum by coadding the individual spectra
and by obtaining higher precision RVs. This process is repeated until convergence is achieved.
All masked pixels are excluded, and telluric-affected pixels are heavily downweighted in the process of
coadding to isolate the stellar spectrum in the template.

The process of improving the template completeness by taking advantage of the relative shift of the mask
is demonstrated in Fig.~\ref{figure:mask}.
The upper and middle panels
show a segment of the telluric mask position of two observations 
after the correction for Earth's barycentric motion.
The lower panel shows the resulting mask of the template, assuming that the intrinsic RV shift of the stellar spectrum
remained small. In the blue shaded regions, the stellar spectrum is shown in one of the
observations, and only the red shaded ranges remain hidden. This results in an improvement of
the template wavelength coverage.

Shifting the spectra to the same reference wavelength results in a wavelength sampling that differs
for data taken at different nights. {\tt serval} avoids resampling the spectral data on a new discrete
wavelength grid and directly carries out a cubic B-spline regression.
In essence, the template is the best-fitting spline through scattered data, and it is a continuous function
that can be evaluated at any point within its boundaries, as illustrated in Fig.~2 of \cite{Zechmeister2018}.
The uniform cubic B-spline has equidistant knots $k$ on a logarithmic wavelength grid $\ln \lambda_k$.
Typically, the number of template knots is comparable to the number of pixels per spectral order.
In the following, $s_k=s(\lambda_k)$ refers to the stellar template flux density at the knot positions.
As an example, a section of the template of Wolf~294 is shown in Fig.\ref{figure:tdtm_vis}a.

As a crucial quantity for the TDTM approach, 
we define $M_k$ as the number of unflagged pixels that contribute
to a knot. $M_k$ is stored for each template knot. 
The link between $M_k$ and the mask can be inferred from the lower panel of Fig.~\ref{figure:mask}.
In this specific example with only two observations, 
$M_k=0$ for template knots falling into the red marked
wavelength range, $M_k=1$ for knots within the blue wavelength range,
and $M_k=2$ for the remaining part of the spectrum.

Therefore, given one particular template knot $k$, three situations are possible:
(1) all pixels contributing to this knot 
contain spectral information (i.e., are not masked),
(2) all pixels are masked,
(3) some pixels contain spectral information,
while the remaining pixels are masked. 
The quantity $M_k$ is maximized in case (1), decreases
for knots that are partly affected by tellurics in case (3), 
and takes a value of $M_k = 0$ for pixels in case (2).
For sampling reasons, spectra can contribute more than one pixel to a template knot, 
so that $M_k$ can exceed the number $N$ of observations.
This is demonstrated in Fig.~\ref{figure:tdtm_vis}a, 
where a few template knots have $M_k \gtrsim 500$,
although the number of observations is $N_{\rm VIS} = 444$.
The red shaded areas show wavelength ranges for which $M_k$ is zero, that is,
the stellar spectrum could not be recovered.

As the total masked template fraction $\gamma$, we
defined the fraction of masked ranges ($M_k = 0$)
compared to the entire wavelength range of the spectrum. 
As a result, the template completeness corresponds to $1-\gamma$
and depends on barycentric RV of the observer.
The wavelength ranges of the stellar spectrum
that are characterized by overlapping 
masked and nonmasked regions, indicated by the blue shaded areas 
in Fig.~\ref{figure:mask} and recovered in the process,
are now additionally available for RV calculation and telluric modeling.
We denote the total fraction of these ranges 
compared to the entire wavelength range of the spectrum 
by $\gamma\,'$.

\subsection{Mask construction}
\label{appendix:mask}
The telluric mask is an essential ingredient of the TDTM technique.
As telluric features are ubiquitous, mask construction needs to balance
strictness and achievable template completeness.
While a restrictive mask that covers all relevant telluric features is crucial
for creating a useful stellar template, an overly strict mask that declares extended chunks of
the spectrum unusable jeopardizes the derivation of a template with any practical value.

To construct masks, we computed a synthetic telluric transmission model 
including H$_2$O, O$_2$, CO$_2$, and CH$_4$ using \texttt{molecfit}.
As input parameters, we used the median observational and atmospheric parameters 
in the data set of each star and adopted
the values from the standard atmosphere profile for the column densities of the atmospheric
constituents. In the resulting transmission models,
we normalized wavelength ranges that were affected by molecular continuum absorption. 
For each target, we finally computed a binary mask by flagging all model features that were deeper than
a specified threshold.
In our analysis, we found that thresholds of $\sim 1\,\%$ provide
a reasonable compromise between capturing
the vast majority of telluric features and removing critical portions of
usable spectrum.

\subsection{Extraction of the telluric spectrum}

To extract the telluric transmission spectrum, we wish to
remove the stellar contribution 
from the individual observations by division. 
Following \citet{Vacca2003}, we therefore approximated the convolution of the product
 in Eq.~(\ref{equation:spec}) by 
\begin{align}
	F(\lambda) &\approx [s(\lambda) \otimes L(\lambda)] \cdot [t(\lambda) \otimes L(\lambda)]\\
	&= S(\lambda) \cdot T(\lambda) \, ,
\label{equation:spec2}
\end{align}
which is now a product of the convolved intrinsic stellar flux density and the 
convolved telluric spectrum,
\begin{align}
S(\lambda) &=  s(\lambda') \otimes L(\lambda, \lambda')\\
T(\lambda) &=  t(\lambda') \otimes L(\lambda, \lambda').
\end{align}
We discuss the validity of this approximation in Appendix~\ref{appendix:convolution}.

Dividing the observed spectra by the appropriately shifted template, we derived
a residual spectrum that is essentially free of stellar features,
\begin{equation}
	\frac{F_{n,i}}{S(\lambda_{n,i})} \approx T(\lambda_{n,i}) \; .
\end{equation}	
As the uncertainty of the template is typically negligible compared to that of the
individual observations as $M_k \gg 1$, we did not propagate the
template uncertainty and used the uncertainty of the observed spectra in
the modeling of the individual residual spectra.

The residual telluric transmission spectrum of Wolf~294 is
shown in Fig.~\ref{figure:tdtm_vis}b.
Again, the red bands represent template knots for which the residual spectrum
could not be constrained.
Consequently, these sections are not used to model the telluric lines.

\subsection{Modeling the telluric lines}
\label{subsection:molecfit}

\begin{table}
\begin{center}
	\caption{\texttt{molecfit} fitting ranges used to calibrate 
	the models for CARMENES VIS and NIR spectra.}
\label{table:fitranges}
\begin{tabular}{ccc} 
\hline\hline 
\noalign{\smallskip}
Channel            & $\Delta \lambda$ [\r{A}] & Main absorber\\
\noalign{\smallskip}
\hline
\noalign{\smallskip}
VIS                & 5375--5530   & H$_2$O \\
VIS                & 5650--6040   & H$_2$O \\
VIS                & 6260--6620   & O$_2$, H$_2$O \\
VIS                & 6860--7450   & O$_2$, H$_2$O \\
VIS                & 7590--7740   & O$_2$ \\
VIS                & 7830--8600   & H$_2$O \\
VIS                & 8800--9270   & H$_2$O \\
\noalign{\smallskip}
\hline
\noalign{\smallskip}
NIR                & 9750--10\,350   & H$_2$O  \\
NIR                & 10\,600--11\,000   & H$_2$O \\
NIR                & 11\,600--12\,400 & H$_2$O \\
NIR                & 12\,400--13\,100 & O$_2$ \\
NIR                & 15\,110--15\,400 & H$_2$O \\
NIR                & 15\,600--15\,800 & CO$_2$ \\
NIR                & 15\,900--16\,200 & CO$_2$ \\
NIR                & 16\,350--16\,660 & CH$_4$ \\
NIR                & 16\,850--17\,100 & CH$_4$, H$_2$O \\
\noalign{\smallskip}
\hline
\end{tabular}
\end{center}
\end{table}

\begin{figure}
\begin{center}
\includegraphics[width=0.49\textwidth]{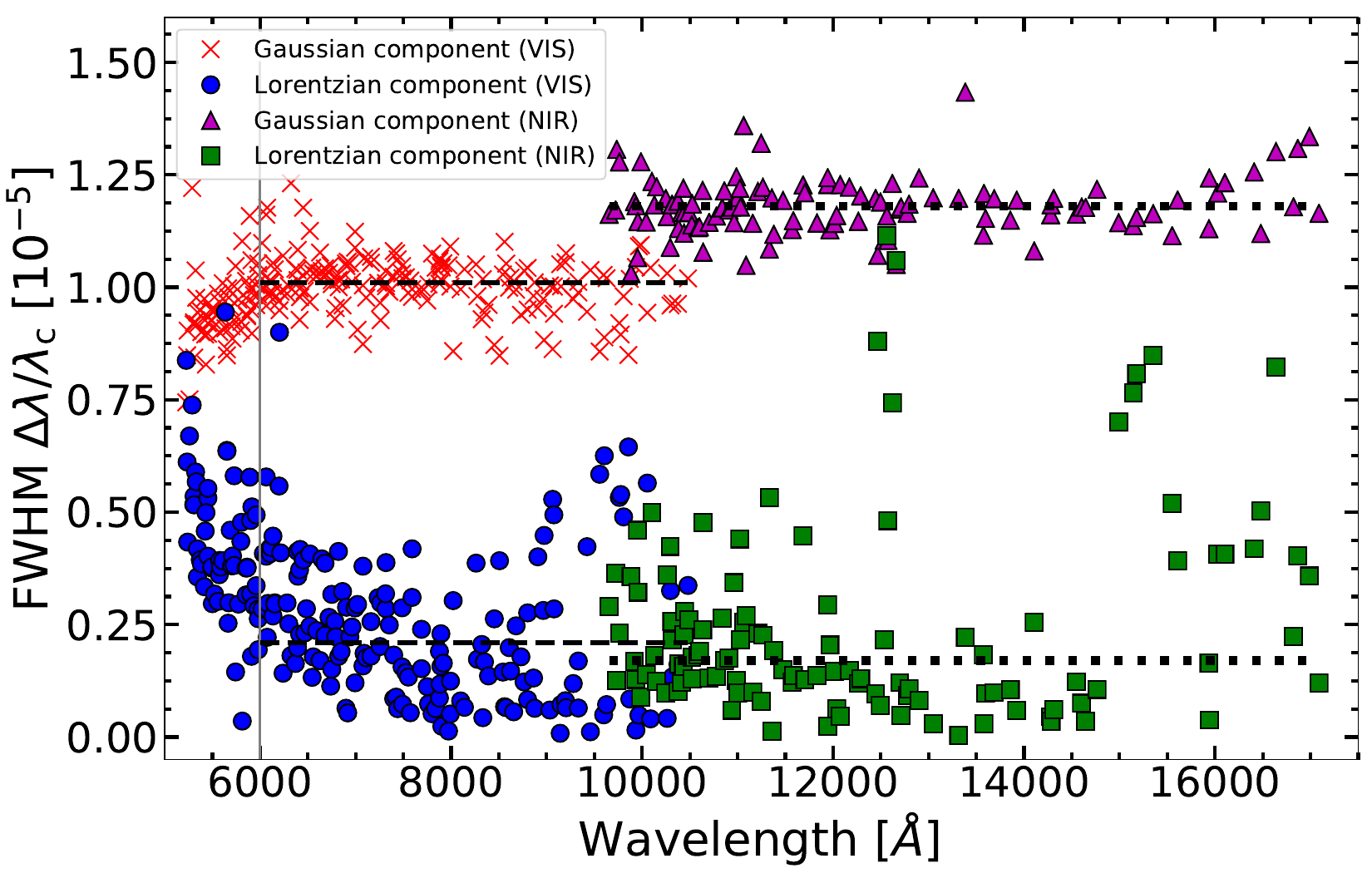}
\caption{\label{figure:lsf} 
Measured Gaussian and Lorentzian FWHM components 
of hollow cathode emission lines as a function of wavelength 
for the VIS and NIR CARMENES spectrographs.
The vertical gray line marks the cutoff wavelength at 6000\,\AA.
The horizontal lines indicate median values of the 
Gaussian and Lorentzian FWHM components 
for the VIS (dashed lines) and NIR channel
(dotted lines).}
\end{center}
\end{figure}

\begin{table}
\begin{center}
\caption{Median averaged Gaussian and Lorentzian FWHMs.}
\label{table:lsf}      
\begin{tabular}{l ccc} 
\hline\hline          
\noalign{\smallskip}
Channel            & Number & $w_{\mathrm{Gauss}}$ & $w_{\mathrm{Lorentz}}$ \\
                & of lines & $[10^{-5} \Delta\lambda/\lambda_c]$ & $[10^{-5} \Delta\lambda/\lambda_c]$ \\
\noalign{\smallskip}
\hline
\noalign{\smallskip}
VIS                & 218 & $1.00$ & $0.28$ \\
VIS ($>6000\,$\r{A}) & 159 & $1.01$ & $0.21$ \\
NIR                & 114 & $1.18$ & $0.17$ \\
\noalign{\smallskip}
\hline
\end{tabular}
\end{center}
\end{table}

The residual telluric spectrum was modeled
using the \texttt{molecfit} package version 1.5.9
\citep{Smette2015, Kausch2015} and
the version 3.6 of the \texttt{aer}\footnote{\url{http://rtweb.aer.com/}} molecular line list.
The altitude stratification of temperature, pressure, and 
molecular abundances served as input for LBLRTM.
To create this profile, \texttt{molecfit} merges information from three
sources: (1) a reference atmospheric profile, (2) global data assimilation system (GDAS) 
profiles\footnote{\url{https://www.ready.noaa.gov/gdas1.php}}, 
and (3) measurements of the ambient conditions obtained during the time of 
observations.
In this study, we used a nightly midlatitude (45\,deg) 
reference model atmosphere\footnote{\url{http://eodg.atm.ox.ac.uk/RFM/atm/}}
as the reference profile.
In addition to the pressure and temperature distribution up to an altitude of 120\,km
as a function of height on a 1\,km grid, the profile also provides 
the abundances of 30 molecules. 
The GDAS profiles describe the pressure, temperature, and relative humidity 
as a function of 23 altitude levels up to roughly 26\,km.

In the spectral modeling with \texttt{molecfit},
we freely varied the abundances of O$_2$, CO$_2$, CH$_4$, and
the atmospheric water vapor content.
We carried out the model fitting over broad wavelength ranges 
that covered large portions of the molecular bands, 
taking full advantage of numerous unsaturated telluric lines 
contained in the CARMENES VIS and NIR channels; 
the selected fitting intervals are given in Table~\ref{table:fitranges}.
In the modeling, the continuum level within the fitting ranges 
was approximated with a low-order polynomial.
Since small errors in the wavelength calibration result in 
large residuals in the corrected spectra, 
we also allowed a Doppler shift of the transmission model to match the observed spectrum.
In this way, instrumental drifts can be accounted for in the modeling.
Wavelength ranges for which no stellar template was available ($M_k = 0$)
were excluded from the fit, along with sections affected by
sky emission features. 

\texttt{molecfit} requires the instrumental LSF in the modeling.
In the case of CARMENES, the LSF can be represented by
a combination of two profiles, one Gaussian and one Lorentzian.
We performed an extensive analysis to derive appropriate parameters for the LSFs of the two CARMENES channels based on calibration data (see Sect.~\ref{appendix:lsf}).
Therefore, the parameters of the LSF were not free in the modeling.

\subsection{CARMENES instrumental line profile}
\label{appendix:lsf}

For a proper modeling of the telluric lines, the
LSF needs to be well known. In order to characterize the LSF, we
analyzed hollow cathode lamp spectra taken
for the purpose of calibration before and after the science observations.

An accurate estimation of the line shape is not trivial because the lines are
sparsely sampled. Within each \'{e}chelle order, 
the coverage of the full width at half maximum (FWHM) of the LSF typically increases from about
2.5 to 4.0 pixels along the main
dispersion direction from the blue to the red part of the order. Due
to several instrumental effects (e.g., temperature variations), 
the spectra move across the detector with time. We therefore combined multiple exposures 
to artificially increase the sampling.
A total of 1455 different lamp exposures in the VIS channel and 2605 in the NIR channel were used for this purpose. For each line we investigated in that
manner ($218$ in the VIS and $114$ in the NIR), the frames were aligned so that the line was centered. 
Before they were combined, they were normalized. 

\begin{figure*}
\begin{center}
\includegraphics[width=0.49\textwidth]{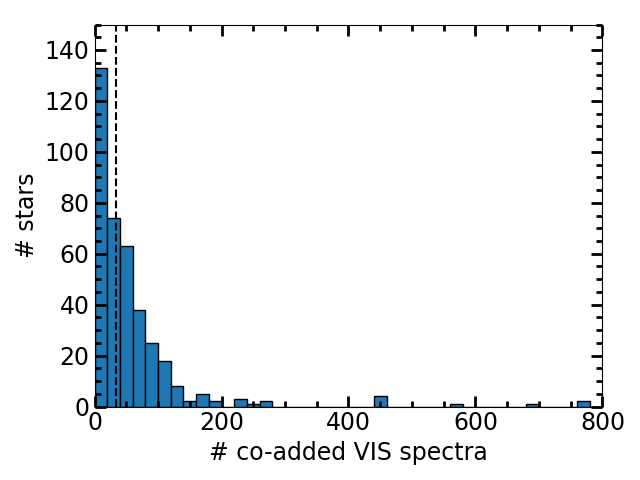}
\includegraphics[width=0.49\textwidth]{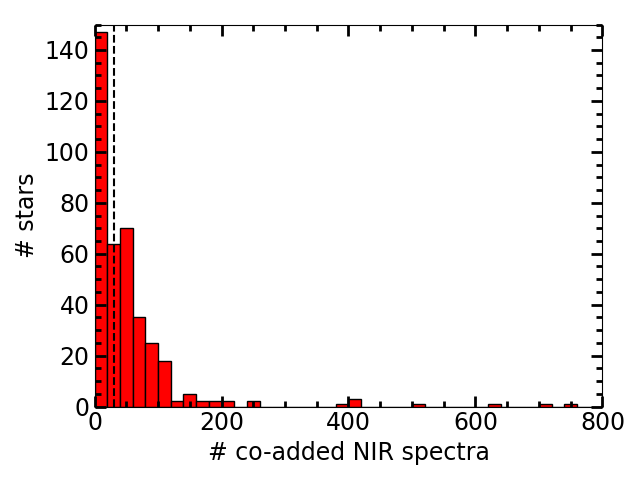}
\caption{\label{figure:hist1}
	Distribution of the number of telluric-corrected spectra we used to build the templates 
	for all sample stars in the VIS (\textit{left}) and NIR (\textit{right}). The vertical dashed lines indicate 
	the median values at 34 in the VIS and 30 in the NIR.
	}
\end{center}
\end{figure*}

\begin{figure*}
\begin{center}
\includegraphics[width=0.49\textwidth]{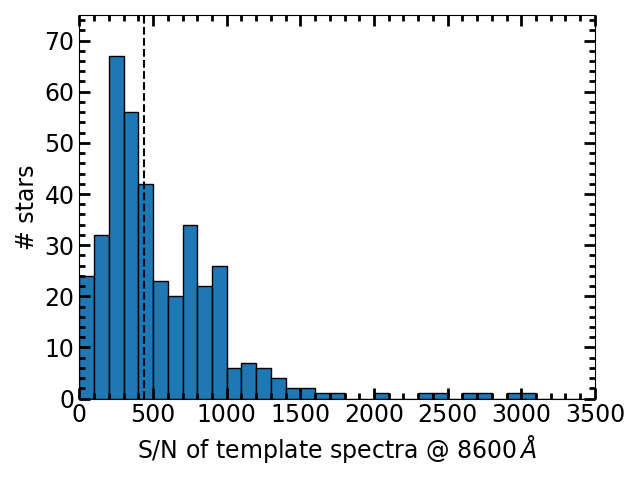}
\includegraphics[width=0.49\textwidth]{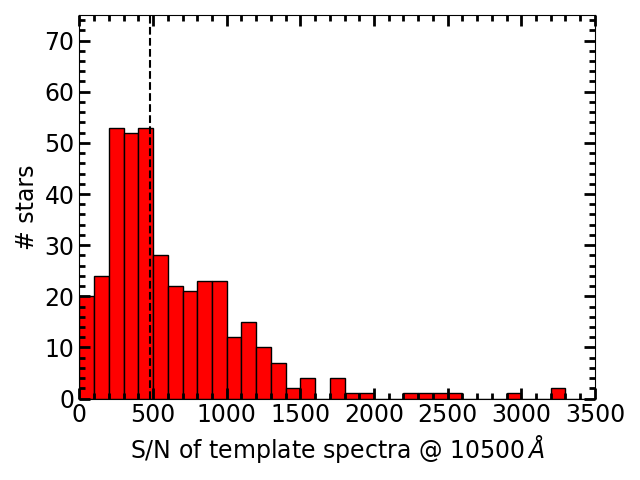}
\caption{\label{figure:hist2}
	Distribution of the S/N of a reference order for the VIS (order 70 around 8\,600\,\r{A}, \textit{left}) 
	and the NIR (order 52 around 10\,500\,\r{A}, \textit{right}). The vertical dashed lines indicate the median values at 438 in the VIS and 482 in the NIR.}
\end{center}
\end{figure*}

These superimposed line profiles are well suited to modeling the shape of the LSF. For this
purpose, we tested different profiles, namely Gaussians and
Lorentzians, generalized Gaussians and Lorentzians with the exponent treated 
as an additional free parameter, as well as Voigt profiles, which are defined by the convolution of Lorentzian and Gaussian distributions. The analysis of all fits
showed that a Voigt profile is the most appropriate for the LSF
modeling. The other profiles were usually not suitable for a proper simultaneous handling
of the line cores and lobes.

In addition, we investigated the variations in the measured line widths, $\Delta\lambda$,
across the detector and found that except for some secondary but
significant features near the detector edges (particularly
in the blue part of the VIS CCD), the line widths still can be
described in a consistent manner. The dominant effect becomes apparent when
the widths of the lines are expressed in terms of wavelength. In the Voigt profile,
the overall width can be described by its Gaussian and 
Lorentzian components. 
If the resolution remains constant, both components should exhibit a linear trend in wavelength. 
However, when they are represented in wavelength-independent units using 
$\Delta\lambda/\lambda_c$, where $\lambda_c$ represents the central
vacuum wavelength of the specific line,
these normalized widths are expected to stay constant if the spectrograph resolution 
is independent of wavelength.

In Fig.~\ref{figure:lsf} we show the measured Gaussian and Lorentzian FWHM components 
of the Voigt profile for both channels. 
The NIR channel has a somewhat lower resolution and therefore presents a wider 
Gaussian component than the VIS channel. However, the widths of the 
Lorentzian component are very similar for both channels redward of 6000\,{\AA}.

To obtain robust estimates of the LSF parameters, we 
used medians for the FWHM of each component and channel
to model the instrumental LSF of CARMENES. 
These parameters are summarized in Table~\ref{table:lsf}. 
Due to the clearly visible deviation in the
blue part of the visual channel, we considered
only lines with a wavelength above 6000\,{\AA}. Moreover, the distribution of the
Lorentzian component, particularly in the NIR part of the spectrum, shows some
conspicuous features.
Some outliers cluster around $12\,500$\,{\AA}, while the scatter seems
generally increased for wavelengths above 15\,000\,{\AA}.
The reasons for these effects could not consistently and objectively be traced back
to line blends or similar effects. 

\begin{figure*}
\begin{center}
\includegraphics[width=0.49\textwidth]{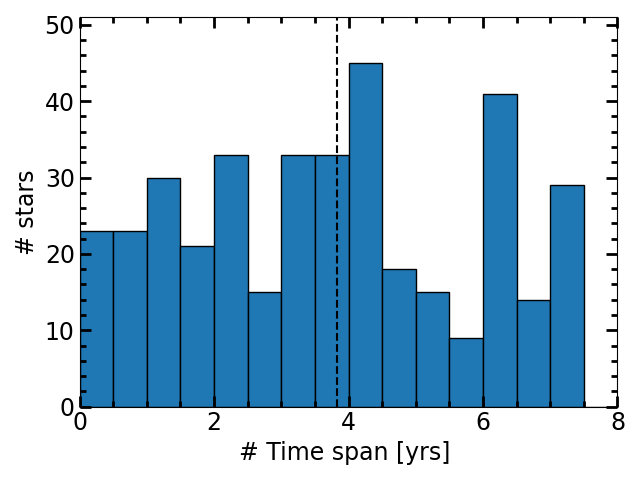}
\includegraphics[width=0.49\textwidth]{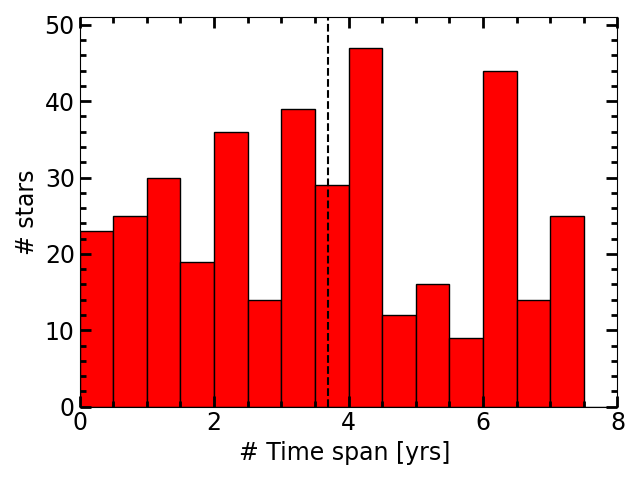}
\caption{\label{figure:hist3}
        Time-span distribution of the observations for the sample stars for the VIS (\textit{left}) and NIR spectra (\textit{right}). 
        The vertical dashed lines indicate the median values at $3.83$ years in the VIS and $3.70$ years in the NIR.}
\end{center}
\end{figure*}

\begin{figure}
\begin{center}
\includegraphics[width=0.49\textwidth]{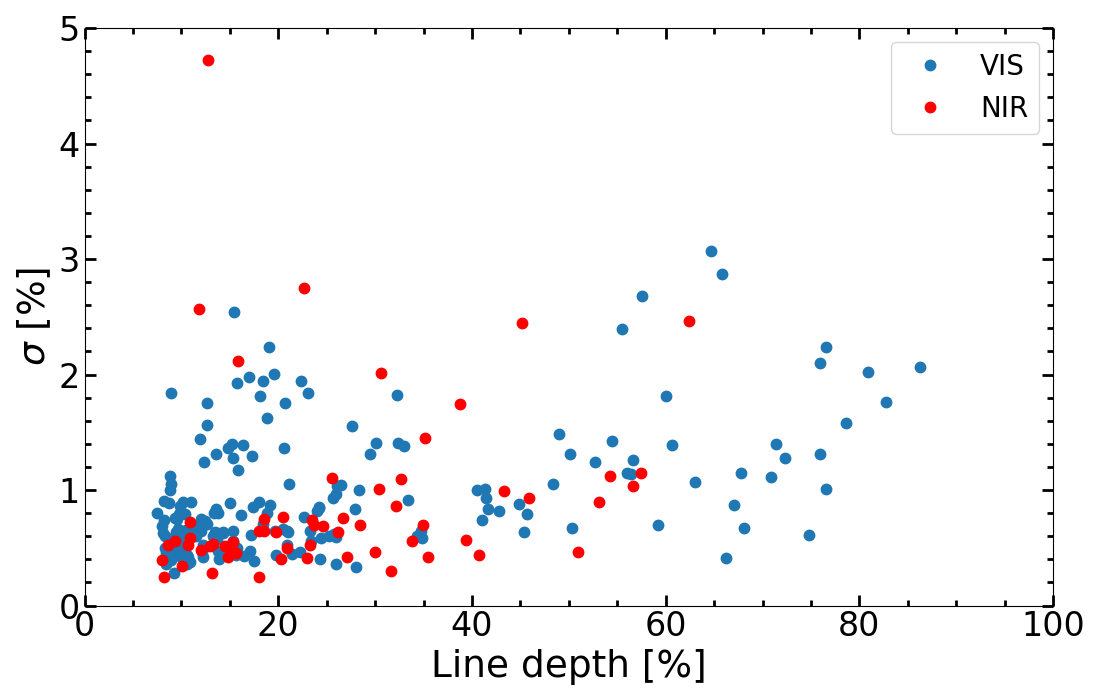}
\includegraphics[width=0.49\textwidth]{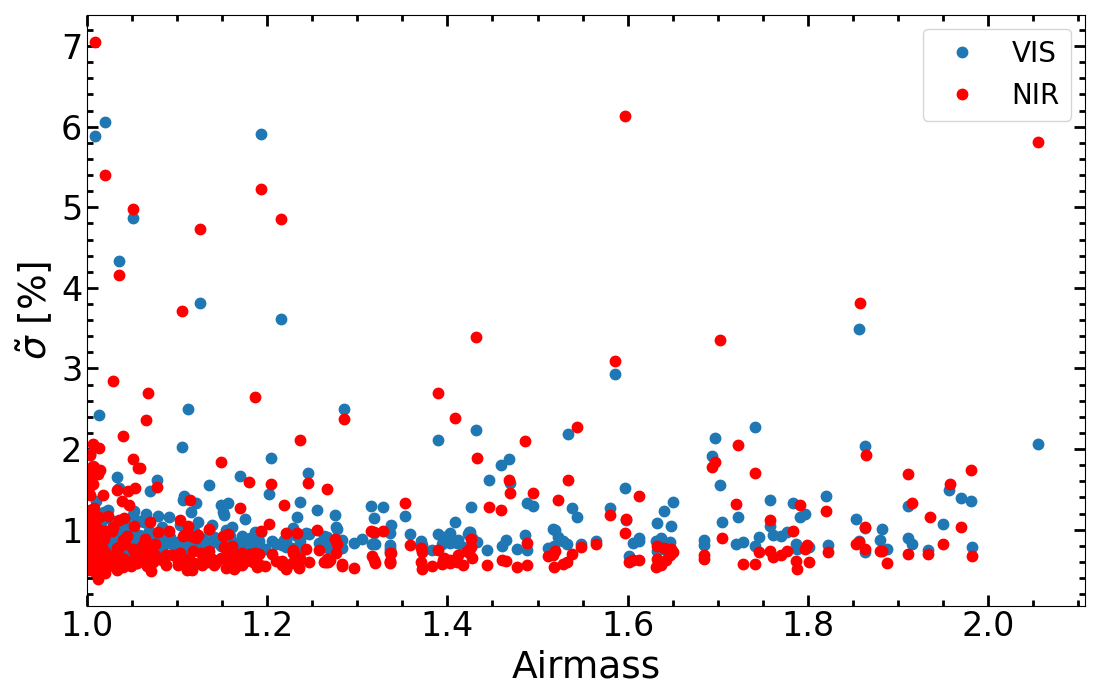}
\includegraphics[width=0.49\textwidth]{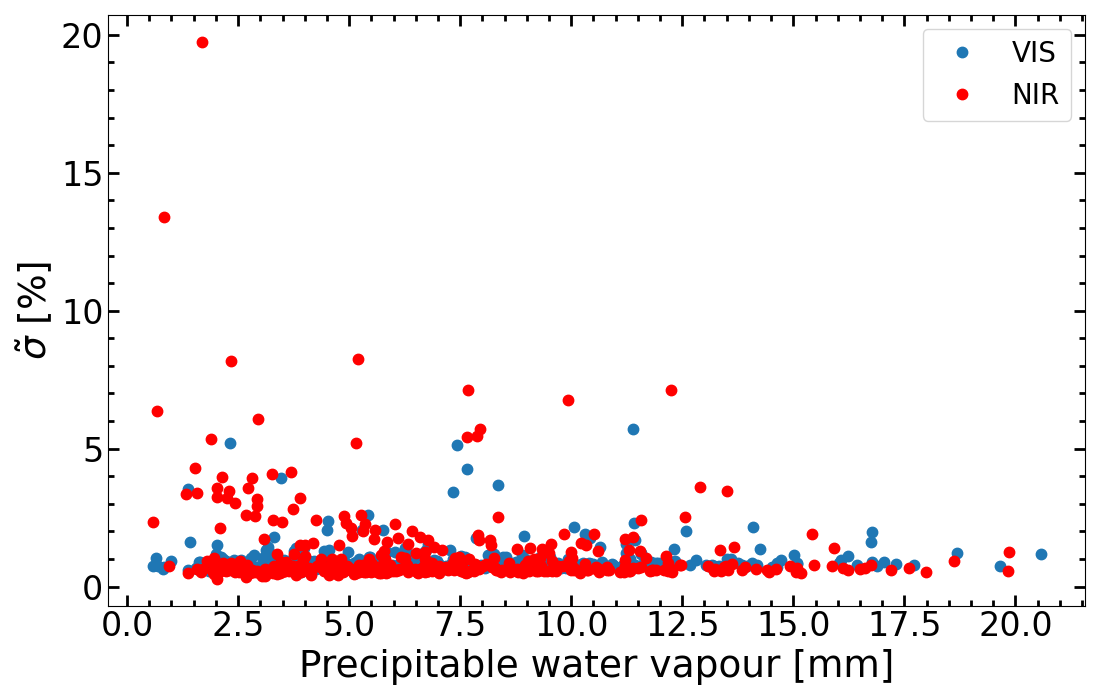}
        \caption{\label{figure:residuals-vs-linedepth}
        Metrics for assessing the quality of the telluric correction.
        \textit{Top panel:} 
        Residual standard deviation ($\sigma$) vs. line depth for a single Wolf~294 VIS (blue) and NIR (red) spectrum. 
        \textit{Middle panel:} Median residual standard deviation ($\tilde{\sigma}$) vs. airmass for all Wolf~294 VIS and NIR spectra. 
        \textit{Bottom panel:} Median $\tilde{\sigma}$ for telluric water bands vs. precipitable water vapor for all Wolf~294 VIS and NIR spectra.}
\end{center}
\end{figure}

\begin{figure*}
\begin{center}
\includegraphics[width=0.49\textwidth]{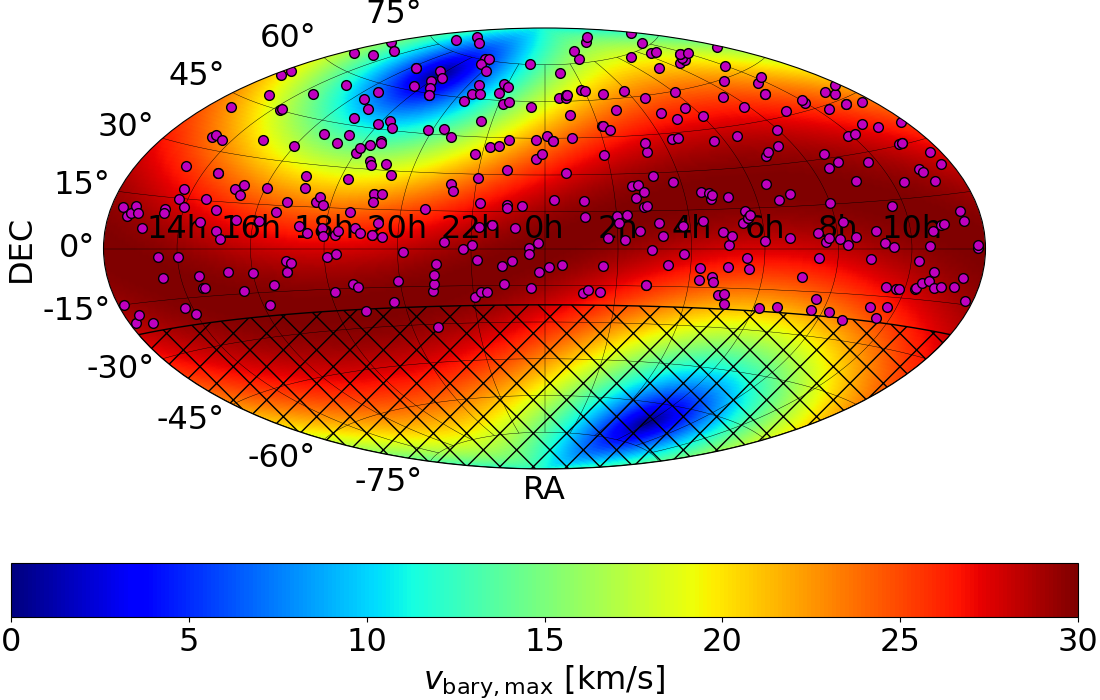}
\includegraphics[width=0.49\textwidth]{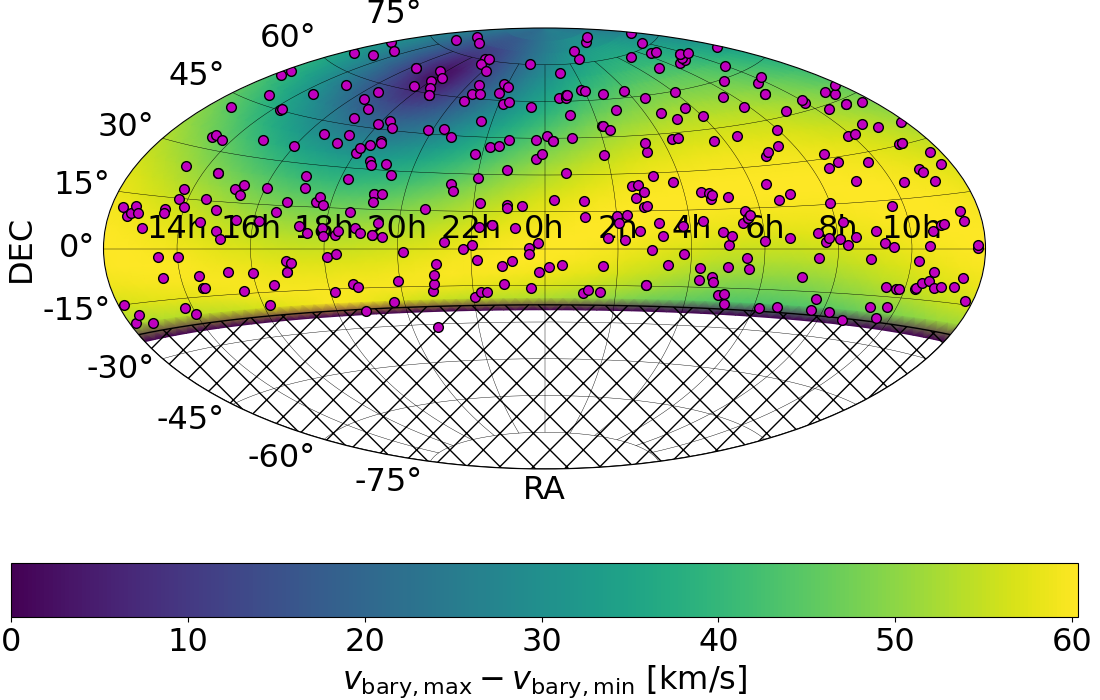}
	\caption{Sky maps in Aitoff projection of the maximum absolute barycentric RV 
	$\varv_{\rm bary,\, max}$ as derived from Eq.~(\ref{equation:baryalphadelta})
	(\textit{left panel}) and full amplitude of the barycentric RV 
	$\varv_{\rm bary,\, max} - \varv_{\rm bary,\, min}$ (\textit{right panel})
	for nights when the star is observable above 30\,deg over the horizon
	for the location of the Calar Alto Observatory. 
	The positions of the 382 sample stars (purple circles) are indicated.
	The hatching pattern indicates the visibility cutoff at $\delta < -23$\,deg 
	of the CARMENES sample.
\label{figure:bervDistmap}}
\end{center}
\end{figure*}

\subsection{Removing the telluric lines}

After fitting a transmission model to the residual telluric spectrum 
with \texttt{molecfit}, we used the best-fit parameters 
as input to compute 
a synthetic transmission model over the entire wavelength range
of the observation with the {\tt calctrans} module of \texttt{molecfit}. 
To derive the spectrum corrected for telluric lines $\hat F_{n,i}$, we finally divided
the CARMENES observation by the transmission model $T(\lambda_{n,i})$,
\begin{equation}
	\hat F_{n,i} = \frac{F_{n,i}}{T(\lambda_{n,i})} \; .
\end{equation}
As an example, we show one observed and its telluric corrected spectrum 
of Wolf~294 in Fig.~\ref{figure:tdtm_vis}d. 

\section{Telluric-corrected M-dwarf template library}
\label{section:results}

\subsection{Sample overview}

We applied the TDTM method to each of the 382 targets in our sample and generated
a spectrum corrected for telluric absorption for each individual VIS and NIR 
CARMENES observation.
The telluric-corrected spectra for each target were subsequently coadded 
to create a telluric-free stellar template spectrum with a high S/N with {\tt serval}. 
We show the example of Wolf~294 in Fig.~\ref{figure:tdtm_vis}e, 
where the displayed order reaches a S/N of 2310. 

To construct the library of telluric-free template spectra with a high S/N, we used a total of 
22\,357 VIS and 20\,314 NIR observations 
obtained between 2016 and 2023 that were 
telluric-corrected with the TDTM method before coaddition.  
In Table~\ref{table:sample} we present the sample of stars we used in this study
and provide information on the targets.
In particular, we tabulate the
CARMENES identifiers, common names, Gliese-Jahreiss numbers, equatorial coordinates at epoch J2000, spectral types,
2MASS $J$ magnitudes \citep{Alonso-Floriano2015, Caballero2016}, 
the number of telluric-corrected VIS and NIR spectra 
used to create the telluric-free stellar templates, and the estimated S/N of a reference order 
at 8\,600\,\r{A} for the VIS templates and at 10\,500\,\r{A} for the NIR templates.
Our template spectra can be downloaded from the CARMENES
GTO Data Archive \citep{Caballero2016b}\footnote{\url{http://carmenes.cab.inta-csic.es}}.

In the VIS, the number of spectra used for the coadding ranged from 5 up to 774, 
with a very similar number of spectra in the NIR (from 4 to 756).
In the reference orders, the range of the S/N in the VIS varies between 9 and 3057, and in the NIR, it varies between 9 
and 3270.
We show the distributions of the number of spectra we used for the coaddition in Fig.~\ref{figure:hist1}.
The median values indicated by the vertical dashed lines in both figures are 34
in the VIS and 30 in the NIR. 
The distributions of the S/N of the template spectra given in the reference orders 
are shown in Fig.~\ref{figure:hist2}, with median values of 438 
in the VIS and 482 in the NIR. 
Finally, we present the time span of the observations covered for each target in Fig.~\ref{figure:hist3},
with median values of 3.83 years in the VIS and 3.70 in the NIR.

\subsection{Correction accuracy}

In the following, we examine the accuracy of the telluric correction applied 
to individual VIS and NIR CARMENES spectra of Wolf~294, using the TDTM approach.
Figures~\ref{figure:accuracy_O2}-\ref{figure:accuracy_CH4} display examples of the corrections 
applied to individual lines within a single spectrum, with a focus on specific molecular constituents. 
For VIS, these molecules include O$_2$ and H$_2$O, while in NIR, they extend to O$_2$, H$_2$O, CO$_2$, and CH$_4$. 
We assessed the accuracy of these corrections by computing the standard deviation $\sigma$ of the residuals. 
In most cases, $\sigma$ was found to be below the 1\,\% level. An exception was observed in the O$_2$ lines 
within the VIS spectrum, which generally fell within the 1--2\,\% range. 
Additionally, some CO$_2$ lines exhibited $\sigma$ values greater than 2\,\%. 
These discrepancies may be attributed to an inadequate removal of the stellar spectrum, 
leading to a distortion in the telluric residual spectrum in the given instance.

To further analyze the variability in the residuals relative to the line depth, 
an automated procedure was implemented. 
This procedure detected lines in the telluric residual spectrum, 
fit a Gaussian model to ascertain the depths, positions ($\mu$), and widths ($\sigma_{\rm Gaussian}$) 
of the telluric lines, and measured the variability in the residuals 
within a wavelength range of \mbox{$\mu \pm 3\,\sigma_{\rm Gaussian}$}. 
The top panel of Fig.~\ref{figure:residuals-vs-linedepth} illustrates the results for a single spectrum of Wolf~294. 
In both CARMENES channels, the $\sigma$ values were confined to a range of 0.3 to 3.0\,\%. 
Optimal outcomes were observed for line depths up to 30\,\%, where $\sigma$ was mostly lower than 1\,\%. 
Our findings indicated a weak or nonexistent correlation between $\sigma$ and line depth.

Subsequently, we explored the variability in the residuals as a function of airmass 
for all VIS and NIR spectra of Wolf~294. 
To this end, we introduced $\tilde{\sigma}$, defined as the median of the measured $\sigma$ 
values within a single Wolf~294 spectrum, and we present the results in the middle panel of Fig.~\ref{figure:residuals-vs-linedepth}. 
Most spectra exhibited a $\tilde{\sigma}$ value of approximately 1--2\,\% in the entire airmass range. 
The discrepancy in $\tilde{\sigma}$ between the VIS and NIR channels is ascribed to the reduced number 
of usable telluric lines in the NIR spectra relative to the VIS spectra (see the top panel of Fig.~\ref{figure:residuals-vs-linedepth}), 
leading to reduced scatter and, consequently, a lower value for $\tilde{\sigma}$. 
However, a minor subset of spectra revealed $\tilde{\sigma}$ values that reached up to 6--7\,\%. 
A visual inspection identified these spectra as characterized by low S/N.

Finally, we calculated $\tilde{\sigma}$ and only included H$_2$O telluric lines. We 
display the results as a function of precipitable water vapor derived with \texttt{molecfit}, 
as illustrated in the bottom panel of Fig.~\ref{figure:residuals-vs-linedepth}. 
Although the $\tilde{\sigma}$ values derived in the NIR exhibit greater scatter than the VIS spectra, 
the findings in both channels agree and did not disclose any discernible trends.

Overall, our results indicate that the correction accuracy mostly depends on the S/N of the data
and to a much lesser extent on the depth of the telluric features. 
Most telluric features can be corrected to basically within
the noise level. Deep lines are problematic for two reasons. First, the S/N decreases in their
cores, and second, lines deeper than $50\,\%$ can
leave comparatively large systematic residuals in the corrected spectra, 
especially in the line cores. 
These residuals may be attributed to the fact that small discrepancies 
between model and observation can result in large residuals, especially in deep lines. 
These discrepancies may result from uncertainties in the
line strengths listed in the HITRAN database, 
leading to inaccurate column density fits \citep{Seifahrt2010, Gordon2011}. 
Potential uncertainties in the line positions affect the wavelength calibration, causing 
P Cygni-like residuals in the worst cases. 
Incomplete corrections may also arise from the instrumental line profile model.
The Gaussian and Lorentzian profile parameters from different lines show some scatter 
and are approximated by a constant value (see Sect.~\ref{appendix:lsf}).
Another source of uncertainty is the approximation of 
Eq.~(\ref{equation:spec}) by Eq.~(\ref{equation:spec2}), which becomes particularly relevant 
for blends between telluric and stellar lines \citep[][their Appendix~\ref{appendix:convolution}]{Sameshima2018}.

We find that most telluric absorption features are corrected 
to within 2\,\% or better of the continuum standard deviation, 
as reported by \citet{Smette2015}. 
For objects with numerous intrinsic features,
the authors proposed to apply \texttt{molecfit} to TSSs observations taken
with the same instrumental setup as the science object to solve for the polynomial continuum coefficients. These results are then used as fixed input to subsequently apply \texttt{molecfit} 
for the science observation's telluric correction.
Our approach, however, is able to accurately and directly extract telluric features 
of various strengths for wavelength ranges with $M_k > 0$,
even for feature-rich objects such as M dwarfs,
whose spectra are dominated by numerous molecular features in the VIS and NIR bands. 

\begin{figure}
\begin{center}
\includegraphics[width=0.46\textwidth]{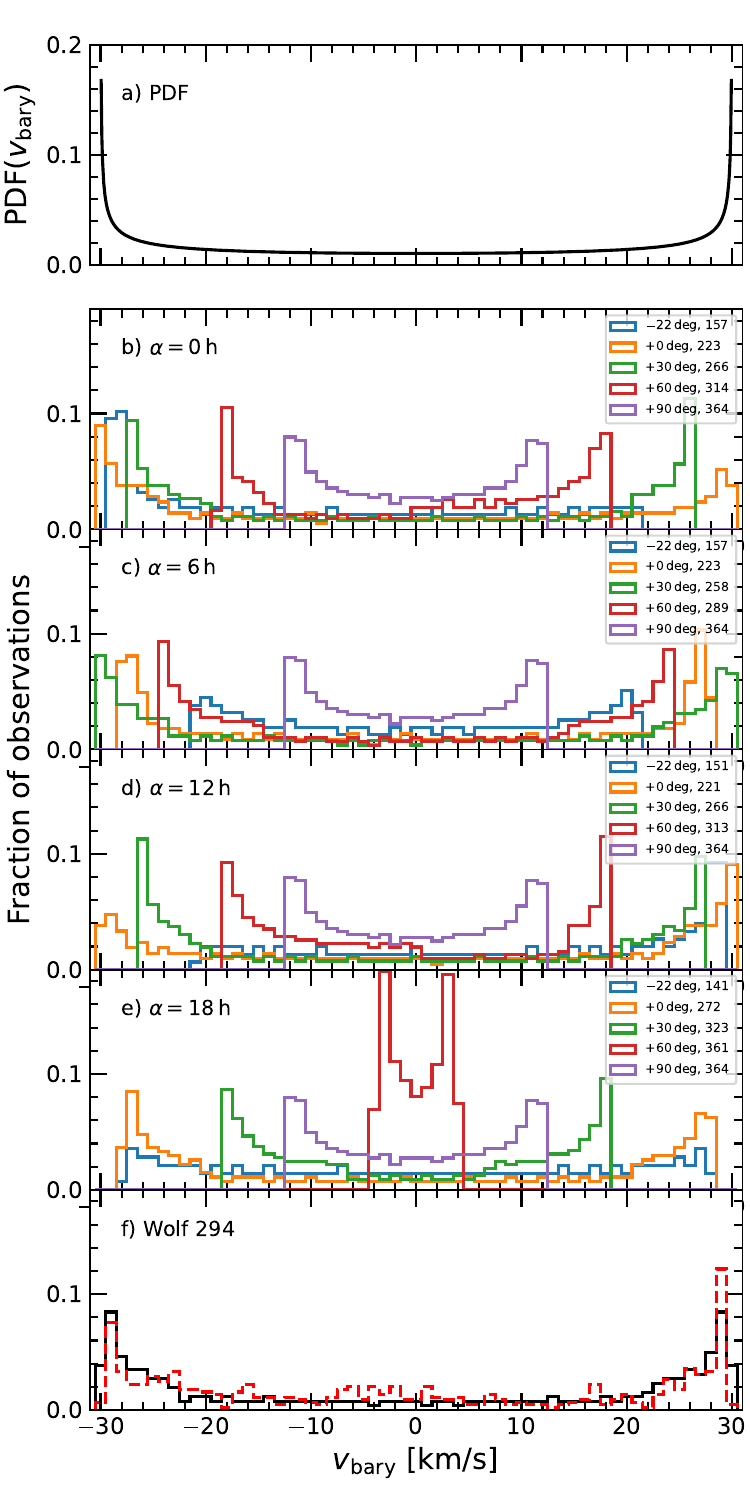}
\caption{\label{figure:berv_dist}
Barycentric velocity distributions corresponding to various coordinate sets.
	\textit{Panel a:}
	Probability density function $\mathrm{PDF}(\varv_{\rm bary})$ as computed with Eq.~(\ref{equation:pdf}).
	\textit{Panels b-e:}
	Simulated barycentric velocity distributions for targets with 
	$\alpha = 0\,$h, 6\,h, 12\,h, and 18\,h, and $\delta = -22\,{\rm deg}, 0\,{\rm deg}, +30\,{\rm deg}, +60\,{\rm deg}, +90\,{\rm deg}$, 
	assuming one daily measurement over a time span of one year and considering the target visibility.
	The second entry in the legend is the number of total observing nights.
\textit{Panel f:} 
	Simulated barycentric velocity distribution of Wolf~294 (black line).
	The barycentric velocity distribution for the datasets of Wolf~294 (dashed red line) is overplotted.}
\end{center}
\end{figure}

\subsection{Visibility constraints}

For the TDTM technique to work properly, a good template is essential. While a lack of S/N
may be addressed by taking more observations, which may pose a practical but not a fundamental problem,
the relative shift between telluric and stellar lines
is also crucial for the template construction. 
In the case of planet-induced reflex motion, the barycentric motion of
the Earth and its rotation dominate the sum of relative shifts.
The maximum absolute barycentric velocity $\varv_{\rm bary,\, max}$ 
depends on the ecliptic latitude $\beta$ of a target,
\begin{equation}
	\varv_{\rm bary,\, max} \approx 30\,\mathrm{km\,s}^{-1} |\cos \beta| \; .
\end{equation}
The transformation from the ecliptic $(\lambda, \beta)$ to the equatorial system $(\alpha, \delta)$
with the axial tilt of the Earth $\varepsilon\approx 23.44$\,deg yields
\begin{equation}
	\sin \beta  = \sin \delta \cos \varepsilon - \cos \delta \sin \alpha \sin \varepsilon \,.
\end{equation}
Thus, $\varv_{\rm bary,\,max}$ as a function of $\alpha$ and $\delta$ reads
\begin{equation}
	\varv_{\rm bary,\, max}(\alpha,\delta) \approx 30\,\mathrm{km\,s}^{-1} \sqrt{1-\sin^2 \beta(\alpha,\delta)}\,,
	\label{equation:baryalphadelta}
\end{equation}
which provides a sky map as in the left panel of Fig.~\ref{figure:bervDistmap}
showing the maximum offset between telluric lines and stellar lines.
Except for objects situated near the ecliptic poles, $\varv_{\rm bary,\, max}$
is larger than the natural line width 
of the tellurics and the instrumental resolution ($\sim 3\,$km\,s$^{-1}$), which is
required to disentangle the stellar and telluric spectra. 

To examine the largest possible improvement on the template completeness,
we carried out a simulation of the full amplitude of the barycentric velocity 
range $\varv_{\rm bary,\, max} - \varv_{\rm bary,\, min}$ 
as a function of ecliptic coordinates down to the visibility cutoff 
for stars with $\delta < 23$\,deg, which is also the CARMENES survey limit
\citep{Garcia-Piquer2017}. In particular,
we computed the dates on which a target is 30\,deg above the horizon 
between the astronomical dusk and dawn for the location of the Calar Alto Observatory,
and calculated the barycentric RV at midnight for these dates. 
In the right panel of Fig.~\ref{figure:bervDistmap}, we show the 
difference between the maximum and minimum barycentric RV 
$\varv_{\rm bary,\, max} - \varv_{\rm bary,\, min}$ 
for each pair of coordinates. 

The observed barycentric velocity range is further constrained by the visibility.
The main contribution to the barycentric velocity is a yearly sinusoid.
We considered an object with a half-year visibility period, during which 
the barycentric Earth RV changes from a maximum to a minimum. In particular,
we considered 
the first half of a cosine
\begin{equation}
	\varv_{\rm bary}(t) = \varv_{\rm bary,\, max} \cdot \cos \left( \frac{2\pi t}{T_{\oplus}}\right) \; ,
\end{equation}
where $T_{\oplus}$ is the orbital period of the Earth. When the target is observed at a random time,
the probability to observe a barycentric RV shift equal to or smaller
than $\varv_{\rm obs}$ is
\begin{equation}
	P(\varv_{\rm bary} \le \varv_{\rm obs}) = 1 - \frac{1}{\pi} \arccos \left(\frac{\varv_{\rm obs}}{\varv_{\rm bary,\, max}}\right) \; . 
\end{equation}
The probability density function, $\mathrm{PDF}(\varv_{\rm bary})$,
to observe the target in some interval of barycentric velocity is
given by
\begin{equation}
	\label{equation:pdf}
	\mathrm{PDF}(\varv_{\rm bary}) = \frac{1}{\pi \; \varv_{\rm bary,\, max}\sqrt{1 - \left(\frac{\varv_{\rm bary}}{\varv_{\rm bary,\, max}}\right)^2 }} \; ,
\end{equation}
which is shown in Fig.~\ref{figure:berv_dist}a.

To study the barycentric velocity distribution as a function of right ascension
and declination over one year for the location of the Calar Alto Observatory, 
we carried out simulations for a set of coordinates with 
$\alpha = 0\,$h, 6\,h, 12\,h, and 18\,h, and 
$\delta = 0\,{\rm deg}, +30\,{\rm deg}, +60\,{\rm deg}$, and $+90\,{\rm deg}$.
In addition, we included a southern coordinate sample with $\delta=-22$\,deg, which is near
the visibility limit of the CARMENES survey.
Assuming that objects can generally only be observed down to an elevation of
$30$\,deg, we determined the nights when the target is observable 
and computed the barycentric velocity at the time between evening and 
morning astronomical twilight.
Our results are presented in Fig.~\ref{figure:berv_dist}b-e.
The simulated distributions show a plateau around $\varv_{\rm bary}=0$ and increasing slopes at 
both ends where $\varv_{\rm bary}\rightarrow \pm \varv_{\rm bary,\, max}$.
This shape is a consequence of the regular sampling of the yearly sinusoidal 
barycentric velocity contribution.
However, the simulations show that all barycentric velocities between 
$-\varv_{\rm berv,\, max}$ and $+\varv_{\rm berv,\, max}$ are covered. 
We finally present the predicted and observed barycentric velocity 
distribution of Wolf~294 (Fig.~\ref{figure:berv_dist}f) of our spectroscopic data set.

\subsection{Total masking fraction}
In the case of Wolf~294, a mask that flags telluric features deeper than 1\,\% 
results in a total masking fraction $\gamma_{\rm max,\,VIS} = 33.5\,\%$ 
for the optical spectral range (0.52--0.96\,$\mu$m) and 
$\gamma_{\rm max,\,NIR} = 62.8\,\%$ for the NIR (0.96--1.71\,$\mu$m) when using one observation.
As the number of observations with different barycentric velocities 
is increased, the total masked fraction decreases 
and converges to a limit of $\gamma_{\rm min, VIS} = 16.5\,\%$
and $\gamma_{\rm min, NIR} = 42.5\,\%$ for the 1\,\% mask. 
These limits are mainly defined by broad telluric features, for example, the strong water bands 
centered around 1.15\,$\mu$m and 1.4\,$\mu$m, which prevent further reduction of 
masked regions. 

The difference between the maximum and minimum masking fraction is the 
total fraction that is gained for the modeling of the telluric spectrum, which is
$\gamma\,'_{\rm max, VIS} = 17.0\,\%$ 
in the optical and $\gamma\,'_{\rm max, NIR} = 20.3\,\%$ 
in the NIR employing the 1\,\% mask.
In the VIS, the useable amount of the stellar spectrum was increased from 66.5\,\% to 83.5\,\%, 
which corresponds to an increase of 25.6\,\%. The increment 
is even larger in the NIR, where the useable range was increased 
from 37.2\,\% to 57.5\,\%, which corresponds to a relative growth of 54.6\,\%.

\begin{figure}
\begin{center}
\includegraphics[width=0.49\textwidth]{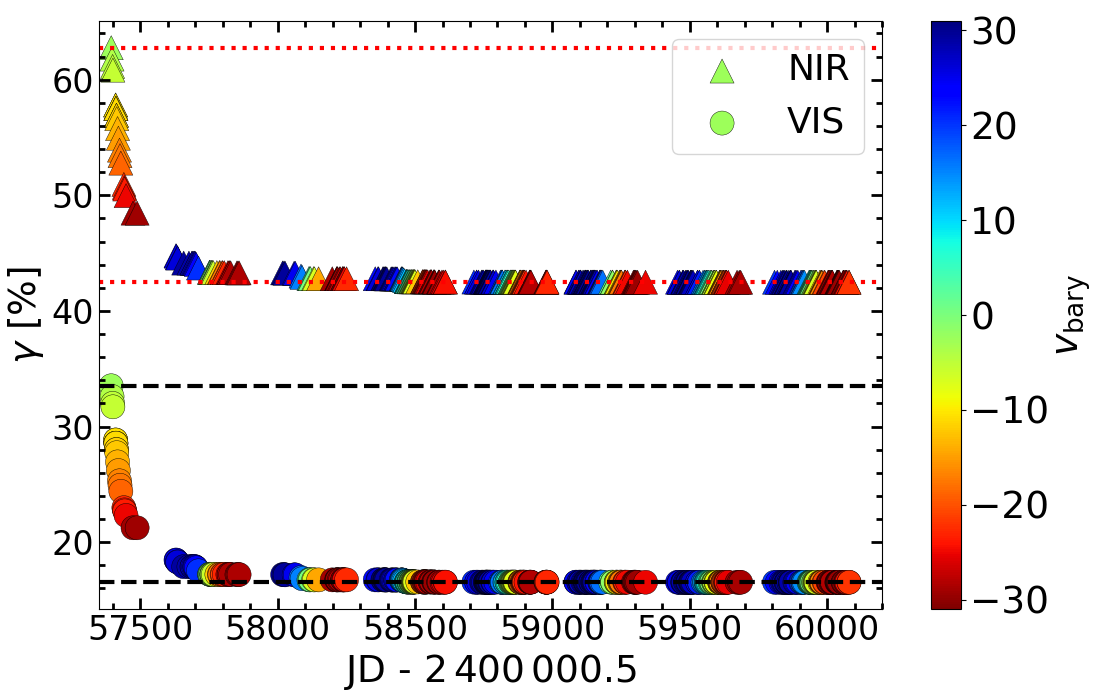}
\caption{\label{figure:maskFraction}
Evolution of the total masked wavelength fraction color-coded with the barycentric 
velocity for Wolf~294. 
The circles represent the VIS channel (\textit{bottom}) and the triangles the NIR channel (\textit{top}).
The upper and lower dashed black lines mark the maximum and minimum limits of the total masked 
fraction in the optical wavelength range. The dotted red lines mark the same limits for the NIR
wavelength range.}
\end{center}
\end{figure}

We show the evolution of the total masked fraction using the 1\,\% mask
for Wolf~294 Fig.~\ref{figure:maskFraction}. 
To reach the lower limit of the total masked fraction, it is necessary to 
cover the full range of barycentric velocities. 
As shown in Fig.~\ref{figure:maskFraction}, 
this requirement was fulfilled for Wolf~294 after the first observing season. 
Additionally, a limited number of observations at approximately $\pm30\,$km\,s$^{-1}$ 
substantially reduces the total masked fraction. 
Our simulations indicate that with six observations that are uniformly distributed across the barycentric velocity space of $\pm30$\,km\,s$^{-1}$, values of 17.9\,\% in the VIS and 40.3\,\% in the NIR can be achieved. However, for proper telluric correction, a barycentric velocity coverage 2-5 times larger than the minimum 3\,km\,s$^{-1}$ required to resolve the lines is essential, which means that velocity spans of 6–15\,km\,s$^{-1}$ are imperative.

\section{Summary and conclusions}
\label{section:conclusion}

We presented the TDTM technique, a combination of data-driven (template construction)
and forward-modeling (spectral fitting) methods
to correct for telluric absorption lines in high-resolution 
VIS and NIR spectra. 
After applying a telluric mask to a time series of spectra, 
we constructed a stellar template with a high S/N and used it to remove
the stellar contribution from the observations. 
We then used \texttt{molecfit} to fit an atmospheric transmission model 
to the resulting telluric spectrum and finally corrected
the target spectrum at all wavelengths. 
While the telluric correction via spectral modeling with \texttt{molecfit} alone
is especially challenging for late-type stars with their high density of molecular lines,
the TDTM technique is applicable to spectra of any spectral type. 
Although we chose the \texttt{molecfit} code to fit the transmission 
model to the telluric spectra in this work, 
other software packages that produce 
synthetic atmospheric transmission spectra, such as 
{\tt TelFit}\footnote{\url{https://pypi.org/project/TelFit/}}
\citep{Gullikson2014} or TAPAS \citep{Bertaux2014},
could be used. 

Telluric correction with our method works best with a high number 
of observations and good coverage of the barycentric velocity space. 
To demonstrate the performance of our correction method,
we applied it to high-resolution optical and NIR
CARMENES observations of the M3.0\,V star Wolf~294. 
We found that \texttt{molecfit} corrects most telluric lines close to noise level 
in the residual telluric transmission spectrum obtained with the help of the template.
Moreover, we did not find a correlation between the correction accuracy and the depth 
of telluric lines.

We applied the TDTM approach to the whole CARMENES survey sample, which comprises 382 
M-dwarf stars. After correcting for telluric lines, we coadded the individual observations
and constructed telluric-corrected high-resolution template spectra with a high S/N
for each of our targets. Our spectral library is publicly available and can be found on 
the CARMENES data archive homepage. 

\begin{acknowledgements}
  We acknowledge the support of the Deutsche Forschungsgemeinschaft (DFG) priority program SPP 1992 ``Exploring the Diversity of Extrasolar Planets'' (CZ 222/3-1, JE 701/5-1).

  CARMENES is an instrument at the Centro Astron\'omico Hispano en Andaluc\'ia (CAHA) at Calar Alto (Almer\'{\i}a, Spain), operated jointly by the Junta de Andaluc\'ia and the Instituto de Astrof\'isica de Andaluc\'ia (CSIC).
    
  CARMENES was funded by the Max-Planck-Gesellschaft (MPG), 
  the Consejo Superior de Investigaciones Cient\'{\i}ficas (CSIC),
  the Ministerio de Econom\'ia y Competitividad (MINECO) and the European Regional Development Fund (ERDF) through projects FICTS-2011-02, ICTS-2017-07-CAHA-4, and CAHA16-CE-3978, 
  and the members of the CARMENES Consortium 
  (Max-Planck-Institut f\"ur Astronomie,
  Instituto de Astrof\'{\i}sica de Andaluc\'{\i}a,
  Landessternwarte K\"onigstuhl,
  Institut de Ci\`encies de l'Espai,
  Institut f\"ur Astrophysik G\"ottingen,
  Universidad Complutense de Madrid,
  Th\"uringer Landessternwarte Tautenburg,
  Instituto de Astrof\'{\i}sica de Canarias,
  Hamburger Sternwarte,
  Centro de Astrobiolog\'{\i}a and
  Centro Astron\'omico Hispano-Alem\'an), 
  with additional contributions by the MINECO, 
  the DFG through the Major Research Instrumentation Programme and Research Unit FOR2544 ``Blue Planets around Red Stars'', 
  the Klaus Tschira Stiftung, 
  the states of Baden-W\"urttemberg and Niedersachsen, 
  and by the Junta de Andaluc\'{\i}a.
  
  We used data from the CARMENES data archive at CAB (CSIC-INTA).
  
  We acknowledge financial support from the Agencia Estatal de Investigaci\'on (AEI/10.13039/501100011033) of the Ministerio de Ciencia e Innovaci\'on and the ERDF ``A way of making Europe'' through projects 
  PID2021-125627OB-C31,		
  PID2019-109522GB-C5[1:4],	
and the Centre of Excellence ``Severo Ochoa'' and ``Mar\'ia de Maeztu'' awards to the Instituto de Astrof\'isica de Canarias (CEX2019-000920-S), Instituto de Astrof\'isica de Andaluc\'ia (CEX2021-001131-S) and Institut de Ci\`encies de l'Espai (CEX2020-001058-M).

  This work was also funded by the Generalitat de Catalunya/CERCA programme, 
and the Israel Science Foundation through grant No. 848/16.

  We thank the anonymous referee for carefully reading our manuscript and for offering constructive comments that substantially helped to improve its quality.
\end{acknowledgements}

\bibliographystyle{aa}
\bibliography{bibs}

\begin{appendix}
\clearpage

\section{Approximation of the convolution equation}
\label{appendix:convolution}

In an actual observation, the product of the stellar spectrum, $s(\lambda)$, and
the telluric transmission spectrum, $t(\lambda)$, is received by the system of telescope and
spectrograph, which we call the instrument. The effect of the instrument on the received
spectrum is modeled by a convolution with the instrumental LSF, $L(\lambda)$, so that
we obtain the left-hand side of the equation,
\begin{equation}
	[s(\lambda)\cdot t(\lambda)] \otimes L(\lambda) \approx [s(\lambda) \otimes L(\lambda)] \cdot[t(\lambda)
	\otimes L(\lambda)] \; .
     \label{equation:conv}
\end{equation}
As the convolution cannot easily be inverted, the left-hand side of
Eq.~(\ref{equation:conv}) is
frequently approximated as a product of two spectra separately convolved
with the LSF. As previously pointed out, for instance, by \citet{Sameshima2018},
the two sides of the equation differ.
In their analysis, \citet{Sameshima2018} found that a telluric-corrected
spectrum
based on this approximation
differs noticeably from the stellar spectrum if the width of the stellar
lines
is comparable to or narrower than the instrumental resolution
and if the stellar lines are heavily blended with telluric features.

We analytically investigated the difference between the two sides of
Eq.~(\ref{equation:conv}), making
the simplifying assumption of normality for the stellar, telluric, and
instrumental line profiles.
In particular, we adopted
\begin{eqnarray}
     s(\lambda) &=& \frac{A_s}{\sqrt{2\pi \sigma_s^2}}
\exp\left(-\frac{(\lambda-\mu_s)^2}{2\sigma_s^2}\right) \nonumber \\
     &=& A_s\, N(\mu_s, \sigma_s^2) \; , \\
     t(\lambda) &=& A_t\, N(\mu_t, \sigma_t^2) \; , \\
     L(\lambda) &=& N(0, \sigma_L^2) \; .
\end{eqnarray}
Substitution of $s(\lambda)$, $t(\lambda)$, and $L(\lambda)$ into the left-hand side
of Eq.~(\ref{equation:conv}) leads to
\begin{eqnarray}
        [s(\lambda) \times t(\lambda)] \otimes L(\lambda) \nonumber \\
     = \frac{A_s A_t}{2 \pi \sigma_z^2}
     \exp \left(-\frac{(\lambda - \mu_t)^2 \sigma_s^2 + (\lambda - \mu_s)^2 \sigma_t^2 + (\mu_s-\mu_t)^2 \sigma_L^2}{2 \sigma_z^4}\right) \label{equation:leftside}\\
	= a \,,
\end{eqnarray}
\noindent where $\sigma_z^4 = \sigma_s^2 \sigma_t^2 + \sigma_s^2 \sigma_L^2 +
\sigma_t^2\sigma_L^2$ and $a$ abbreviates Eq.~(\ref{equation:leftside}). Using
\begin{eqnarray}
     s(\lambda) \otimes L(\lambda) &=& A_s\, N(\mu_s, \sigma_s^2 +
\sigma_L^2) \\
     t(\lambda) \otimes L(\lambda) &=& A_t\, N(\mu_t, \sigma_t^2 +
\sigma_L^2)\,,
\end{eqnarray}
we find that
the substitution of $s(\lambda)$, $t(\lambda)$, and $L(\lambda)$ into
the right-hand side
of Eq.~(\ref{equation:conv}) leads to
\begin{eqnarray}
	[s(\lambda) \otimes L(\lambda)] \cdot [t(\lambda) \otimes L(\lambda)] &=&
\nonumber \\
	\frac{A_s A_t}{2 \pi \sqrt{\sigma_z^4 + \sigma_L^4}} \exp\left(-\frac{(\lambda - \mu_s)^2}{2 (\sigma_s^2 + \sigma_L^2)} - \frac{(\lambda - \mu_t)^2}{2(\sigma_t^2 + \sigma_L^2)}\right) \label{eq:right} = b\,,
\end{eqnarray}
where $b$ again abbreviates our findings.

\subsection{Estimating the difference}

The details of the differences between the two sides of 
Eq.~(\ref{equation:conv}) depend on the summands represented by 
$a$ and $b$ and, therefore, the parameters $\mu_{s,t}$, $A_{s,t}$, and $\sigma_{s,t,L}$. 
Some estimates can be obtained in special cases.

\begin{figure}
\begin{center}
\includegraphics[width=0.49\textwidth]{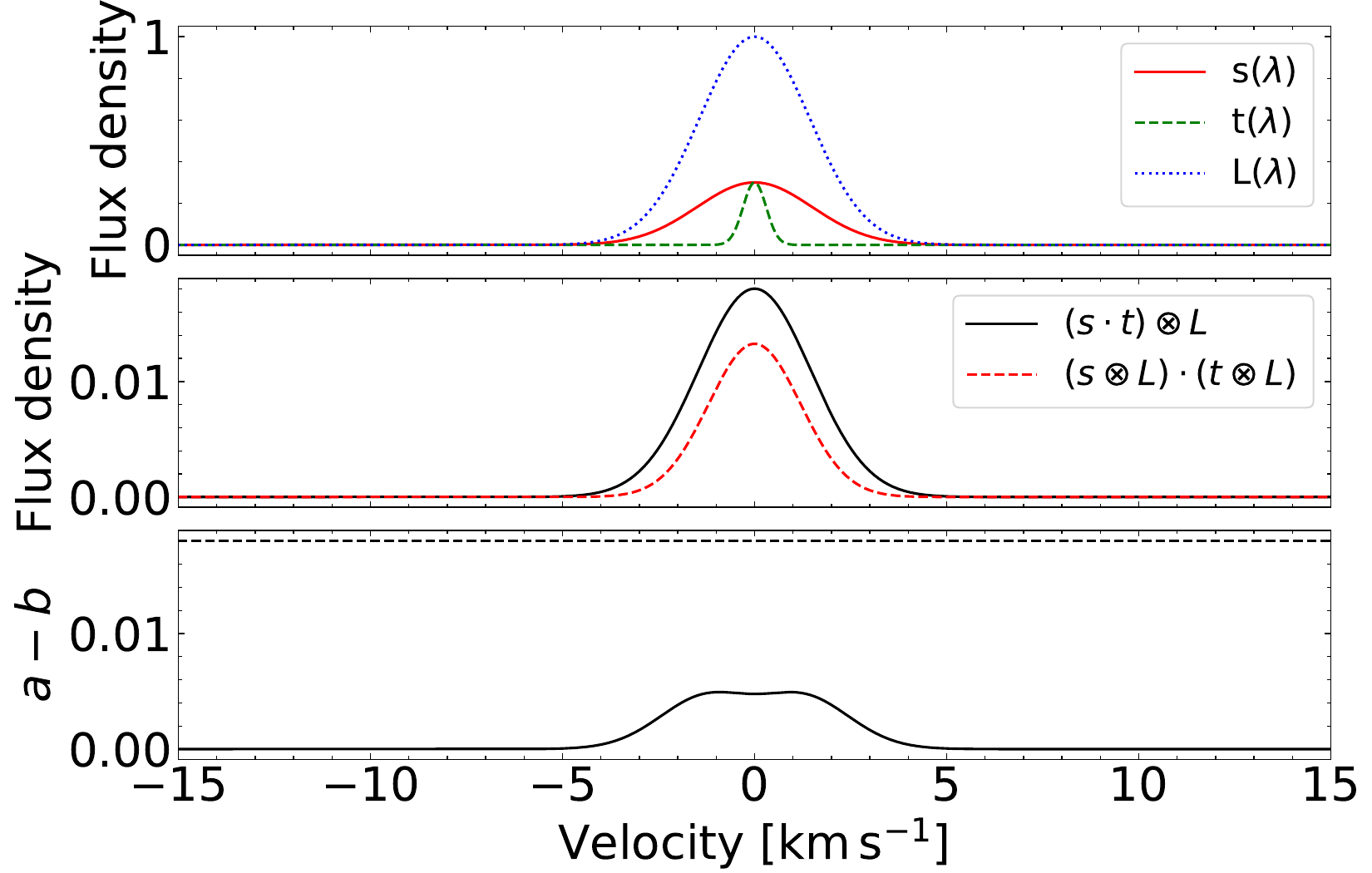}
	\caption{
	Approximation of the convolution function for the special case where
        a stellar line is blended by a telluric line ($\mu_s=\mu_t$).
	\textit{Upper panel:} 
	Artificial line profiles representing a stellar line (red line), 
	a telluric line (dashed green line),	and the instrumental line profile (dotted blue line,
	normalized for the sake of clarity).
	\textit{Middle panel:} Left and right side of Eq.~(\ref{equation:conv}).
        \textit{Lower panel:} Difference between the left and right side of 
	Eq.~(\ref{equation:conv}) expressed by $a-b$. The dashed line indicates the limit
	set by the product of the line depths, 
 $d_s d_t \sigma_s \sigma_t \sigma_z^{-2}$.        \label{figure:lsf_example}}
\end{center}
\end{figure}

Assuming a perfectly blended stellar and telluric line $\mu_s=\mu_t=\lambda$, 
we obtain for the difference between Eq.~(\ref{equation:leftside}) and Eq.~(\ref{eq:right})
\begin{eqnarray}
	a - b &=& \frac{A_s A_t}{2 \pi \sigma_z^2} \left( 1 - \frac{1}{\sqrt{1 + \frac{\sigma_L^4}{\sigma_z^4}}}\right) \nonumber \\
	&=& d_s d_t \frac{\sigma_s \sigma_t}{\sigma_z^2} \left(1 - \frac{1}{\sqrt{1 + \frac{\sigma_L^4}{\sigma_z^4}}} \right)\; , \label{eq:ab}
\end{eqnarray}
where  $d_{s,t}$ denote the line depths,
\begin{equation}
     d_{s,t} = \frac{A_{s,t}}{\sqrt{2\pi \sigma_{s,t}^2}} \; .
\end{equation}
In general, $\sigma_z^2 \ge \sigma_s \sigma_t$, and we find $a - b \ge 0$,
from which it follows that $a - b < d_s d_t \sigma_s \sigma_t \sigma_z^{-2}$.
Under the stated assumptions, we therefore find that the accuracy of the
approximation depends on the product of the line depths. 
This also follows from the linearity of the convolution operator 
in Eq.~(\ref{equation:conv}).

In Fig.~\ref{figure:lsf_example},
we illustrate the differences between the two sides of
Eq.~(\ref{equation:conv})
for the special case of a central blend of a stellar and a telluric line
(i.e., $\mu_s=\mu_t$) for parameters appropriate for the
CARMENES spectrograph.
In our approximation, we consider only normal line profiles.
We choose the standard deviation of the instrumental line profile,
$\sigma_L$, such that its FWHM matches that of the true
VIS channel Voigt line profile of CARMENES (Table~\ref{table:lsf}),
which yields $\sigma_L=0.048\,$\r{A}.
We assumed stellar and telluric lines located at
$\mu_s=\mu_t=10\,000\,$\r{A}
with line depths of $d_s = 0.3$ and $d_t = 0.3$.
The intrinsic stellar and telluric line widths
(before convolution by the instrumental LSF) were estimated from
a synthetic PHOENIX spectrum \citep{Husser2013} and a telluric transmission
model, which yields $\sigma_s=0.05\,$\r{A} and $\sigma_t = 0.01\,$\r{A}.
We present the resulting Gaussian profiles in the upper panel
of Fig.~\ref{figure:lsf_example}, the left and right side of
Eq.~(\ref{equation:conv})
in the middle panel of Fig.~\ref{figure:lsf_example},
and the difference $a-b$ as well as the limit expressed by
$d_s d_t \sigma_s \sigma_t \sigma_z^{-2}$ 
in the lower panel of Fig.~\ref{figure:lsf_example}.
For the adopted parameters,
the difference between the left and right side
of Eq.~(\ref{equation:conv}) is approximately 0.5\,\%. This is in the range
of $2$\,\% cited as the correction accuracy of \texttt{molecfit}. For shallower lines, 
the effect is less pronounced.
In the extreme case of a very high instrument resolution represented by the
limit $\sigma_L \rightarrow 0$,
we find that $\sigma_z^2$ approaches $\sigma_g\sigma_h$ so that 
Eq.~(\ref{eq:ab}) approaches zero and
the difference vanishes. 

Our results allow obtaining an estimate of the accuracy of the
approximation in Eq.~(\ref{equation:conv})
based on a normal approximation of the line profiles. They are also
consistent with the findings of \citet{Sameshima2018}.

\clearpage
\onecolumn

\section{Additional figures}

\begin{figure*}[h]
\begin{center}
\includegraphics[width=\textwidth]{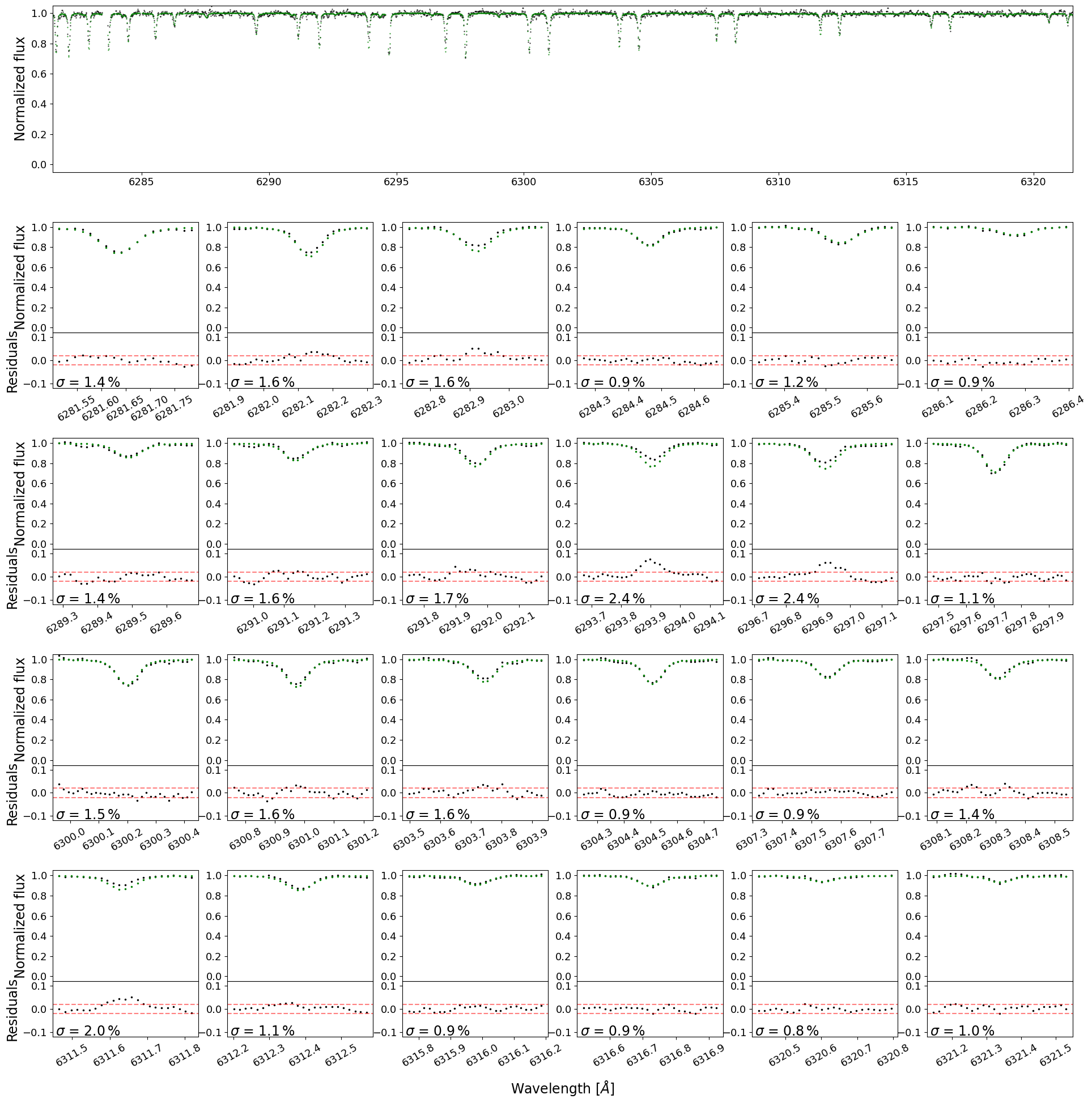}
	\caption{
            Example of telluric line fits for the O$_2$ band in a single Wolf~294 spectrum.
            The top panel displays the residual telluric spectrum (black dots) 
            after dividing the science spectrum by the stellar template, 
            and the best-fit telluric model (green dots) derived with \texttt{molecfit}. 
            The subplots illustrate the individual telluric lines (top) and 
            the residuals (bottom, black dots), with dashed red lines marking 
            a $2\,\%$ deviation between data and model. The standard deviation $\sigma$ 
            of the residuals is also shown in each subplot.
            \label{figure:accuracy_O2}}
\end{center}
\end{figure*}

\begin{figure*}
\begin{center}
\includegraphics[width=\textwidth]{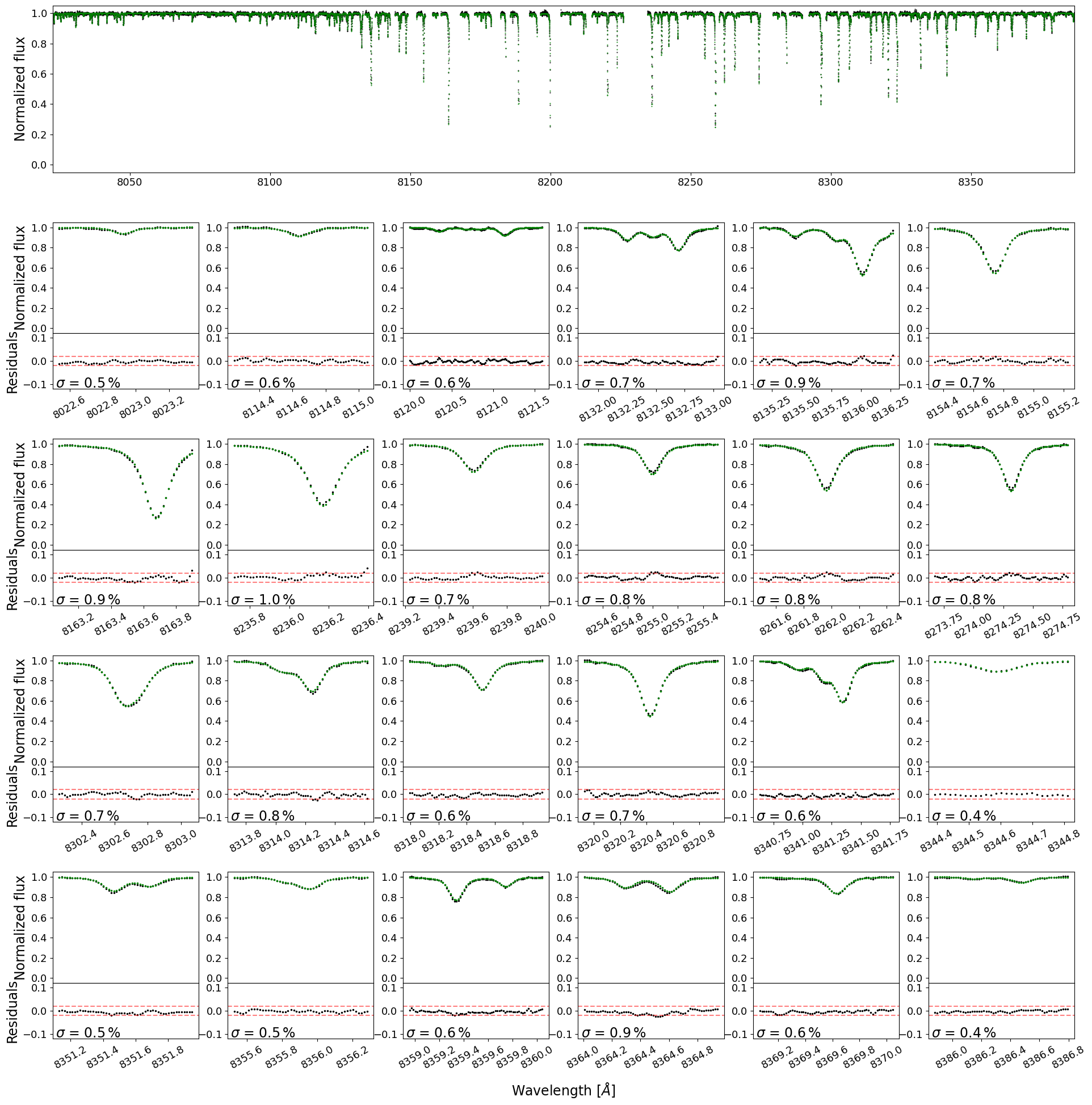}
	\caption{
            Same as Fig.~\ref{figure:accuracy_O2}, but for the H$_2$O band around $8\,200\,$\r{A}.
	\label{figure:accuracy_H2O}}
\end{center}
\end{figure*}

\begin{figure*}
\begin{center}
\includegraphics[width=\textwidth]{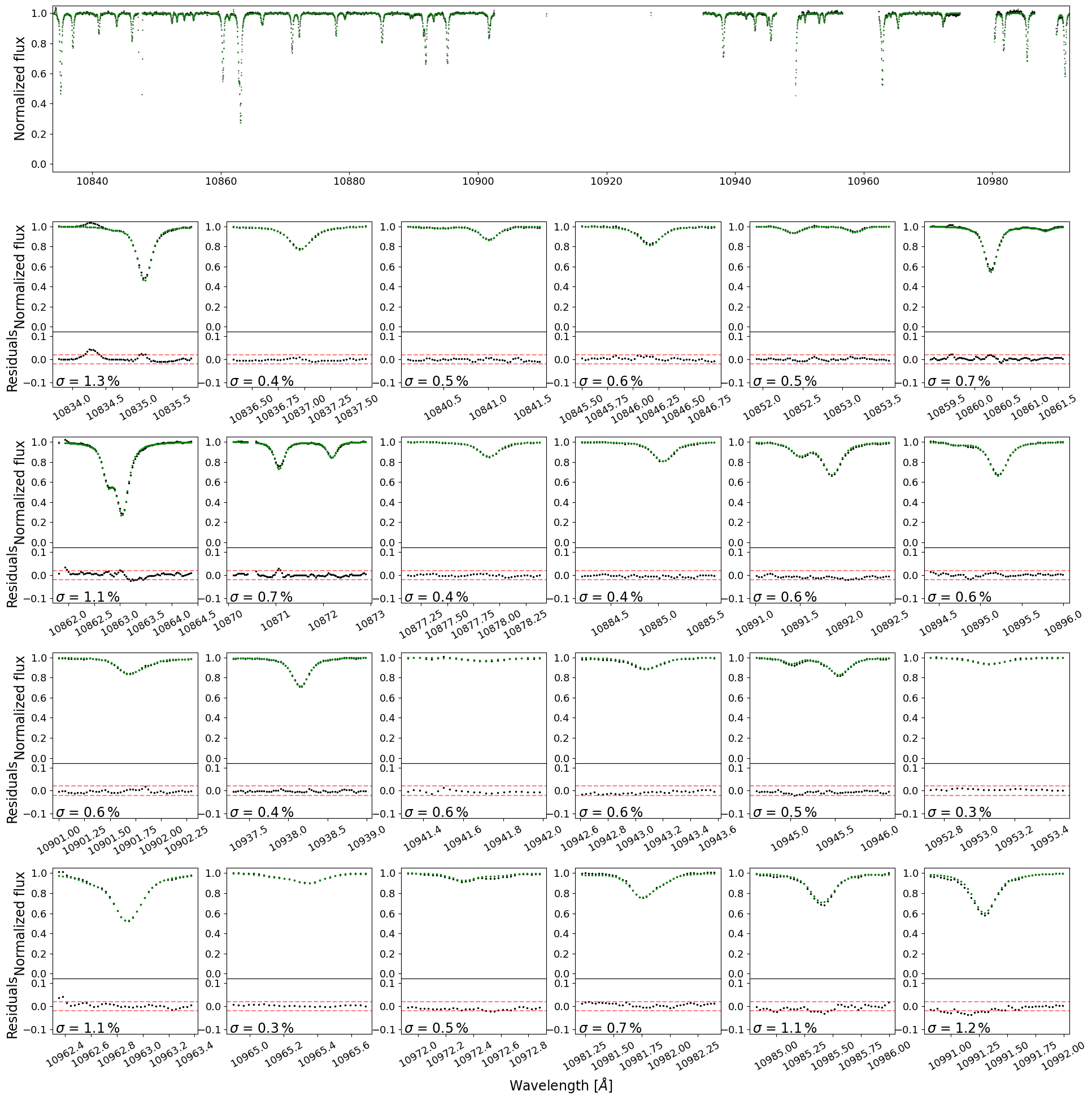}
	\caption{
            Same as Fig.~\ref{figure:accuracy_O2}, but for a H$_2$O band around $11\,000\,$\r{A}.
	\label{figure:accuracy_H2O_nir}}
\end{center}
\end{figure*}

\newpage

\begin{figure*}
\begin{center}
\includegraphics[width=\textwidth]{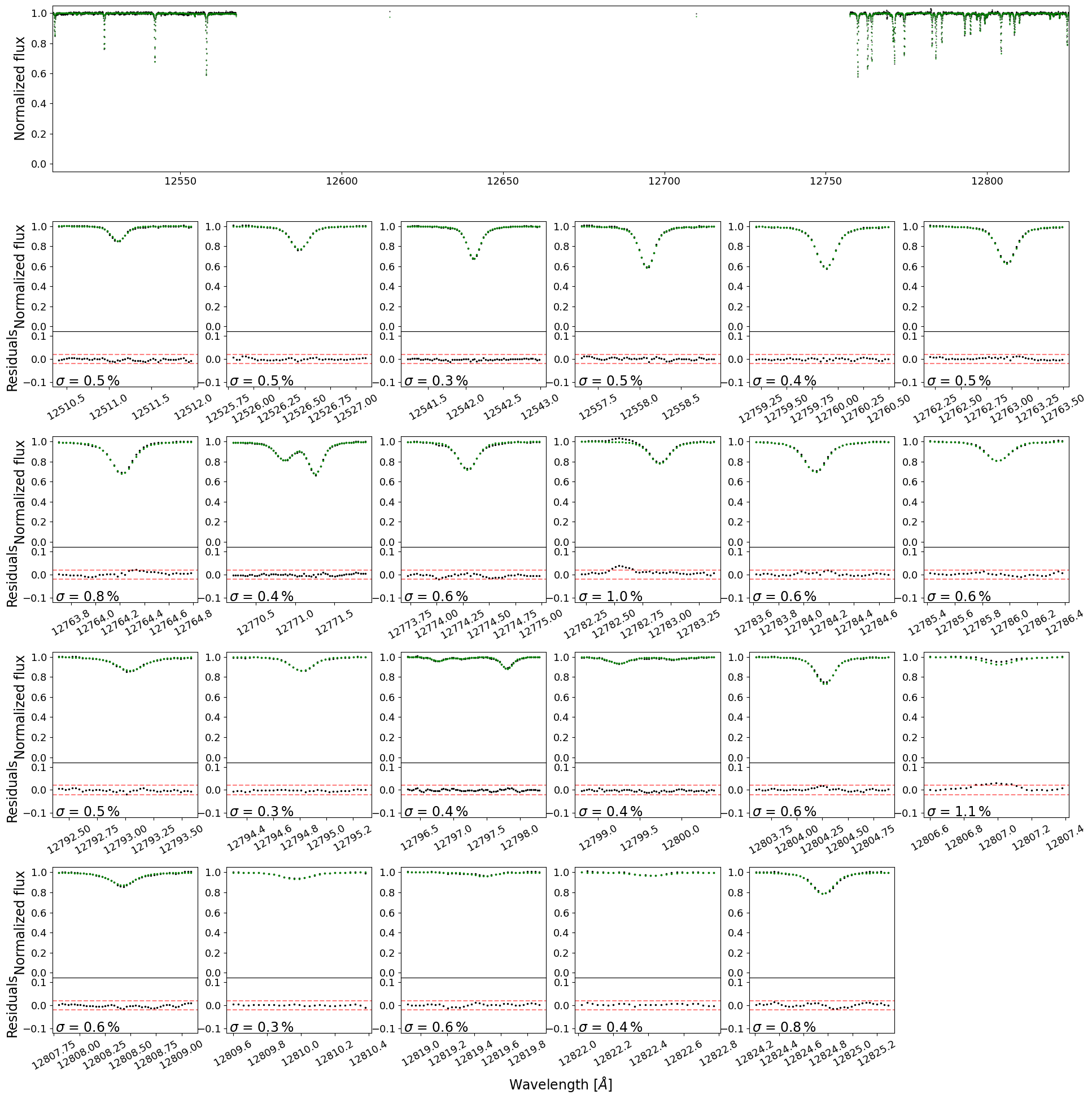}
	\caption{
            Same as Fig.~\ref{figure:accuracy_O2}, but for a O$_2$ band around $12\,600\,$\r{A}.
	\label{figure:accuracy_O2_nir}}
\end{center}
\end{figure*}

\newpage

\begin{figure*}
\begin{center}
\includegraphics[width=\textwidth]{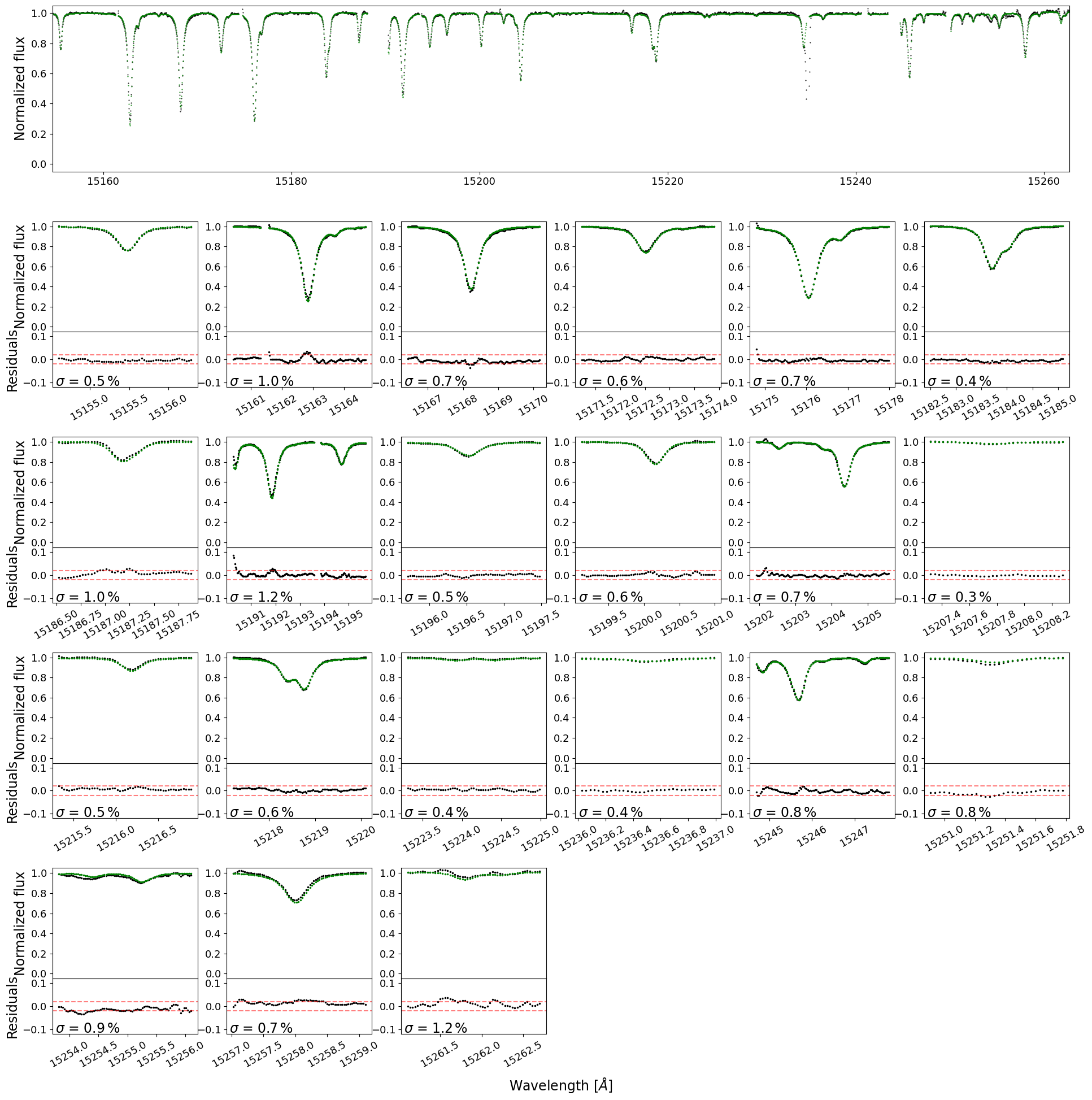}
	\caption{
            Same as Fig.~\ref{figure:accuracy_O2}, but for a H$_2$O band around $15\,000\,$\r{A}.
	\label{figure:accuracy_H2O_nir2}}
\end{center}
\end{figure*}

\newpage

\begin{figure*}
\begin{center}
\includegraphics[width=\textwidth]{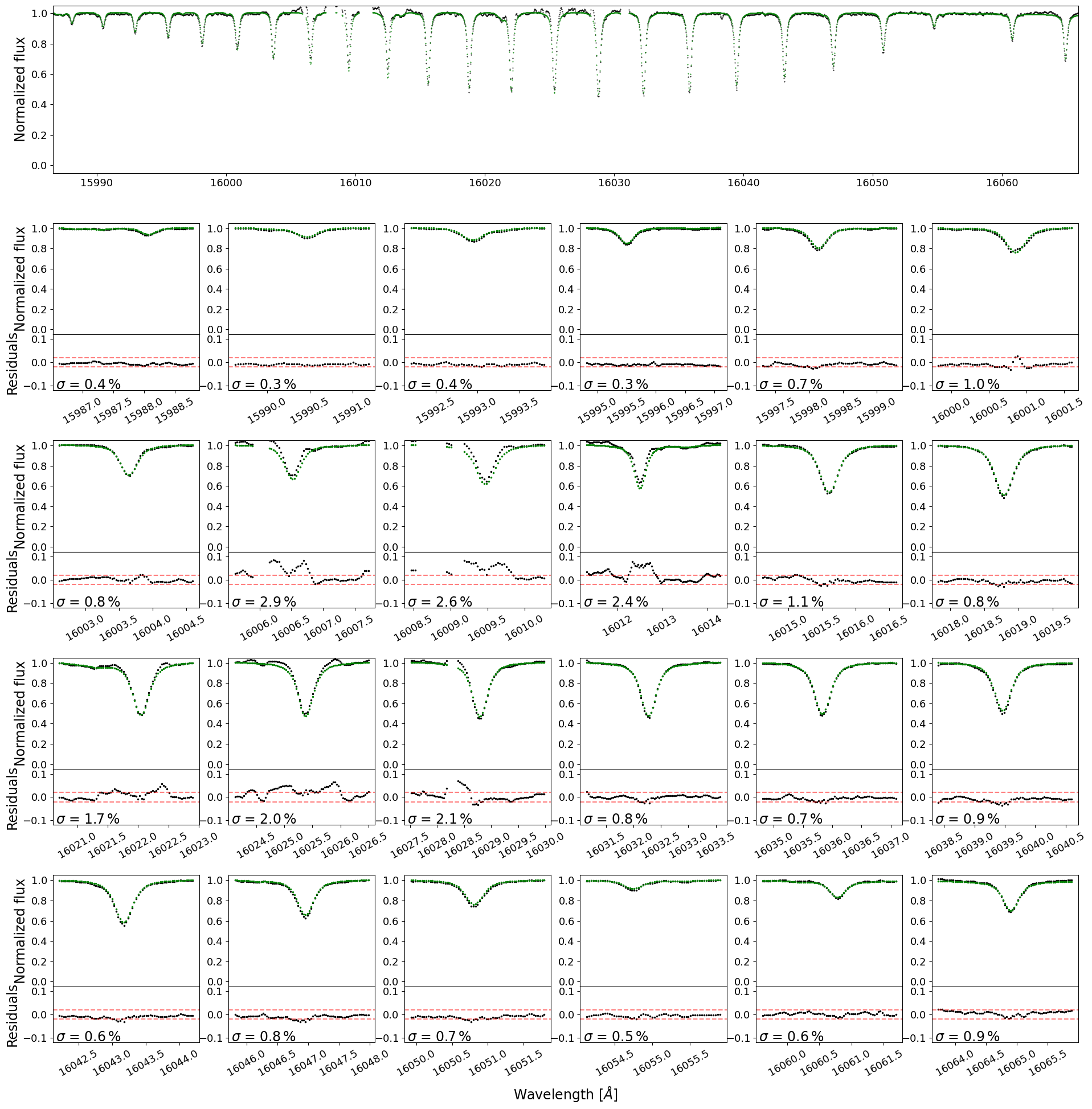}
	\caption{
            Same as Fig.~\ref{figure:accuracy_O2}, but for a CO$_2$ band around $15\,800\,$\r{A}.
	\label{figure:accuracy_CO2}}
\end{center}
\end{figure*}

\newpage

\begin{figure*}
\begin{center}
\includegraphics[width=\textwidth]{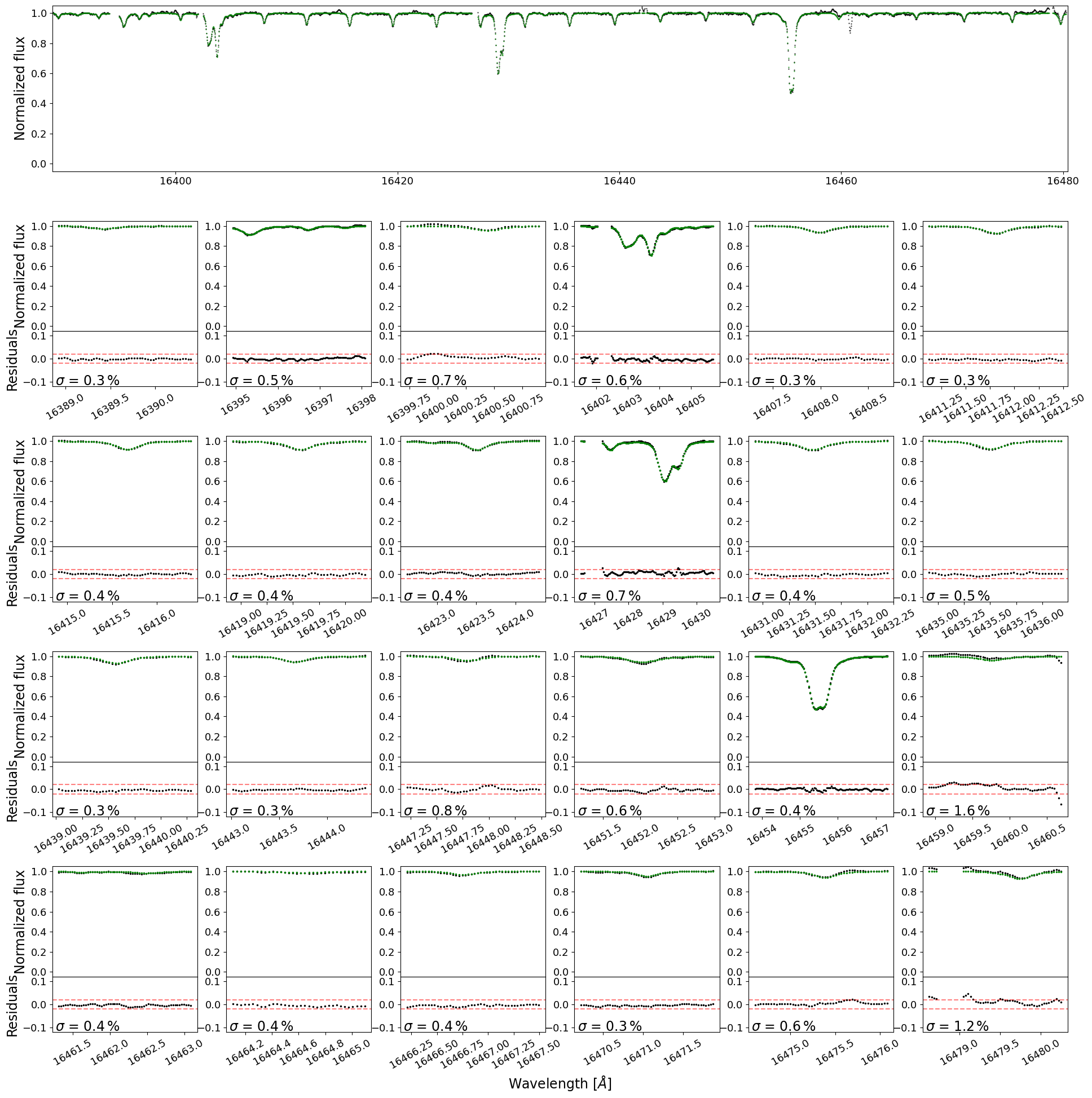}
	\caption{
            Same as Fig.~\ref{figure:accuracy_O2}, but for a CH$_4$ band around $16\,400\,$\r{A}.
	\label{figure:accuracy_CH4}}
\end{center}
\end{figure*}

\newpage


\begin{small}
\begin{landscape}
\section{Table of sample stars}

\begin{table}[h]
\begin{center}
\caption{CARMENES M-dwarf sample of high S/N template spectra corrected for telluric absorption.}
\label{table:sample}
\begin{tabular}{rlll cc cc rrrr}
\hline\hline
\noalign{\smallskip}
$\#$ & Karmn & Name & GJ & $\alpha$ & $\delta$ & Spectral type & $J$ [mag] & $\#$ VIS obs & S$/$N & $\#$ NIR obs & S$/$N \\
\noalign{\smallskip}
\hline
\noalign{\smallskip}
1 & J00051+457 & BD+44 4548 & 2 & 00 05 10.89 & +45 47 11.6 & M1.0 & 6.70 & 53 & 833 & 44 & 756 \\ 
2 & J00067-075 & GJ 1002 & 1002 & 00 06 43.20 & -07 32 17.0 & M5.5 & 8.32 & 91 & 641 & 80 & 948 \\ 
3 & J00162+198E & LP 404-062 & 1006B & 00 16 16.15 & +19 51 50.5 & M4.0 & 8.89 & 19 & 295 & 17 & 333 \\ 
4 & J00183+440 & GX And & 15A & 00 18 22.88 & +44 01 22.6 & M1.0 & 4.85 & 222 & 2040 & 196 & 2230 \\ 
5 & J00184+440 & GQ And & 15B & 00 18 25.82 & +44 01 38.1 & M3.5 & 6.79 & 195 & 1539 & 179 & 1972 \\ 
6 & J00286-066 & GJ 1012 & 1012 & 00 28 39.46 & -06 39 49.2 & M4.0 & 8.04 & 50 & 714 & 43 & 803 \\ 
7 & J00389+306 & Wolf 1056 & 26 & 00 38 59.04 & +30 36 58.4 & M2.5 & 7.45 & 61 & 911 & 55 & 944 \\ 
8 & J00403+612 & 2MASS J00402129+6112490 &   & 00 40 21.29 & +61 12 49.1 & M1.5 & 10.15 & 46 & 233 & 42 & 182 \\ 
9 & J00570+450 & G 172-030 &   & 00 57 02.69 & +45 05 09.8 & M3.0 & 8.10 & 20 & 432 & 17 & 495 \\ 
10 & J01013+613 & GJ 47 & 47 & 01 01 20.05 & +61 21 56.7 & M2.0 & 7.27 & 10 & 307 & 9 & 251 \\ 
11 & J01019+541 & G 218-020 & 3069 & 01 01 59.49 & +54 10 57.7 & M5.0 & 9.78 & 21 & 202 & 18 & 279 \\ 
12 & J01025+716 & Ross 318 & 48 & 01 02 32.24 & +71 40 47.3 & M3.0 & 6.30 & 120 & 1183 & 103 & 1231 \\ 
13 & J01026+623 & BD+61 195 & 49 & 01 02 38.87 & +62 20 42.2 & M1.5 & 6.23 & 83 & 927 & 80 & 969 \\ 
14 & J01033+623 & V388 Cas & 51 & 01 03 19.84 & +62 21 55.8 & M5.0 & 8.61 & 26 & 302 & 23 & 368 \\ 
15 & J01048-181 & GJ 1028 & 1028 & 01 04 53.80 & -18 07 28.6 & M5.0 & 9.39 & 112 & 389 & 98 & 438 \\ 
16 & J01066+192 & LSPM J0106+1913 &   & 01 06 36.98 & +19 13 33.2 & M3.0 & 9.34 & 66 & 428 & 59 & 403 \\ 
17 & J01078+128 & G 2-21 &   & 01 07 52.23 & +12 52 51.8 & M1.5 & 8.79 & 17 & 328 & 14 & 254 \\ 
18 & J01125-169 & YZ Cet & 54.1 & 01 12 30.64 & -16 59 56.4 & M4.5 & 7.26 & 113 & 974 & 101 & 1197 \\ 
19 & J01339-176 & LP 768-113 &   & 01 33 58.00 & -17 38 23.8 & M4.0 & 8.84 & 29 & 342 & 27 & 440 \\ 
20 & J01352-072 & Barta 161 12 &   & 01 35 13.92 & -07 12 51.5 & M4.0 & 8.96 & 11 & 218 & 10 & 290 \\ 
21 & J01433+043 & GJ 70 & 70 & 01 43 20.18 & +04 19 18.0 & M2.0 & 7.37 & 35 & 695 & 30 & 725 \\ 
22 & J01518+644 & G 244-037 & 3117 & 01 51 51.13 & +64 26 05.9 & M2.5 & 7.84 & 60 & 780 & 55 & 765 \\ 
23 & J01550+379 & LSR J0155+3758 &   & 01 55 02.33 & +37 58 02.3 & M5.0 & 10.47 & 9 & 70 & 9 & 77 \\ 
24 & J02002+130 & TZ Ari & 83.1 & 02 00 12.96 & +13 03 07.0 & M3.5 & 7.51 & 92 & 975 & 85 & 1309 \\ 
25 & J02015+637 & G 244-047 & 3126 & 02 01 35.34 & +63 46 12.0 & M3.0 & 7.26 & 50 & 763 & 45 & 832 \\ 
26 & J02022+103 & LP 469-067 & 3128 & 02 02 16.24 & +10 20 13.9 & M5.5 & 9.84 & 12 & 91 & 12 & 104 \\ 
27 & J02070+496 & G 173-037 &   & 02 07 03.86 & +49 38 43.5 & M3.5 & 8.37 & 31 & 460 & 26 & 517 \\ 
28 & J02088+494 & G 173-039 & 3136 & 02 08 53.62 & +49 26 56.4 & M3.5 & 8.42 & 17 & 341 & 14 & 407 \\ 
29 & J02123+035 & BD+02 348 & 87 & 02 12 20.99 & +03 34 32.2 & M1.5 & 6.83 & 66 & 929 & 53 & 889 \\ 
30 & J02164+135 & LP 469-206 & 3146 & 02 16 29.85 & +13 35 12.8 & M5.5 & 9.87 & 8 & 57 & 8 & 87 \\ 
31 & J02222+478 & BD+47 612 & 96 & 02 22 14.64 & +47 52 48.1 & M0.5 & 6.38 & 55 & 696 & 41 & 669 \\ 
32 & J02336+249 & GJ 102 & 102 & 02 33 37.18 & +24 55 37.7 & M4.0 & 8.47 & 30 & 442 & 28 & 520 \\ 
33 & J02358+202 & BD+19 381 & 104 & 02 35 53.31 & +20 13 11.6 & M2.0 & 7.21 & 39 & 724 & 32 & 689 \\ 
34 & J02362+068 & BX Cet & 105 & 02 36 15.27 & +06 52 17.9 & M4.0 & 7.33 & 50 & 723 & 42 & 825 \\ 
35 & J02442+255 & VX Ari & 109 & 02 44 15.51 & +25 31 24.1 & M3.0 & 6.75 & 52 & 722 & 50 & 882 \\ 
36 & J02465+164 & LP 411-006 & 3181 & 02 46 34.71 & +16 25 10.2 & M6.0 & 10.97 & 18 & 54 & 18 & 74 \\ 
37 & J02486+621 & 2MASS J02483695+6211228 &   & 02 48 36.99 & +62 11 22.9 & M5.0 & 12.51 & 6 & 9 & 6 & 9 \\ 
38 & J02489-145E & PM J02489-1432E &   & 02 48 59.83 & -14 32 16.2 & M2.5 & 9.73 & 24 & 188 & 22 & 192 \\ 
39 & J02489-145W & PM J02489-1432W &   & 02 48 59.27 & -14 32 14.9 & M2.0 & 9.53 & 34 & 241 & 30 & 218 \\ 
40 & J02519+224 & RBS 365 &   & 02 51 54.10 & +22 27 29.9 & M4.0 & 8.92 & 15 & 272 & 15 & 352 \\ 
41 & J02530+168 & Teegarden's Star &   & 02 53 00.89 & +16 52 52.6 & M7.0 & 8.39 & 272 & 973 & 251 & 1569 \\ 
42 & J02560-006 & LP 591-156 &   & 02 56 03.90 & -00 36 33.1 & M5.0 & 10.42 & 6 & 49 & 6 & 50 \\ 
43 & J02565+554W & Ross 364 & 119A & 02 56 34.42 & +55 26 14.1 & M1.0 & 7.42 & 10 & 271 & 6 & 112 \\ 
44 & J02573+765 & G 245-61 &   & 02 57 18.26 & +76 33 11.3 & M3.0 & 9.62 & 58 & 304 & 56 & 323 \\ 
45 & J03090+100 & GJ 1055 & 1055 & 03 09 00.17 & +10 01 25.7 & M5.0 & 9.93 & 8 & 93 & 6 & 80 \\ 
\noalign{\smallskip}
\hline
\end{tabular}
\end{center}
\tablefoot{
In the last four columns, we show the number of spectra we used to build the
templates as well as the S/N per pixel in order 70 ($\sim 8600$\,\AA) for the VIS and in order 52 ($\sim 10\,500$\,\AA) for the NIR. 
Equatorial coordinates, spectral types (class V), and $J$ magnitudes were gathered
from the CARMENES Cool dwarf Information and daTa Archive (Carmencita, see \citealt{Alonso-Floriano2015}, \citealt{Caballero2016b}, and references therein). 
}
\end{table}
\end{landscape}

\newpage
\begin{landscape}
\begin{table}
\ContinuedFloat
\caption{Continued.}
\begin{center}
\begin{tabular}{rlll cc cc rrrr}
\hline\hline
\noalign{\smallskip}
$\#$ & Karmn & Name & GJ & $\alpha$ & $\delta$ & Spectral type & $J$ [mag] & $\#$ VIS obs & S$/$N & $\#$ NIR obs & S$/$N \\
\noalign{\smallskip}
\hline
\noalign{\smallskip}
46 & J03133+047 & CD Cet & 1057 & 03 13 22.92 & +04 46 29.3 & M5.0 & 8.78 & 110 & 651 & 105 & 920 \\ 
47 & J03142+286 & LP 299-036 & 3208 & 03 14 12.47 & +28 40 39.6 & M6.0 & 10.99 & 10 & 38 & 9 & 43 \\ 
48 & J03181+382 & HD 275122 & 134 & 03 18 07.48 & +38 15 07.4 & M1.5 & 7.02 & 56 & 905 & 50 & 942 \\ 
49 & J03213+799 & GJ 133 & 133 & 03 21 21.78 & +79 58 02.2 & M2.0 & 7.70 & 33 & 563 & 31 & 632 \\ 
50 & J03217-066 & G 77-046 & 3218 & 03 21 46.92 & -06 40 24.2 & M2.0 & 7.86 & 15 & 399 & 13 & 426 \\ 
51 & J03230+420 & GJ 1059 & 1059 & 03 23 01.76 & +42 00 26.8 & M5.0 & 10.39 & 12 & 77 & 12 & 79 \\ 
52 & J03463+262 & HD 23453 & 154 & 03 46 20.14 & +26 12 55.8 & M0.0 & 6.69 & 49 & 781 & 46 & 930 \\ 
53 & J03473+086 & LTT 11262 & 3250 & 03 47 20.89 & +08 41 47.1 & M4.5 & 9.85 & 9 & 96 & 8 & 74 \\ 
54 & J03473-019 & G 80-021 &   & 03 47 23.34 & -01 58 19.9 & M3.0 & 7.80 & 11 & 278 & 9 & 346 \\ 
55 & J03531+625 & Ross 567 &   & 03 53 10.44 & +62 34 08.1 & M3.0 & 7.78 & 39 & 662 & 37 & 776 \\ 
56 & J04153-076 & omi02 Eri C & 166C & 04 15 21.54 & -07 39 20.7 & M4.5 & 6.75 & 53 & 550 & 54 & 759 \\ 
57 & J04167-120 & LP 714-47 &   & 04 16 45.60 & -12 05 02.5 & M0.0 & 9.49 & 35 & 271 & 33 & 227 \\ 
58 & J04198+425 & LSR J0419+4233 &   & 04 19 52.13 & +42 33 30.4 & M8.5 & 11.09 & 35 & 51 & 42 & 112 \\ 
59 & J04225+105 & LSPM J0422+1031 &   & 04 22 31.99 & +10 31 18.8 & M3.5 & 8.47 & 18 & 327 & 14 & 396 \\ 
60 & J04290+219 & BD+21 652 & 169 & 04 29 00.12 & +21 55 21.7 & M0.5 & 5.67 & 170 & 1491 & 151 & 1398 \\ 
61 & J04311+589 & G 175-34 & 169.1A & 04 31 11.51 & +58 58 37.5 & M4.0 & 6.62 & 11 & 312 & 9 & 214 \\ 
62 & J04343+430 & PM J04343+4302 &   & 04 34 22.50 & +43 02 14.7 & M2.5 & 9.62 & 56 & 317 & 55 & 283 \\ 
63 & J04376+528 & BD+52 857 & 172 & 04 37 40.93 & +52 53 37.0 & M0.0 & 5.87 & 126 & 1273 & 97 & 1332 \\ 
64 & J04376-110 & BD-11 916 & 173 & 04 37 41.87 & -11 02 20.0 & M1.5 & 6.94 & 43 & 726 & 39 & 681 \\ 
65 & J04406-128 & TOI-2457 &   & 04 40 40.14 & -12 53 26.9 & M0.0 & 9.74 & 53 & 266 & 53 & 228 \\ 
66 & J04429+189 & HD 285968 & 176 & 04 42 55.77 & +18 57 29.4 & M2.0 & 6.46 & 23 & 451 & 16 & 424 \\ 
67 & J04429+214 & 2MASS J04425586+2128230 &   & 04 42 55.86 & +21 28 23.0 & M3.5 & 7.96 & 16 & 327 & 15 & 353 \\ 
68 & J04472+206 & RX J0447.2+2038 &   & 04 47 12.25 & +20 38 10.8 & M5.0 & 9.38 & 11 & 153 & 11 & 226 \\ 
69 & J04520+064 & Wolf 1539 & 179 & 04 52 05.73 & +06 28 35.6 & M3.5 & 7.81 & 10 & 269 & 10 & 313 \\ 
70 & J04538-177 & GJ 180 & 180 & 04 53 49.98 & -17 46 24.3 & M2.0 & 7.41 & 25 & 578 & 22 & 663 \\ 
71 & J04588+498 & BD+49 1280 & 181 & 04 58 50.57 & +49 50 57.4 & M0.0 & 6.92 & 56 & 900 & 49 & 1011 \\ 
72 & J05019+011 & 1RXS J050156.7+010845 &   & 05 01 56.66 & +01 08 42.9 & M4.0 & 8.53 & 19 & 305 & 19 & 425 \\ 
73 & J05019-069 & LP 656-038 & 3323 & 05 01 57.43 & -06 56 46.4 & M4.0 & 7.62 & 8 & 261 & 7 & 276 \\ 
74 & J05033-173 & LP 776-046 & 3325 & 05 03 20.08 & -17 22 24.7 & M3.0 & 7.82 & 68 & 795 & 81 & 799 \\ 
75 & J05062+046 & RX J0506.2+0439 &   & 05 06 12.93 & +04 39 27.2 & M4.0 & 8.91 & 13 & 201 & 11 & 289 \\ 
76 & J05084-210 & 2MASS J05082729-2101444 &   & 05 08 27.30 & -21 01 44.4 & M5.0 & 9.72 & 39 & 161 & 36 & 244 \\ 
77 & J05127+196 & GJ 192 & 192 & 05 12 42.23 & +19 39 56.4 & M2.0 & 7.30 & 43 & 715 & 36 & 695 \\ 
78 & J05280+096 & Ross 41 & 203 & 05 28 00.15 & +09 38 38.2 & M3.5 & 8.31 & 17 & 335 & 14 & 416 \\ 
79 & J05314-036 & HD 36395 & 205 & 05 31 27.40 & -03 40 38.0 & M1.5 & 4.72 & 95 & 1175 & 77 & 1175 \\ 
80 & J05348+138 & Ross 46 & 3356 & 05 34 52.12 & +13 52 46.7 & M3.5 & 7.78 & 21 & 507 & 17 & 577 \\ 
81 & J05360-076 & Wolf 1457 & 3357 & 05 36 00.08 & -07 38 58.4 & M4.0 & 8.46 & 39 & 451 & 32 & 523 \\ 
82 & J05365+113 & V2689 Ori & 208 & 05 36 30.99 & +11 19 40.3 & M0.0 & 6.13 & 131 & 1398 & 113 & 1507 \\ 
83 & J05366+112 & PM J05366+1117 &   & 05 36 38.46 & +11 17 48.8 & M4.0 & 8.27 & 16 & 370 & 13 & 460 \\ 
84 & J05394+406 & LSR J0539+4038 &   & 05 39 24.80 & +40 38 42.8 & M8.0 & 11.11 & 21 & 53 & 29 & 114 \\ 
85 & J05415+534 & HD 233153 & 212 & 05 41 30.73 & +53 29 23.3 & M1.0 & 6.59 & 97 & 1176 & 81 & 1203 \\ 
86 & J05421+124 & V1352 Ori & 213 & 05 42 09.27 & +12 29 21.6 & M4.0 & 7.12 & 51 & 731 & 46 & 893 \\ 
87 & J06000+027 & G 99-049 & 3379 & 06 00 03.50 & +02 42 23.6 & M4.0 & 6.91 & 14 & 303 & 13 & 411 \\ 
88 & J06011+595 & G 192-013 & 3378 & 06 01 11.05 & +59 35 49.9 & M3.5 & 7.46 & 80 & 900 & 74 & 1107 \\ 
89 & J06024+498 & G 192-015 & 3380 & 06 02 29.19 & +49 51 56.2 & M5.0 & 9.35 & 125 & 459 & 117 & 503 \\ 
90 & J06103+821 & GJ 226 & 226 & 06 10 19.85 & +82 06 24.3 & M2.0 & 6.87 & 57 & 809 & 55 & 999 \\ 
\noalign{\smallskip}
\hline
\end{tabular}
\end{center}
\end{table}
\end{landscape}
\newpage
\begin{landscape}
\begin{table}
\ContinuedFloat
\caption{Continued.}
\begin{center}
\begin{tabular}{rlll cc cc rrrr}
\hline\hline
\noalign{\smallskip}
$\#$ & Karmn & Name & GJ & $\alpha$ & $\delta$ & Spectral type & $J$ [mag] & $\#$ VIS obs & S$/$N & $\#$ NIR obs & S$/$N \\
\noalign{\smallskip}
\hline
\noalign{\smallskip}
91 & J06105-218 & HD  42581 & 229A & 06 10 34.61 & -21 51 52.7 & M0.5 & 5.10 & 51 & 742 & 45 & 821 \\ 
92 & J06246+234 & Ross 64 & 232 & 06 24 41.28 & +23 25 58.9 & M4.0 & 8.66 & 9 & 215 & 9 & 263 \\ 
93 & J06318+414 & LP 205-044 & 3396 & 06 31 50.74 & +41 29 45.5 & M5.0 & 9.68 & 37 & 237 & 36 & 362 \\ 
94 & J06371+175 & HD 260655 & 239 & 06 37 10.80 & +17 33 53.3 & M0.0 & 6.67 & 125 & 1253 & 112 & 1331 \\ 
95 & J06396-210 & LP 780-032 &   & 06 39 37.43 & -21 01 34.1 & M4.0 & 8.51 & 58 & 446 & 55 & 576 \\ 
96 & J06421+035 & G 108-021 & 3404A & 06 42 11.19 & +03 34 52.6 & M3.5 & 8.17 & 19 & 453 & 13 & 487 \\ 
97 & J06548+332 & Wolf 294 & 251 & 06 54 48.96 & +33 16 05.4 & M3.0 & 6.10 & 444 & 2310 & 398 & 2383 \\ 
98 & J06574+740 & 2MASS J06572616+7405265 &   & 06 57 26.12 & +74 05 26.6 & M4.0 & 8.93 & 13 & 228 & 12 & 300 \\ 
99 & J06594+193 & GJ 1093 & 1093 & 06 59 28.82 & +19 20 55.9 & M5.0 & 9.16 & 30 & 285 & 26 & 419 \\ 
100 & J07033+346 & LP 255-011 & 3423 & 07 03 23.17 & +34 41 51.4 & M4.0 & 8.77 & 14 & 302 & 14 & 375 \\ 
101 & J07044+682 & GJ 258 & 258 & 07 04 25.94 & +68 17 19.7 & M3.0 & 8.17 & 18 & 378 & 13 & 415 \\ 
102 & J07051-101 & 2MASS J07051194-1007528 &   & 07 05 11.96 & -10 07 52.8 & M5.0 & 10.20 & 11 & 62 & 10 & 80 \\ 
103 & J07274+052 & Luyten's Star & 273 & 07 27 24.50 & +05 13 32.8 & M3.5 & 5.71 & 774 & 1217 & 756 & 1725 \\ 
104 & J07287-032 & GJ 1097 & 1097 & 07 28 45.44 & -03 17 53.3 & M3.0 & 7.54 & 25 & 517 & 24 & 651 \\ 
105 & J07319+362N & V* BL Lyn & 277B & 07 31 57.32 & +36 13 47.4 & M3.5 & 7.57 & 50 & 749 & 48 & 911 \\ 
106 & J07353+548 & GJ 3452 & 3452 & 07 35 21.88 & +54 50 59.0 & M2.0 & 7.77 & 10 & 320 & 10 & 312 \\ 
107 & J07386-212 & LP 763-001 & 3459 & 07 38 40.96 & -21 13 28.5 & M3.0 & 7.85 & 7 & 224 & 7 & 298 \\ 
108 & J07393+021 & BD+02 1729 & 281 & 07 39 23.04 & +02 11 01.2 & M0.0 & 6.77 & 51 & 823 & 43 & 829 \\ 
109 & J07403-174 & LP 783-002 & 283B & 07 40 19.37 & -17 24 45.9 & M6.0 & 10.15 & 53 & 125 & 54 & 218 \\ 
110 & J07446+035 & YZ CMi & 285 & 07 44 40.17 & +03 33 08.9 & M4.5 & 6.58 & 51 & 638 & 39 & 803 \\ 
111 & J07472+503 & 2MASS J07471385+5020386 &   & 07 47 13.87 & +50 20 38.5 & M4.0 & 8.86 & 15 & 293 & 16 & 410 \\ 
112 & J07558+833 & GJ 1101 & 1101 & 07 55 53.93 & +83 23 04.9 & M4.5 & 8.74 & 14 & 234 & 14 & 331 \\ 
113 & J07582+413 & GJ 1105 & 1105 & 07 58 12.70 & +41 18 13.3 & M3.5 & 7.73 & 28 & 479 & 27 & 610 \\ 
114 & J07590+153 & LP 424-004 & 3470 & 07 59 05.84 & +15 23 29.2 & M1.5 & 8.79 & 107 & 449 & 65 & 420 \\ 
115 & J08023+033 & G 50-16 A & 3473 & 08 02 22.88 & +03 20 19.7 & M4.0 & 9.63 & 67 & 307 & 61 & 293 \\ 
116 & J08119+087 & Ross 619 & 299 & 08 11 57.57 & +08 46 23.0 & M4.5 & 8.42 & 16 & 341 & 15 & 432 \\ 
117 & J08126-215 & GJ 300 & 300 & 08 12 40.89 & -21 33 07.0 & M4.0 & 7.60 & 18 & 366 & 16 & 472 \\ 
118 & J08161+013 & GJ 2066 & 2066 & 08 16 07.98 & +01 18 09.3 & M2.0 & 6.62 & 75 & 974 & 70 & 1136 \\ 
119 & J08293+039 & 2MASS J08292191+0355092 &   & 08 29 21.91 & +03 55 09.3 & M2.5 & 7.93 & 15 & 445 & 12 & 497 \\ 
120 & J08298+267 & DX Cnc & 1111 & 08 29 49.35 & +26 46 33.6 & M6.5 & 8.23 & 35 & 382 & 29 & 632 \\ 
121 & J08315+730 & LP 035-219 &   & 08 31 30.12 & +73 03 45.8 & M4.0 & 8.78 & 19 & 305 & 18 & 375 \\ 
122 & J08358+680 & G 234-037 & 3506 & 08 35 49.04 & +68 04 09.1 & M2.5 & 7.86 & 11 & 303 & 11 & 361 \\ 
123 & J08402+314 & LSPM J0840+3127 &   & 08 40 15.98 & +31 27 06.8 & M3.5 & 8.12 & 8 & 281 & 6 & 336 \\ 
124 & J08409-234 & LP 844-008 & 317 & 08 40 59.21 & -23 27 22.6 & M3.5 & 7.93 & 36 & 439 & 35 & 488 \\ 
125 & J08413+594 & LP 090-018 & 3512 & 08 41 20.13 & +59 29 50.4 & M5.5 & 9.62 & 222 & 460 & 207 & 703 \\ 
126 & J08526+283 & rho Cnc B & 324B & 08 52 40.86 & +28 18 58.8 & M4.5 & 8.56 & 11 & 264 & 10 & 324 \\ 
127 & J08536-034 & LP 666-009 & 3517 & 08 53 36.16 & -03 29 32.2 & M9.0 & 11.21 & 45 & 32 & 43 & 82 \\ 
128 & J08599+729 & LP 036-098 & 3520 & 08 59 56.20 & +72 57 36.3 & M4.0 & 9.73 & 18 & 140 & 17 & 116 \\ 
129 & J09003+218 & LP 368-128 &   & 09 00 23.55 & +21 50 04.9 & M6.5 & 9.44 & 25 & 156 & 20 & 263 \\ 
130 & J09005+465 & GJ 1119 & 1119 & 09 00 32.48 & +46 35 11.1 & M4.5 & 8.60 & 8 & 201 & 8 & 269 \\ 
131 & J09028+680 & LP 060-179 & 3526 & 09 02 52.87 & +68 03 46.6 & M4.0 & 8.45 & 10 & 276 & 11 & 393 \\ 
132 & J09033+056 & NLTT 20861 &   & 09 03 20.96 & +05 40 14.6 & M7.0 & 10.77 & 31 & 80 & 27 & 119 \\ 
133 & J09133+688 & G 234-57A &   & 09 13 23.86 & +68 52 31.0 & M2.5 & 7.78 & 16 & 397 & 16 & 472 \\ 
134 & J09143+526 & HD 79210 & 338A & 09 14 22.77 & +52 41 11.8 & M0.0 & 4.89 & 72 & 747 & 65 & 868 \\ 
135 & J09144+526 & HD 79211 & 338B & 09 14 24.68 & +52 41 10.9 & M0.0 & 4.78 & 158 & 1363 & 128 & 1384 \\ 
\noalign{\smallskip}
\hline
\end{tabular}
\end{center}
\end{table}
\end{landscape}
\newpage
\begin{landscape}
\begin{table}
\ContinuedFloat
\caption{Continued.}
\begin{center}
\begin{tabular}{rlll cc cc rrrr}
\hline\hline
\noalign{\smallskip}
$\#$ & Karmn & Name & GJ & $\alpha$ & $\delta$ & Spectral type & $J$ [mag] & $\#$ VIS obs & S$/$N & $\#$ NIR obs & S$/$N \\
\noalign{\smallskip}
\hline
\noalign{\smallskip}
136 & J09161+018 & RX J0916.1+0153 &   & 09 16 10.18 & +01 53 08.8 & M4.0 & 8.77 & 9 & 199 & 9 & 277 \\ 
137 & J09163-186 & LP 787-052 & 3543 & 09 16 20.64 & -18 37 32.9 & M1.5 & 7.35 & 5 & 249 & 4 & 115 \\ 
138 & J09286-121 & G 161-32 &   & 09 28 41.58 & -12 09 55.1 & M2.5 & 8.84 & 7 & 156 & 7 & 124 \\ 
139 & J09307+003 & GJ 1125 & 1125 & 09 30 44.59 & +00 19 21.6 & M3.5 & 7.70 & 22 & 472 & 22 & 584 \\ 
140 & J09360-216 & GJ 357 & 357 & 09 36 01.64 & -21 39 38.9 & M2.5 & 7.34 & 26 & 621 & 25 & 520 \\ 
141 & J09411+132 & Ross 85 & 361 & 09 41 10.36 & +13 12 34.4 & M1.5 & 6.97 & 49 & 824 & 44 & 807 \\ 
142 & J09423+559 & GJ 363 & 363 & 09 42 23.19 & +55 59 01.3 & M3.5 & 8.37 & 10 & 275 & 10 & 377 \\ 
143 & J09425+700 & GJ 360 & 360 & 09 42 34.84 & +70 02 02.0 & M2.0 & 6.92 & 49 & 703 & 42 & 737 \\ 
144 & J09428+700 & GJ 362 & 362 & 09 42 51.73 & +70 02 21.9 & M3.0 & 7.33 & 51 & 754 & 50 & 787 \\ 
145 & J09439+269 & Ross 93 & 3564 & 09 43 55.61 & +26 58 08.4 & M3.5 & 8.04 & 6 & 229 & 7 & 316 \\ 
146 & J09447-182 & GJ 1129 & 1129 & 09 44 47.35 & -18 12 49.0 & M4.0 & 8.12 & 12 & 303 & 11 & 395 \\ 
147 & J09449-123 & G 161-071 &   & 09 44 54.19 & -12 20 54.4 & M5.0 & 8.50 & 10 & 181 & 10 & 290 \\ 
148 & J09468+760 & BD+76 3952 & 366 & 09 46 48.49 & +76 02 38.1 & M1.5 & 7.44 & 26 & 580 & 23 & 550 \\ 
149 & J09511-123 & BD-11 2741 & 369 & 09 51 09.64 & -12 19 47.5 & M0.5 & 6.99 & 25 & 598 & 20 & 485 \\ 
150 & J09535+355 & Wolf 327 &   & 09 53 30.91 & +35 34 16.7 & M2.5 & 9.31 & 24 & 246 & 19 & 216 \\ 
151 & J09561+627 & BD+63 869 & 373 & 09 56 08.67 & +62 47 18.5 & M0.0 & 6.03 & 69 & 898 & 65 & 898 \\ 
152 & J09597+472 & G 146-005 & 3579 & 09 59 45.99 & +47 12 11.3 & M4.0 & 9.76 & 11 & 115 & 11 & 97 \\ 
153 & J10023+480 & BD+48 1829 & 378 & 10 02 21.75 & +48 05 19.7 & M1.0 & 6.95 & 22 & 515 & 17 & 541 \\ 
154 & J10087+355 & Wolf 346 &   & 15 .1 6308 & 35. 47 62 & M1.5 & 9.17 & 13 & 180 & 13 & 157 \\ 
155 & J10088+692 & TYC 4384-1735-1 &   & 10 08 51.81 & +69 16 35.6 & M0.5 & 8.71 & 41 & 498 & 36 & 491 \\ 
156 & J10122-037 & AN Sex & 382 & 10 12 17.67 & -03 44 44.4 & M1.5 & 5.89 & 76 & 944 & 60 & 954 \\ 
157 & J10125+570 & LP 092-048 &   & 10 12 34.78 & +57 03 49.0 & M3.5 & 7.76 & 10 & 305 & 9 & 337 \\ 
158 & J10167-119 & GJ 386 & 386 & 10 16 45.95 & -11 57 42.4 & M3.0 & 7.32 & 20 & 510 & 19 & 448 \\ 
159 & J10185-117 & LP  729-54 &   & 10 18 35.14 & -11 43 00.2 & M4.0 & 9.01 & 51 & 407 & 51 & 450 \\ 
160 & J10196+198 & AD Leo & 388 & 10 19 36.28 & +19 52 12.0 & M3.0 & 5.45 & 85 & 615 & 79 & 654 \\ 
161 & J10238+438 & LP 212-062 &   & 10 23 51.86 & +43 53 33.1 & M5.0 & 10.04 & 11 & 78 & 9 & 66 \\ 
162 & J10251-102 & BD-09 3070 & 390 & 10 25 10.84 & -10 13 43.3 & M1.0 & 6.89 & 29 & 652 & 28 & 620 \\ 
163 & J10289+008 & BD+01 2447 & 393 & 10 28 55.55 & +00 50 27.6 & M2.0 & 6.18 & 85 & 990 & 78 & 1060 \\ 
164 & J10350-094 & LP 670-017 &   & 10 35 01.12 & -09 24 38.6 & M3.0 & 8.28 & 16 & 361 & 13 & 459 \\ 
165 & J10360+051 & RY Sex & 398 & 10 36 01.22 & +05 07 12.8 & M3.5 & 8.46 & 7 & 201 & 7 & 297 \\ 
166 & J10396-069 & GJ 399 & 399 & 10 39 40.56 & -06 55 25.5 & M2.5 & 7.66 & 7 & 277 & 7 & 337 \\ 
167 & J10416+376 & GJ 1134 & 1134 & 10 41 37.88 & +37 36 39.2 & M4.5 & 8.49 & 11 & 255 & 9 & 287 \\ 
168 & J10482-113 & LP 731-058 & 3622 & 10 48 12.61 & -11 20 09.6 & M6.5 & 8.86 & 79 & 334 & 72 & 630 \\ 
169 & J10508+068 & EE Leo & 402 & 10 50 52.03 & +06 48 29.3 & M4.0 & 7.32 & 53 & 806 & 49 & 1007 \\ 
170 & J10564+070 & CN Leo & 406 & 10 56 28.92 & +07 00 53.0 & M6.0 & 7.08 & 78 & 644 & 63 & 1034 \\ 
171 & J10584-107 & LP 731-076 &   & 10 58 27.99 & -10 46 30.5 & M5.0 & 9.51 & 53 & 252 & 49 & 344 \\ 
172 & J11000+228 & Ross 104 & 408 & 11 00 04.26 & +22 49 58.6 & M2.5 & 6.31 & 60 & 793 & 50 & 869 \\ 
173 & J11026+219 & DS Leo & 410 & 11 02 38.34 & +21 58 01.7 & M1.0 & 6.52 & 53 & 791 & 43 & 858 \\ 
174 & J11033+359 & Lalande 21185 & 411 & 11 03 20.19 & +35 58 11.5 & M1.5 & 4.10 & 565 & 2731 & 505 & 2940 \\ 
175 & J11044+304 & LSPM J1104+3027 &   & 11 04 28.55 & +30 27 31.5 & M3.0 & 10.93 & 94 & 149 & 91 & 157 \\ 
176 & J11054+435 & BD+44 2051A & 412A & 11 05 28.58 & +43 31 36.4 & M1.0 & 5.54 & 118 & 1130 & 106 & 1246 \\ 
177 & J11055+435 & WX UMa & 412B & 11 05 30.89 & +43 31 17.9 & M5.5 & 8.74 & 60 & 356 & 55 & 513 \\ 
178 & J11108+479 & G 176-015 & 3646 & 11 10 51.52 & +47 57 02.0 & M4.0 & 10.08 & 11 & 99 & 8 & 79 \\ 
179 & J11110+304E & HD  97101A & 414A & 11 11 05.17 & +30 26 45.7 & K7.0 & 5.76 & 13 & 465 & 12 & 563 \\ 
180 & J11110+304W & HD  97101B & 414B & 11 11 02.54 & +30 26 41.3 & M2.0 & 6.59 & 51 & 760 & 48 & 832 \\ 
\noalign{\smallskip}
\hline
\end{tabular}
\end{center}
\end{table}
\end{landscape}
\newpage
\begin{landscape}
\begin{table}
\ContinuedFloat
\caption{Continued.}
\begin{center}
\begin{tabular}{rlll cc cc rrrr}
\hline\hline
\noalign{\smallskip}
$\#$ & Karmn & Name & GJ & $\alpha$ & $\delta$ & Spectral type & $J$ [mag] & $\#$ VIS obs & S$/$N & $\#$ NIR obs & S$/$N \\
\noalign{\smallskip}
\hline
\noalign{\smallskip}
181 & J11126+189 & StKM 1-928 & 3649 & 11 12 38.97 & +18 56 05.4 & M1.5 & 7.45 & 20 & 542 & 16 & 616 \\ 
182 & J11195+466 & LP 169-022 &   & 11 19 30.61 & +46 41 43.1 & M5.5 & 10.09 & 5 & 44 & 4 & 42 \\ 
183 & J11201-104 & LP 733-099 &   & 11 20 06.10 & -10 29 46.7 & M2.0 & 7.81 & 27 & 496 & 25 & 545 \\ 
184 & J11289+101 & Wolf 398 & 3666 & 11 28 56.26 & +10 10 39.2 & M3.5 & 8.48 & 9 & 252 & 8 & 282 \\ 
185 & J11302+076 & K2-18 &   & 11 30 14.52 & +07 35 18.3 & M2.5 & 9.76 & 63 & 305 & 61 & 394 \\ 
186 & J11306-080 & LP 672-042 &   & 11 30 41.83 & -08 05 43.0 & M3.5 & 8.03 & 15 & 358 & 13 & 421 \\ 
187 & J11417+427 & Ross 1003 & 1148 & 11 41 44.64 & +42 45 07.1 & M4.0 & 7.61 & 80 & 925 & 70 & 1206 \\ 
188 & J11421+267 & Ross 905 & 436 & 11 42 11.09 & +26 42 23.7 & M2.5 & 6.90 & 444 & 1404 & 400 & 1827 \\ 
189 & J11423+230 & LP 375-23 &   & 11 42 18.37 & +23 01 36.7 & M0.5 & 8.65 & 86 & 671 & 77 & 606 \\ 
190 & J11467-140 & GJ 443 & 443 & 11 46 42.91 & -14 00 51.8 & M3.0 & 7.96 & 15 & 386 & 15 & 492 \\ 
191 & J11474+667 & 1RXS J114728.8+664405 &   & 11 47 28.57 & +66 44 02.7 & M5.0 & 9.68 & 54 & 221 & 49 & 292 \\ 
192 & J11476+002 & LP 613-49A & 3685A & 11 47 40.75 & +00 15 20.1 & M4.0 & 8.99 & 7 & 150 & 7 & 186 \\ 
193 & J11476+786 & GJ 445 & 445 & 11 47 41.39 & +78 41 28.2 & M3.5 & 6.72 & 66 & 774 & 53 & 942 \\ 
194 & J11477+008 & FI Vir & 447 & 11 47 44.40 & +00 48 16.4 & M4.0 & 6.50 & 56 & 704 & 50 & 880 \\ 
195 & J11509+483 & GJ 1151 & 1151 & 11 50 57.72 & +48 22 38.6 & M4.5 & 8.49 & 117 & 670 & 106 & 757 \\ 
196 & J11511+352 & BD+36 2219 & 450 & 11 51 07.34 & +35 16 19.2 & M1.5 & 6.42 & 112 & 1123 & 101 & 1345 \\ 
197 & J12054+695 & Ross 689 & 3704 & 12 05 29.68 & +69 32 22.6 & M4.0 & 8.74 & 7 & 180 & 6 & 217 \\ 
198 & J12100-150 & LP 734-032 & 3707 & 12 10 05.60 & -15 04 17.0 & M3.5 & 7.77 & 67 & 670 & 61 & 724 \\ 
199 & J12111-199 & LTT 4562 & 3708A & 12 11 11.76 & -19 57 38.1 & M3.0 & 7.89 & 19 & 395 & 18 & 444 \\ 
200 & J12123+544S & HD 238090 & 458A & 12 12 20.86 & +54 29 08.7 & M0.0 & 6.88 & 111 & 1221 & 100 & 1408 \\ 
201 & J12156+526 & StKM 2-809 &   & 12 15 39.36 & +52 39 08.8 & M4.0 & 8.59 & 13 & 247 & 12 & 310 \\ 
202 & J12189+111 & GL Vir & 1156 & 12 18 59.40 & +11 07 33.8 & M5.0 & 8.53 & 13 & 233 & 10 & 344 \\ 
203 & J12230+640 & Ross 690 & 463 & 12 23 00.16 & +64 01 51.0 & M3.0 & 7.94 & 160 & 884 & 147 & 1505 \\ 
204 & J12248-182 & Ross 695 & 465 & 12 24 52.50 & -18 14 32.3 & M2.0 & 7.73 & 15 & 370 & 14 & 436 \\ 
205 & J12312+086 & BD+09 2636 & 471 & 12 31 15.81 & +08 48 38.2 & M0.5 & 6.78 & 48 & 804 & 41 & 899 \\ 
206 & J12350+098 & GJ 476 & 476 & 12 35 00.71 & +09 49 42.6 & M2.5 & 8.00 & 10 & 307 & 8 & 321 \\ 
207 & J12373-208 & LP 795-038 &   & 12 37 21.57 & -20 52 35.5 & M4.0 & 8.97 & 19 & 193 & 16 & 213 \\ 
208 & J12388+116 & Wolf 433 & 480 & 12 38 52.44 & +11 41 46.1 & M3.0 & 7.58 & 11 & 383 & 11 & 364 \\ 
209 & J12428+418 & G 123-055 &   & 12 42 49.88 & +41 53 47.1 & M4.0 & 8.12 & 9 & 254 & 8 & 291 \\ 
210 & J12479+097 & Wolf 437 & 486 & 12 47 56.62 & +09 45 05.0 & M3.5 & 7.20 & 105 & 1045 & 97 & 1257 \\ 
211 & J13005+056 & FN Vir & 493.1 & 13 00 33.52 & +05 41 08.0 & M4.5 & 8.55 & 12 & 183 & 9 & 200 \\ 
212 & J13102+477 & G 177-025 &   & 13 10 12.63 & +47 45 18.7 & M5.0 & 9.58 & 36 & 243 & 34 & 327 \\ 
213 & J13119+658 & PM J13119+6550 &   & 13 11 59.56 & +65 50 01.7 & M3.0 & 9.71 & 12 & 140 & 11 & 147 \\ 
214 & J13196+333 & Ross 1007 & 507.1 & 13 19 40.13 & +33 20 47.5 & M1.5 & 7.27 & 9 & 278 & 6 & 299 \\ 
215 & J13209+342 & BD+35 2439 & 508.2 & 13 20 58.05 & +34 16 44.2 & M1.0 & 7.40 & 16 & 469 & 15 & 415 \\ 
216 & J13229+244 & Ross 1020 & 3779 & 13 22 56.75 & +24 28 03.6 & M4.0 & 8.73 & 112 & 712 & 105 & 959 \\ 
217 & J13255+688 & 2MASS J13253177+6850106 &   & 13 25 31.78 & +68 50 10.6 & M0.0 & 10.04 & 70 & 265 & 65 & 239 \\ 
218 & J13283-023W & Ross 486A & 512A & 13 28 21.08 & -02 21 37.1 & M3.0 & 7.51 & 12 & 344 & 10 & 322 \\ 
219 & J13293+114 & GJ 513 & 513 & 13 29 21.31 & +11 26 26.7 & M3.5 & 8.37 & 6 & 201 & 6 & 215 \\ 
220 & J13299+102 & BD+11 2576 & 514 & 13 29 59.79 & +10 22 37.8 & M0.5 & 5.90 & 447 & 2431 & 417 & 2425 \\ 
221 & J13300-087 & Ross 476 & 514.1 & 13 30 02.80 & -08 42 25.5 & M4.0 & 9.60 & 8 & 104 & 6 & 62 \\ 
222 & J13427+332 & Ross 1015 & 3801 & 13 42 43.27 & +33 17 24.3 & M3.5 & 7.79 & 19 & 433 & 15 & 509 \\ 
223 & J13450+176 & BD+18 2776 & 525 & 13 45 05.08 & +17 47 07.6 & M0.0 & 7.00 & 31 & 686 & 28 & 707 \\ 
224 & J13457+148 & HD 119850 & 526 & 13 45 43.78 & +14 53 29.5 & M1.5 & 5.18 & 250 & 1796 & 199 & 1785 \\ 
225 & J13458-179 & LP 798-034 & 3804 & 13 45 50.71 & -17 58 05.6 & M3.5 & 7.75 & 11 & 268 & 10 & 334 \\ 
\noalign{\smallskip}
\hline
\end{tabular}
\end{center}
\end{table}
\end{landscape}
\newpage
\begin{landscape}
\begin{table}
\ContinuedFloat
\caption{Continued.}
\begin{center}
\begin{tabular}{rlll cc cc rrrr}
\hline\hline
\noalign{\smallskip}
$\#$ & Karmn & Name & GJ & $\alpha$ & $\delta$ & Spectral type & $J$ [mag] & $\#$ VIS obs & S$/$N & $\#$ NIR obs & S$/$N \\
\noalign{\smallskip}
\hline
\noalign{\smallskip}
226 & J13536+776 & RX J1353.6+7737 &   & 13 53 38.79 & +77 37 08.2 & M4.0 & 8.63 & 25 & 331 & 24 & 440 \\ 
227 & J13582+125 & Ross 837 & 3817 & 13 58 13.92 & +12 34 43.9 & M3.0 & 8.27 & 10 & 282 & 8 & 265 \\ 
228 & J13591-198 & LP 799-007 & 3820 & 13 59 10.41 & -19 50 03.7 & M4.0 & 8.33 & 17 & 230 & 17 & 317 \\ 
229 & J14010-026 & HD 122303 & 536 & 14 01 03.19 & -02 39 17.5 & M1.0 & 6.52 & 27 & 490 & 25 & 602 \\ 
230 & J14082+805 & BD+81 465 & 540 & 14 08 12.98 & +80 35 50.2 & M1.0 & 7.18 & 37 & 661 & 32 & 638 \\ 
231 & J14152+450 & Ross 992 & 3836 & 14 15 16.98 & +45 00 53.3 & M3.0 & 8.01 & 9 & 297 & 8 & 327 \\ 
232 & J14173+454 & RX J1417.3+4525 &   & 14 17 22.10 & +45 25 46.0 & M5.0 & 9.47 & 12 & 154 & 9 & 209 \\ 
233 & J14251+518 & tet Boo B & 549B & 14 25 11.58 & +51 49 53.1 & M2.5 & 7.88 & 13 & 390 & 12 & 475 \\ 
234 & J14257+236E & BD+24 2733B & 548B & 14 25 46.65 & +23 37 13.7 & M0.5 & 6.89 & 49 & 810 & 41 & 774 \\ 
235 & J14257+236W & BD+24 2733A & 548A & 14 25 43.47 & +23 37 01.5 & M0.0 & 6.77 & 64 & 902 & 57 & 993 \\ 
236 & J14294+155 & Ross 130 & 552 & 14 29 29.70 & +15 31 57.5 & M2.0 & 7.23 & 6 & 234 & 5 & 258 \\ 
237 & J14307-086 & BD-07 3856 & 553 & 14 30 47.72 & -08 38 46.8 & M0.5 & 6.62 & 93 & 1101 & 85 & 1230 \\ 
238 & J14310-122 & Wolf 1478 & 553.1 & 14 31 01.16 & -12 17 46.0 & M3.5 & 7.80 & 6 & 218 & 5 & 232 \\ 
239 & J14321+081 & LP 560-035 &   & 14 32 08.51 & +08 11 31.2 & M6.0 & 10.11 & 58 & 169 & 58 & 259 \\ 
240 & J14342-125 & HN Lib & 555 & 14 34 16.81 & -12 31 10.4 & M4.0 & 6.84 & 98 & 974 & 84 & 1258 \\ 
241 & J14524+123 & G 66-37 & 3871 & 14 52 28.53 & +12 23 32.8 & M2.0 & 7.97 & 26 & 491 & 20 & 480 \\ 
242 & J14544+355 & Ross 1041 & 3873 & 14 54 27.91 & +35 32 57.0 & M3.5 & 8.24 & 33 & 544 & 32 & 688 \\ 
243 & J14578+566 & GJ 1187 & 1187 & 14 57 53.73 & +56 39 24.5 & M5.5 & 10.21 & 7 & 67 & 7 & 64 \\ 
244 & J15013+055 & G 15-2 & 3885 & 15 01 20.11 & +05 32 55.5 & M3.0 & 8.33 & 19 & 393 & 17 & 419 \\ 
245 & J15095+031 & Ross 1047 & 3892 & 15 09 35.59 & +03 10 00.6 & M3.0 & 7.72 & 17 & 409 & 16 & 408 \\ 
246 & J15100+193 & G 136-072 & 3893 & 15 10 04.82 & +19 21 27.5 & M4.0 & 9.06 & 10 & 183 & 10 & 146 \\ 
247 & J15194-077 & HO Lib & 581 & 15 19 26.83 & -07 43 20.2 & M3.0 & 6.71 & 52 & 794 & 45 & 954 \\ 
248 & J15218+209 & OT Ser & 9520 & 15 21 52.93 & +20 58 39.9 & M1.5 & 6.61 & 53 & 766 & 48 & 757 \\ 
249 & J15238+174 & Ross 508 & 585 & 15 23 51.14 & +17 27 57.4 & M4.5 & 9.11 & 6 & 142 & 6 & 126 \\ 
250 & J15305+094 & NLTT 40406 &   & 15 30 30.33 & +09 26 01.4 & M5.5 & 9.57 & 14 & 140 & 12 & 215 \\ 
251 & J15369-141 & Ross 802 & 592 & 15 36 58.62 & -14 08 01.8 & M4.0 & 8.43 & 9 & 193 & 5 & 188 \\ 
252 & J15499+796 & LP 022-420 &   & 15 49 55.14 & +79 39 51.6 & M5.0 & 9.72 & 15 & 126 & 14 & 206 \\ 
253 & J15583+354 & G 180-018 & 3929 & 15 58 18.80 & +35 24 24.3 & M3.5 & 8.69 & 82 & 568 & 74 & 585 \\ 
254 & J15598-082 & BD-07 4156 & 606 & 15 59 53.38 & -08 15 11.5 & M1.0 & 7.18 & 24 & 582 & 19 & 477 \\ 
255 & J16028+205 & GJ 609 & 609 & 16 02 50.94 & +20 35 21.0 & M4.0 & 8.13 & 26 & 458 & 24 & 527 \\ 
256 & J16092+093 & G 137-084 &   & 16 09 16.27 & +09 21 07.5 & M3.0 & 7.97 & 9 & 281 & 6 & 176 \\ 
257 & J16102-193 & K2-33 &   & 16 10 14.74 & -19 19 09.4 & M3.0 & 11.10 & 27 & 41 & 26 & 59 \\ 
258 & J16167+672N & EW Dra & 617B & 16 16 45.31 & +67 15 22.5 & M3.0 & 6.91 & 107 & 1060 & 103 & 1121 \\ 
259 & J16167+672S & HD 147379 & 617A & 16 16 42.75 & +67 14 19.8 & M0.0 & 5.78 & 188 & 1667 & 179 & 1704 \\ 
260 & J16254+543 & GJ 625 & 625 & 16 25 24.62 & +54 18 14.8 & M1.5 & 6.61 & 32 & 577 & 28 & 737 \\ 
261 & J16303-126 & V2306 Oph & 628 & 16 30 18.06 & -12 39 45.3 & M3.5 & 5.95 & 92 & 994 & 91 & 1204 \\ 
262 & J16313+408 & G 180-060 & 3959 & 16 31 18.79 & +40 51 51.7 & M5.0 & 9.46 & 13 & 144 & 11 & 200 \\ 
263 & J16327+126 & GJ 1203 & 1203 & 16 32 45.19 & +12 36 46.0 & M3.0 & 8.43 & 13 & 285 & 12 & 312 \\ 
264 & J16343+571 & CM Dra & 630.1A & 16 34 20.33 & +57 09 44.4 & M4.5 & 8.50 & 45 & 201 & 43 & 277 \\ 
265 & J16462+164 & LP 446-006 & 3972 & 16 46 13.74 & +16 28 40.8 & M2.5 & 7.95 & 17 & 443 & 16 & 442 \\ 
266 & J16554-083N & Wolf 629 & 643 & 16 55 25.22 & -08 19 21.3 & M3.5 & 7.55 & 31 & 540 & 27 & 675 \\ 
267 & J16555-083 & VB 8 & 644C & 16 55 35.26 & -08 23 40.8 & M7.0 & 9.78 & 128 & 241 & 110 & 398 \\ 
268 & J16570-043 & LP 686-027 & 1207 & 16 57 05.74 & -04 20 56.3 & M3.5 & 7.97 & 15 & 344 & 12 & 443 \\ 
269 & J16578+473 & HD 153557B & 649.1B & 16 57 53.61 & +47 22 02.4 & M1.5 & 6.87 & 9 & 440 & 8 & 467 \\ 
270 & J16581+257 & BD+25 3173 & 649 & 16 58 08.85 & +25 44 39.0 & M1.0 & 6.45 & 55 & 883 & 49 & 1017 \\ 
\noalign{\smallskip}
\hline
\end{tabular}
\end{center}
\end{table}
\end{landscape}
\newpage
\begin{landscape}
\begin{table}
\ContinuedFloat
\caption{Continued.}
\begin{center}
\begin{tabular}{rlll cc cc rrrr}
\hline\hline
\noalign{\smallskip}
$\#$ & Karmn & Name & GJ & $\alpha$ & $\delta$ & Spectral type & $J$ [mag] & $\#$ VIS obs & S$/$N & $\#$ NIR obs & S$/$N \\
\noalign{\smallskip}
\hline
\noalign{\smallskip}
271 & J17033+514 & G 203-042 & 3988 & 17 03 23.88 & +51 24 22.9 & M4.5 & 8.77 & 72 & 526 & 65 & 552 \\ 
272 & J17052-050 & Wolf 636 & 654 & 17 05 13.78 & -05 05 39.2 & M1.5 & 6.78 & 50 & 776 & 46 & 838 \\ 
273 & J17071+215 & Ross 863 & 655 & 17 07 07.49 & +21 33 14.5 & M3.0 & 7.88 & 15 & 413 & 15 & 422 \\ 
274 & J17115+384 & Wolf 654 & 3992 & 17 11 34.75 & +38 26 33.9 & M3.5 & 7.63 & 77 & 942 & 66 & 949 \\ 
275 & J17166+080 & GJ 2128 & 2128 & 17 16 40.98 & +08 03 30.2 & M2.0 & 7.93 & 19 & 457 & 17 & 469 \\ 
276 & J17198+417 & GJ 671 & 671 & 17 19 52.71 & +41 42 49.7 & M2.5 & 7.71 & 20 & 467 & 17 & 489 \\ 
277 & J17303+055 & BD+05 3409 & 678.1 & 17 30 22.73 & +05 32 54.7 & M0.0 & 6.24 & 55 & 821 & 57 & 987 \\ 
278 & J17338+169 & 1RXS J173353.5+165515 &   & 17 33 53.18 & +16 55 13.1 & M5.5 & 8.89 & 12 & 181 & 13 & 252 \\ 
279 & J17355+616 & BD+61 1678 & 685 & 17 35 34.48 & +61 40 53.6 & M0.5 & 6.88 & 26 & 609 & 24 & 729 \\ 
280 & J17364+683 & BD+68 946 & 687 & 17 36 25.90 & +68 20 20.9 & M3.0 & 5.33 & 41 & 631 & 36 & 645 \\ 
281 & J17378+185 & BD+18 3421 & 686 & 17 37 53.35 & +18 35 30.2 & M1.0 & 6.36 & 106 & 1241 & 104 & 1471 \\ 
282 & J17481+159 & 2MASS J17481125+1558465 &   & 17 48 11.25 & +15 58 46.7 & M3.0 & 9.61 & 30 & 216 & 29 & 176 \\ 
283 & J17542+073 & GJ 1222 & 1222 & 17 54 17.12 & +07 22 44.8 & M4.0 & 8.77 & 12 & 213 & 11 & 216 \\ 
284 & J17578+046 & Barnard's Star & 699 & 17 57 48.50 & +04 41 36.3 & M3.5 & 5.24 & 766 & 2920 & 708 & 3270 \\ 
285 & J17578+465 & G 204-039 & 4040 & 17 57 50.97 & +46 35 19.1 & M2.5 & 7.85 & 26 & 541 & 22 & 578 \\ 
286 & J18012+355 & G 182-034 &   & 18 01 16.10 & +35 35 50.5 & M3.5 & 9.70 & 35 & 238 & 33 & 168 \\ 
287 & J18022+642 & LP 071-082 &   & 18 02 16.60 & +64 15 44.2 & M5.0 & 8.54 & 28 & 344 & 26 & 487 \\ 
288 & J18027+375 & GJ 1223 & 1223 & 18 02 46.26 & +37 31 03.0 & M5.0 & 9.72 & 122 & 395 & 114 & 406 \\ 
289 & J18051-030 & HD 165222 & 701 & 18 05 07.58 & -03 01 52.7 & M1.0 & 6.16 & 56 & 839 & 57 & 1013 \\ 
290 & J18075-159 & GJ 1224 & 1224 & 18 07 32.84 & -15 57 47.1 & M4.5 & 8.64 & 15 & 239 & 15 & 283 \\ 
291 & J18131+260 & LP 390-16 & 4044 & 18 13 06.57 & +26 01 51.9 & M4.0 & 8.90 & 16 & 243 & 12 & 254 \\ 
292 & J18165+048 & G 140-51 &   & 18 16 31.54 & +04 52 45.8 & M5.0 & 9.80 & 52 & 262 & 46 & 345 \\ 
293 & J18174+483 & TYC 3529-1437-1 &   & 18 17 25.13 & +48 22 02.3 & M2.0 & 7.77 & 72 & 849 & 65 & 1061 \\ 
294 & J18180+387E & G 204-58 & 4048A & 18 18 04.23 & +38 46 32.6 & M3.0 & 8.04 & 18 & 366 & 17 & 378 \\ 
295 & J18189+661 & LP 71-165 & 4053 & 18 18 57.23 & +66 11 33.3 & M4.5 & 8.74 & 13 & 242 & 12 & 344 \\ 
296 & J18198-019 & HD 168442 & 710 & 18 19 50.84 & -01 56 19.0 & K7.0 & 7.08 & 147 & 1596 & 141 & 1708 \\ 
297 & J18221+063 & Ross 136 & 712 & 18 22 06.68 & +06 20 37.6 & M4.0 & 8.67 & 17 & 296 & 15 & 298 \\ 
298 & J18224+620 & GJ 1227 & 1227 & 18 22 27.09 & +62 03 01.7 & M4.0 & 8.64 & 61 & 530 & 58 & 669 \\ 
299 & J18319+406 & G 205-028 & 4062 & 18 31 58.38 & +40 41 11.0 & M3.5 & 8.06 & 20 & 438 & 17 & 460 \\ 
300 & J18346+401 & LP 229-017 & 4063 & 18 34 36.65 & +40 07 26.4 & M3.5 & 7.18 & 79 & 960 & 73 & 1162 \\ 
301 & J18353+457 & BD+45 2743 & 720A & 18 35 18.39 & +45 44 38.5 & M0.5 & 6.88 & 16 & 484 & 16 & 579 \\ 
302 & J18356+329 & LSR J1835+3259 &   & 18 35 37.88 & +32 59 53.3 & M8.5 & 10.27 & 59 & 87 & 63 & 202 \\ 
303 & J18363+136 & Ross 149 & 4065 & 18 36 19.23 & +13 36 26.4 & M4.0 & 8.19 & 26 & 418 & 22 & 457 \\ 
304 & J18409-133 & BD-13 5069 & 724 & 18 40 57.31 & -13 22 46.6 & M1.0 & 7.40 & 84 & 1006 & 81 & 814 \\ 
305 & J18419+318 & Ross 145 & 4070 & 18 41 59.04 & +31 49 49.8 & M3.0 & 7.52 & 24 & 505 & 22 & 484 \\ 
306 & J18427+596N & HD 173739 & 725A & 18 42 46.70 & +59 37 49.4 & M3.0 & 5.19 & 71 & 928 & 70 & 1006 \\ 
307 & J18427+596S & HD 173740 & 725B & 18 42 46.89 & +59 37 36.7 & M3.5 & 5.72 & 77 & 966 & 72 & 933 \\ 
308 & J18480-145 & G 155-042 & 4077 & 18 48 01.27 & -14 34 51.2 & M2.5 & 8.38 & 21 & 359 & 19 & 427 \\ 
309 & J18482+076 & G 141-036 &   & 18 48 17.54 & +07 41 21.2 & M5.0 & 8.85 & 54 & 423 & 46 & 518 \\ 
310 & J18498-238 & V1216 Sgr & 729 & 18 49 49.36 & -23 50 10.4 & M3.5 & 6.22 & 56 & 788 & 51 & 817 \\ 
311 & J18580+059 & BD+05 3993 & 740 & 18 58 00.14 & +05 54 29.2 & M0.5 & 6.24 & 34 & 608 & 30 & 673 \\ 
312 & J19025+754 & LSPM J1902+7525 &   & 19 02 31.93 & +75 25 07.0 & M2.5 & 9.80 & 46 & 261 & 44 & 244 \\ 
313 & J19070+208 & Ross 730 & 745A & 19 07 05.56 & +20 53 16.9 & M2.0 & 7.29 & 38 & 703 & 33 & 681 \\ 
314 & J19072+208 & HD 349726 & 745B & 19 07 13.20 & +20 52 37.3 & M2.0 & 7.28 & 44 & 770 & 39 & 810 \\ 
315 & J19084+322 & G 207-019 & 4098 & 19 08 29.93 & +32 16 51.6 & M3.0 & 7.91 & 24 & 484 & 23 & 529 \\ 
\noalign{\smallskip}
\hline
\end{tabular}
\end{center}
\end{table}
\end{landscape}
\newpage
\begin{landscape}
\begin{table}
\ContinuedFloat
\caption{Continued.}
\begin{center}
\begin{tabular}{rlll cc cc rrrr}
\hline\hline
\noalign{\smallskip}
$\#$ & Karmn & Name & GJ & $\alpha$ & $\delta$ & Spectral type & $J$ [mag] & $\#$ VIS obs & S$/$N & $\#$ NIR obs & S$/$N \\
\noalign{\smallskip}
\hline
\noalign{\smallskip}
316 & J19098+176 & GJ 1232 & 1232 & 19 09 50.87 & +17 40 06.4 & M4.5 & 8.82 & 23 & 313 & 17 & 342 \\ 
317 & J19169+051N & V1428 Aql & 752A & 19 16 55.26 & +05 10 08.0 & M2.5 & 5.58 & 129 & 1304 & 114 & 1278 \\ 
318 & J19169+051S & V1298 Aql & 752B & 19 16 57.61 & +05 09 01.6 & M8.0 & 9.91 & 51 & 142 & 44 & 241 \\ 
319 & J19206+731S & 2MASS J19204172+7311434 &   & 19 20 41.73 & +73 11 43.5 & M4.5 & 10.60 & 22 & 117 & 23 & 132 \\ 
320 & J19216+208 & GJ 1235 & 1235 & 19 21 38.70 & +20 52 03.3 & M4.5 & 8.80 & 25 & 332 & 23 & 374 \\ 
321 & J19242+755 & GJ 1238 & 1238 & 19 24 16.31 & +75 33 11.8 & M5.5 & 9.91 & 222 & 356 & 218 & 420 \\ 
322 & J19251+283 & Ross 164 & 4109 & 19 25 08.48 & +28 21 13.8 & M3.0 & 8.44 & 26 & 439 & 25 & 492 \\ 
323 & J19255+096 & LSPM J1925+0938 &   & 19 25 30.91 & +09 38 23.3 & M8.0 & 11.21 & 100 & 67 & 97 & 168 \\ 
324 & J19346+045 & BD+04 4157 & 763 & 19 34 39.84 & +04 34 57.0 & M0.0 & 6.71 & 53 & 836 & 49 & 981 \\ 
325 & J19422-207 & 2MASS J19421282-2045477 &   & 19 42 12.82 & -20 45 48.0 & M5.1 & 9.60 & 27 & 170 & 24 & 182 \\ 
326 & J19511+464 & G 208-042 & 1243 & 19 51 09.32 & +46 29 00.2 & M4.0 & 8.59 & 14 & 272 & 9 & 306 \\ 
327 & J19573-125 & HD 188807 B & 773B & 19 57 23.80 & -12 33 50.2 & M5.0 & 10.21 & 8 & 36 & 8 & 50 \\ 
328 & J20093-012 & 2MASS J20091824-0113377 &   & 20 09 18.25 & -01 13 38.3 & M5.0 & 9.40 & 13 & 172 & 13 & 238 \\ 
329 & J20109+708 & TYC 4450-1440-1 &   & 20 10 57.12 & +70 52 09.8 & K5.0 & 9.71 & 78 & 400 & 68 & 348 \\ 
330 & J20227+473 & Ross 176 &   & 20 22 45.19 & +47 18 27.8 & K5.0 & 9.24 & 102 & 598 & 84 & 435 \\ 
331 & J20260+585 & Wolf 1069 & 1253 & 20 26 05.30 & +58 34 22.7 & M5.0 & 9.03 & 269 & 901 & 248 & 1120 \\ 
332 & J20305+654 & GJ 793 & 793 & 20 30 32.05 & +65 26 58.4 & M2.5 & 6.74 & 53 & 756 & 47 & 948 \\ 
333 & J20336+617 & GJ 1254 & 1254 & 20 33 40.32 & +61 45 13.6 & M4.0 & 8.29 & 52 & 581 & 45 & 757 \\ 
334 & J20405+154 & GJ 1256 & 1256 & 20 40 33.86 & +15 29 58.7 & M4.5 & 8.64 & 26 & 358 & 22 & 358 \\ 
335 & J20450+444 & BD+44 3567 & 806 & 20 45 04.10 & +44 29 56.6 & M1.5 & 7.33 & 97 & 1084 & 91 & 1150 \\ 
336 & J20451-313 & AU Mic & 803 & 20 45 09.53 & -31 20 27.2 & M0.5 & 5.44 & 98 & 896 & 94 & 570 \\ 
337 & J20525-169 & LP 816-060 &   & 20 52 33.02 & -16 58 29.0 & M4.0 & 7.09 & 45 & 691 & 41 & 789 \\ 
338 & J20533+621 & HD 199305 & 809 & 20 53 19.79 & +62 09 15.8 & M1.0 & 5.43 & 160 & 1344 & 143 & 1511 \\ 
339 & J20556-140S & GJ 810 B & 810B & 20 55 37.12 & -14 03 54.9 & M5.0 & 9.72 & 54 & 256 & 50 & 339 \\ 
340 & J20567-104 & Wolf 896 & 811.1 & 20 56 46.60 & -10 26 54.7 & M2.5 & 7.77 & 19 & 496 & 19 & 537 \\ 
341 & J21019-063 & Wolf 906 & 816 & 21 01 58.64 & -06 19 07.5 & M2.5 & 7.56 & 67 & 885 & 65 & 903 \\ 
342 & J21152+257 & LP 397-041 & 4184 & 21 15 12.60 & +25 47 45.5 & M3.0 & 8.40 & 22 & 423 & 22 & 510 \\ 
343 & J21164+025 & LSPM J2116+0234 &   & 21 16 27.28 & +02 34 51.4 & M3.0 & 8.22 & 85 & 879 & 81 & 1124 \\ 
344 & J21221+229 & TYC 2187-512-1 &   & 21 22 06.28 & +22 55 53.1 & M1.0 & 7.40 & 94 & 1201 & 89 & 1387 \\ 
345 & J21348+515 & Wolf 926 & 4205 & 21 34 50.34 & +51 32 13.6 & M3.0 & 8.04 & 71 & 904 & 66 & 1129 \\ 
346 & J21463+382 & LSPM J2146+3813 &   & 21 46 22.07 & +38 13 05.0 & M4.0 & 7.95 & 48 & 668 & 42 & 779 \\ 
347 & J21466+668 & G 264-012 &   & 21 46 40.24 & +66 48 10.6 & M4.0 & 8.84 & 162 & 853 & 153 & 1011 \\ 
348 & J21466-001 & Wolf 940 & 1263A & 21 46 40.42 & -00 10 23.8 & M4.0 & 8.36 & 22 & 426 & 22 & 534 \\ 
349 & J21474+627 & TYC 4266-736-1 &   & 21 47 24.79 & +62 45 13.9 & M0.0 & 8.77 & 60 & 554 & 54 & 531 \\ 
350 & J22012+283 & V374 Peg & 4247 & 22 01 13.12 & +28 18 24.9 & M4.0 & 7.63 & 13 & 312 & 12 & 364 \\ 
351 & J22020-194 & LP 819-017 & 843 & 22 02 00.79 & -19 28 59.2 & M3.5 & 8.05 & 10 & 274 & 9 & 307 \\ 
352 & J22021+014 & BD+00 4810 & 846 & 22 02 10.28 & +01 24 00.8 & M0.5 & 6.20 & 77 & 1004 & 66 & 1055 \\ 
353 & J22057+656 & G 264-18 A & 4258 & 22 05 45.36 & +65 38 55.5 & M1.5 & 8.42 & 92 & 882 & 87 & 1103 \\ 
354 & J22096-046 & BD-05 5715 & 849 & 22 09 40.34 & -04 38 26.7 & M3.5 & 6.51 & 61 & 826 & 58 & 990 \\ 
355 & J22102+587 & UCAC4 744-073158 &   & 22 10 15.14 & +58 42 22.2 & M2.0 & 9.86 & 117 & 399 & 110 & 352 \\ 
356 & J22114+409 & 1RXS J221124.3+410000 &   & 22 11 24.16 & +40 59 58.7 & M5.5 & 9.72 & 57 & 231 & 55 & 346 \\ 
357 & J22115+184 & Ross 271 & 851 & 22 11 30.09 & +18 25 34.3 & M2.0 & 6.72 & 68 & 988 & 61 & 1100 \\ 
358 & J22125+085 & Wolf 1014 & 9773 & 22 12 35.94 & +08 33 11.6 & M3.0 & 8.28 & 117 & 984 & 107 & 1173 \\ 
359 & J22137-176 & LP 819-052 & 1265 & 22 13 42.86 & -17 41 08.7 & M4.5 & 8.96 & 89 & 490 & 87 & 714 \\ 
360 & J22231-176 & LP 820-012 & 4274 & 22 23 07.00 & -17 36 26.3 & M4.5 & 8.24 & 13 & 225 & 10 & 275 \\ 
\noalign{\smallskip}
\hline
\end{tabular}
\end{center}
\end{table}
\end{landscape}
\newpage
\begin{landscape}
\begin{table}
\ContinuedFloat
\caption{Continued.}
\begin{center}
\begin{tabular}{rlll cc cc rrrr}
\hline\hline
\noalign{\smallskip}
$\#$ & Karmn & Name & GJ & $\alpha$ & $\delta$ & Spectral type & $J$ [mag] & $\#$ VIS obs & S$/$N & $\#$ NIR obs & S$/$N \\
\noalign{\smallskip}
\hline
\noalign{\smallskip}
361 & J22252+594 & G 232-070 & 4276 & 22 25 17.07 & +59 24 49.8 & M4.0 & 8.74 & 103 & 706 & 96 & 951 \\ 
362 & J22298+414 & G 215-050 & 1270 & 22 29 48.99 & +41 28 48.6 & M4.0 & 8.85 & 25 & 343 & 18 & 370 \\ 
363 & J22330+093 & BD+08 4887 & 863 & 22 33 02.23 & +09 22 40.7 & M1.0 & 7.21 & 82 & 1151 & 75 & 1045 \\ 
364 & J22468+443 & EV Lac & 873 & 22 46 49.73 & +44 20 02.4 & M3.5 & 6.11 & 107 & 1045 & 97 & 1197 \\ 
365 & J22503-070 & BD-07 5871 & 875 & 22 50 19.42 & -07 05 24.4 & M0.5 & 6.93 & 54 & 889 & 43 & 757 \\ 
366 & J22518+317 & GT Peg & 875.1 & 22 51 53.54 & +31 45 15.2 & M3.0 & 7.70 & 12 & 292 & 10 & 345 \\ 
367 & J22526+750 & NLTT 55174 &   & 22 52 39.70 & +75 04 18.8 & M4.5 & 9.09 & 10 & 153 & 11 & 178 \\ 
368 & J22532-142 & IL Aqr & 876 & 22 53 16.73 & -14 15 49.3 & M4.0 & 5.93 & 70 & 763 & 62 & 927 \\ 
369 & J22559+178 & StKM 1-2065 & 4306 & 22 55 59.85 & +17 48 39.8 & M1.0 & 7.32 & 11 & 387 & 10 & 388 \\ 
370 & J22565+165 & HD 216899 & 880 & 22 56 34.80 & +16 33 12.4 & M1.5 & 5.36 & 697 & 3057 & 631 & 3220 \\ 
371 & J23064-050 & 2MUCD 12171 &   & 23 06 29.37 & -05 02 29.0 & M8.0 & 11.35 & 163 & 55 & 130 & 72 \\ 
372 & J23113+085 & NLTT 56083 &   & 23 11 23.79 & +08 31 01.4 & M3.5 & 8.47 & 98 & 722 & 89 & 857 \\ 
373 & J23216+172 & LP 462-027 & 4333 & 23 21 37.45 & +17 17 25.4 & M4.0 & 7.39 & 65 & 892 & 61 & 1104 \\ 
374 & J23245+578 & BD+57 2735 & 895 & 23 24 30.51 & +57 51 15.5 & M1.0 & 6.79 & 61 & 928 & 53 & 1012 \\ 
375 & J23340+001 & Wolf 1039 & 899 & 23 34 03.33 & +00 10 45.9 & M2.5 & 7.66 & 38 & 743 & 41 & 920 \\ 
376 & J23351-023 & GJ 1286 & 1286 & 23 35 10.46 & -02 23 20.6 & M5.5 & 9.15 & 72 & 401 & 67 & 454 \\ 
377 & J23381-162 & G 273-093 & 4352 & 23 38 08.16 & -16 14 10.2 & M2.0 & 7.81 & 55 & 773 & 51 & 894 \\ 
378 & J23419+441 & HH And & 905 & 23 41 55.04 & +44 10 38.8 & M5.0 & 6.88 & 99 & 903 & 82 & 1255 \\ 
379 & J23431+365 & GJ 1289 & 1289 & 23 43 06.31 & +36 32 13.1 & M4.0 & 8.11 & 28 & 469 & 25 & 595 \\ 
380 & J23492+024 & BR Psc & 908 & 23 49 12.53 & +02 24 04.4 & M1.0 & 5.83 & 453 & 2606 & 416 & 2512 \\ 
381 & J23505-095 & LP 763-012 & 4367 & 23 50 31.64 & -09 33 32.7 & M4.0 & 8.94 & 72 & 509 & 64 & 598 \\ 
382 & J23548+385 & RX J2354.8+3831 &   & 23 54 51.46 & +38 31 36.2 & M4.0 & 8.94 & 13 & 246 & 10 & 301 \\ 
\noalign{\smallskip}
\hline
\end{tabular}
\end{center}
\tablefoot{
\tablefoottext{a}{In the last four columns we show the number of spectra used to build the
templates as well as the S/N per pixel in order 70 ($\sim 8600$\,\AA) for the VIS and in order 52 ($\sim 10\,500$\,\AA) for the NIR. 
Equatorial coordinates, spectral types (class V), and $J$ magnitudes were gathered
from the CARMENES Cool dwarf Information and daTa Archive (Carmencita, see \citealt{Alonso-Floriano2015}, \citealt{Caballero2016b}, and references therein).}
}

\end{table}
\end{landscape}

\end{small}

\end{appendix}
\end{document}